\documentclass[a4paper,11pt]{article}
\usepackage{heppub}
\pdfoutput=1
\graphicspath{{plots/}}

\newcommand{\abs}[1]{\lvert#1\rvert}
\newcommand{\ABS}[1]{\biggl\lvert#1\biggr\rvert}
\newcommand{\ord}[1]{\mathcal{O}(#1)}

\newcommand{\ORd}[1]{\mathcal{O}\Bigl(#1\Bigr)}

\newcommand{\df}{\mathrm{d}}
\newcommand{\img}{\mathrm{i}}

\newcommand{\eps}{\varepsilon}

\newcommand{\GeV}{\,\mathrm{GeV}}
\newcommand{\TeV}{\,\mathrm{TeV}}

\newcommand{\nn}{\nonumber}

\newcommand{\bt}{{\vec b}_T}
\newcommand{\qt}{{\vec q}_T}

\newcommand{\tB}{\tilde{B}}
\newcommand{\tS}{\tilde{S}}

\newcommand{\tf}{\tilde{f}}
\newcommand{\talpha}{\tilde{\alpha}}

\newcommand{\as}{\alpha_s}
\newcommand{\aem}{\alpha_\mathrm{em}}
\newcommand{\Ecm}{E_\mathrm{cm}}
\newcommand{\lqcd}{\Lambda_\mathrm{QCD}}
\newcommand{\MSbar}{$\overline{\text{MS}}$}
\newcommand{\thetani}{\theta_{n,i}}
\newcommand{\f}{f} % matrix-element constants

\newcommand{\zero}{{(0)}}
\newcommand{\cusp}{\mathrm{cusp}}

\newcommand{\WidthTwoSubfigs}{0.49\textwidth}

\title{Beyond Scale Variations: Perturbative Theory Uncertainties from Nuisance Parameters}

\author{Frank J. Tackmann}

\affiliation{Deutsches Elektronen-Synchrotron DESY, Notkestr. 85, 22607 Hamburg, Germany}

\emailAdd{frank.tackmann@desy.de}

\abstract{%
We develop a new approach to estimate the uncertainty due to missing higher
orders in perturbative predictions (the perturbative ``theory uncertainty''),
which overcomes many inherent limitations of the currently prevalent methods
based on varying unphysical renormalization scales. In our approach, the true
underlying sources of the theory uncertainty, namely the missing higher-order
terms, are identified and parameterized in terms of mutually independent theory
nuisance parameters (TNPs). The TNPs are true parameters of the calculation,
i.e., they have a well-defined true value that is not or only imprecisely known.
This approach affords the theory uncertainty all benefits of a truly parametric
uncertainty: It provides correct correlations and allows for consistent error
propagation and combination. Furthermore, the TNPs can be profiled in fits,
allowing the data to reduce the theory uncertainties. On the theory side, it
allows maximally exploiting all available higher-order information to reduce the
theory uncertainty, such as partial higher-order results or any nontrivial
knowledge of the higher-order or all-order structure.

We first discuss the method in general as it can be applied across the board of
perturbative calculations. As a concrete application, we then discuss the
resummed transverse momentum ($q_T$) spectrum in Drell-Yan production, and how
TNP-based uncertainties can correctly capture the correlations across the $q_T$
spectrum and between $Z$ and $W$ production. This application is the basis of
the theory model enabling the recent precise measurement of the $W$-boson mass
by the CMS experiment. In a forthcoming paper, we use it to study the theory
uncertainties in extracting the strong coupling constant $\alpha_s$ from the $Z$
$q_T$ spectrum.
}
\date{November 27, 2024}

\preprint{\vbox{%
\hbox{DESY-19-021}}
}

\begin{document}

\maketitle

%%%%%%%%%%%%%%%%%%%%%%%%%%%%%%%%%%%%%%%%%%%%%%%%%%%%%%%%%%%%%%%%%%%%%%%%%%%%%%%%
\section{Introduction}
\label{sec:intro}
%%%%%%%%%%%%%%%%%%%%%%%%%%%%%%%%%%%%%%%%%%%%%%%%%%%%%%%%%%%%%%%%%%%%%%%%%%%%%%%%

The interpretation of precision measurements requires equally precise theoretical predictions.
Just as for experimental measurements, theoretical predictions are ultimately only as
useful as their uncertainties are meaningful. We are specifically interested in theory predictions
based on perturbation theory and their uncertainty due to missing higher-order corrections,
which we will refer to as the perturbative ``theory uncertainty''.
For a theory uncertainty to be meaningful it must realistically reflect our degree of knowledge.
This not only means that it has a realistic size but also that it provides correct correlations
and allows for some form of statistical interpretation.

The prevalent
traditional approach for estimating perturbative theory uncertainties based on
scale variations is neither particularly reliable nor does it provide correlations
let alone a meaningful statistical interpretation. These limitations are in principle well known.
They have been a long-standing bottleneck in our ability to interpret experimental
measurements using theoretical predictions, which is only becoming more severe as experimental
measurements become ever more precise. The approach put forward in this paper allows us to
address this issue by equipping perturbative predictions with meaningful theory uncertainties.

When designing measurement and interpretation strategies we optimize
the total uncertainty budget, and the theory
uncertainty is part of this budget. Currently, this optimization often gets skewed toward
reducing as much as possible the impact of unreliable theory uncertainties.
This inevitably leads to sacrificing experimental
precision. Reliable, meaningful theory uncertainties make such sacrifices
unnecessary and thus allow improving the overall uncertainty budget beyond just
the immediate effect of improving the theory prediction itself. They can also
enable entirely new measurement strategies that would otherwise be
unfeasible. An example is the recent precision measurement of the $W$-boson mass
by CMS~\cite{CMS:2024lrd}.
Thus, meaningful theory uncertainties greatly facilitate our ability to
fully exploit the potential of existing and future precision measurements.

The above requirements for a meaningful theory uncertainty are elaborated on
in \sec{philosophy}. The key points are: First, the theory uncertainty is a property
of the current prediction that should reflect its intrinsic precision.
In particular, it is \emph{not} meant or defined to be the unknown difference to the all-order
result (or some formally more accurate result standing in for the all-order one).
Second, ``theory correlations'' -- the correlations in the theory uncertainties
of different predictions -- are required as soon as more than a single theory prediction
is used at a time.
An important example is considering a differential spectrum, as
each of its bins has a priori its own theory prediction. The bin-by-bin correlations
are essential when one is interested in shape effects, since a shape uncertainty
is basically a statement about how the uncertainties at different points in
the spectrum are correlated. Theory correlations are thus critical if one wants
to distinguish the shape effect induced
by a to-be-determined parameter of interest from that caused by theory
uncertainties.
Third, a theory prediction simply cannot be used for interpreting
experimental measurements without any quantitative statistical meaning for its uncertainty.

The limitations of scale variations are discussed
in more detail in \sec{scale_var}. In short, their lack of correlations basically stems
from the fact that their variation cannot be interpreted like that of a normal
parameter whose uncertainty is being propagated.
Hence, they are notoriously unreliable
for estimating shape uncertainties, which unfortunately is exactly what they are often
used for (primarily due to the lack of alternatives).
This is becoming a severe limitation in many precision studies.
Presently, to be on the safe side we like to avoid attaching any statistical meaning to
theory uncertainties derived from scale variations.
However, this is not helpful at all. It just skirts the issue and puts the burden
on the users of our predictions since they are now forced to attach some ad hoc
quantitative statistical meaning to them.
This state of affairs is clearly unsatisfactory and frankly speaking
rather embarrassing.

Some frequentist statistical models attached to theory uncertainties are discussed
for example in \refscite{Charles:2016qtt, Cowan:2018lhq}.
There have been various proposals to obtain theory uncertainty estimates with
a more meaningful statistical interpretation via a Bayesian
model~\cite{Cacciari:2011ze, Bagnaschi:2014wea, Bonvini:2020xeo, Duhr:2021mfd},
or series acceleration~\cite{David:2013gaa}, or based on a set of reference
processes~\cite{Ghosh:2022lrf}.
These methods go in the right direction by trying to more directly
estimate the size of missing higher-order corrections and by more explicitly
exposing the assumptions made.
However, like scale variations they base the uncertainty estimate on the known
lower-order terms without parameterizing the actual underlying source of uncertainty
and as a result share many of the limitations of scale variations. They have a
similar level of arbitrariness and reliability, and in particular they also lack
theory correlations.

A theory uncertainty is a form of systematic (epistemic) uncertainty and as such we cannot
hope to render it as robust as a purely statistical (aleatoric) uncertainty. However, the same
requirements to be meaningful are shared by experimental systematic uncertainties.
Our approach thus treats theory uncertainties
following the same
logic that is routinely applied for experimental systematic uncertainties
to cast them into parametric uncertainties. This is the key to render
them meaningful and is discussed
in \sec{tnp_method} and \sec{tnps}.
In a nutshell, we identify the underlying sources of
uncertainty, namely the relevant missing perturbative ingredients, and
parameterize them in terms of unknown parameters, which are the ``theory
nuisance parameters'' (TNPs). Predictions
for different cross sections that depend on the same perturbative ingredient will
share a common TNP and the associated uncertainty will be
100\% correlated among them. The TNPs have true values,
which can in principle be determined from a higher-order calculation, but which
are a priori unknown (or treated as such).
Without explicit knowledge of their true value, we can use auxiliary
information at our disposal to constrain their allowed ranges.
The TNPs are then explicitly varied
or floated in fits within their allowed ranges to account for the theory
uncertainties and propagate them with correct correlations to subsequent interpretations.

Whilst constraining the TNPs based on auxiliary theoretical information
necessarily involves making some educated choices, this can
be thought of as an imagined auxiliary measurement. Furthermore, depending on the context,
such theoretical constraints can be supplemented or even
replaced by constraining the TNPs with real auxiliary measurements or in situ in the interpretation
of the nominal measurement itself. As a result,
the TNP-based theory uncertainties admit an analogous statistical treatment and
interpretation as experimental systematics based on nuisance parameters
constrained by (real or imagined) auxiliary control measurements
(see e.g.~\refscite{Cowan:2010js, Cousins:2024gkj}).
Finally, even if individual TNPs may not necessarily have a precisely known
probability distribution, since the total theory uncertainty will typically arise
as the combination of a number of TNPs, the central-limit theorem ensures that it will be
asymptotically Gaussian distributed.

Another key advantage of our approach is that it overcomes the
paradigm of only being able to systematically improve theory predictions in large
discrete steps based on completely known formal orders. The desire
to utilize available higher-order information for actual phenomenological benefit,
i.e.\ to reduce theory uncertainties, without having to wait until the formally
complete next order eventually becomes available is more than obvious. In fact,
likely sooner than later this is going to become an actual requirement for making progress,
because as we push to higher and higher orders, formally complete orders for final
predictions are increasingly difficult to achieve and might eventually become out of reach.
For this reason, more and more predictions are appearing
that are formally ``approximate'' in some way ranging from very unjustified
to very well justified.
The underlying issue is of course that at present we lack meaningful theory uncertainties,
so the primary guiding principle are formally complete orders.

We believe that in the long run an essential benefit of our approach will be to
allow our community to break out of this rigid paradigm. Meaningful theory
uncertainties are by construction a much better judge of the actual precision
than the formal accuracy. Our approach naturally allows for predictions that are
formally incomplete in a fully justified, systematic, and formally consistent
manner. It ultimately allows for an (almost) continuous integration of newly
available higher-order results into final theory predictions, taking full and
immediate advantage of them for reducing theory uncertainties and thereby for
maximal and immediate phenomenological impact. Moreover, our approach makes it
very transparent which missing perturbative ingredients are causing the largest
uncertainties at any given stage, allowing one to anticipate already beforehand
the impact of explicitly calculating a certain higher-order ingredient. This can
greatly help to guide efforts and to provide clear and tangible justification
for allocating resources.

The approach of this paper was first advocated in \refcite{TNPtalkSCET}, and has
already been used since in several instances~\cite{McGowan:2022nag, Dehnadi:2022prz,
Cal:2024yjz}. In these cases, the TNPs serve to estimate the uncertainty due to
still missing ingredients at the nominal, approximate working order.
The application of our approach to the resummed transverse momentum ($q_T$) spectrum
of $W$ and $Z$ bosons produced in hadronic collisions as discussed in \sec{qT}
forms the basis of the theoretical
modelling that has enabled a high-precision measurement of the $W$-boson
mass by the CMS experiment~\cite{CMS:2024lrd}. In a forthcoming
paper~\cite{TNPalphas}, we use it to study the expected theory uncertainties in the extraction of the
strong coupling constant $\alpha_s$ from the $Z$ $p_T$ spectrum.
A promising first application of our approach to a variety of fixed-order
single-differential distributions has been carried out in \refcite{Lim:2024nsk}.

At a basic level, it is of course not a new idea to estimate a missing
higher-order coefficient and the uncertainty caused by it.
For example, in the past resummed calculations at N$^3$LL and beyond
(see e.g.~\refscite{Moch:2005ba, Becher:2008cf, Abbate:2010xh, Becher:2013xia,
Bonvini:2014joa, Hoang:2014wka})
have used Pad{\'e} approximations for varying the 4-loop cusp anomalous
dimension and other 3-loop ingredients missing at the time.
In high-precision QED and electroweak calculations, scale variations
are less prevalent than for QCD calculations, and theory uncertainties are more
commonly estimated by explicit, more-or-less ad hoc estimates of the expected size of
missing higher-order terms (see e.g.~\refcite{Mohr:2012tt}) including attempts
to constrain them from measurements (see e.g.~\refcite{Sturm:2014bla}).
The methods of \refscite{Cacciari:2011ze, David:2013gaa, Bagnaschi:2014wea, Bonvini:2020xeo, Duhr:2021mfd}
amount to modelling the size of missing terms based on the size of the
known terms.

While the main focus of our discussion is on perturbative predictions in QCD,
our approach in principle applies to any other systematic, truncated expansion
and its truncation uncertainty, such as the power
expansions performed in effective field theories.
For example, a similar strategy can be followed to account for the truncation
uncertainty in the SMEFT, see e.g.~\refscite{Berthier:2015gja, Alte:2017pme, Trott:2021vqa}.

This paper is organized as follows. As already mentioned, in \sec{pert_theory_unc}
we discuss general aspects of perturbative theory uncertainties.
\Sec{tnps} gives a general discussion of the approach of theory nuisance
parameters and is intended for all audiences. \Sec{tnp_method_continued}
gives a general overview  of the approach, while the remaining subsections
discuss several specific aspects. Readers interested in an executive 5-page
summary of our approach can just read \secs{tnp_method}{tnp_method_continued}.
\Sec{parameterization} provides a guide for how to derive suitable
parameterizations in terms of TNPs.
It is intended for readers who wish to implement TNP-based
uncertainties into their predictions, providing
principles and strategies to follow as well as several examples for illustration.
In \sec{tnp_scalar} we then focus on TNPs for scalar series in QCD and discuss how we
can obtain robust theory constraints on them based on our theoretical knowledge
and available information from existing higher-order calculations.
In \sec{qT}, we present an explicit full-fledged example application of our
approach for the case of $q_T$ resummation for $pp\to Z/\gamma^*$
and $pp\to W$ production. We conclude in \sec{conclusions}.

%%%%%%%%%%%%%%%%%%%%%%%%%%%%%%%%%%%%%%%%%%%%%%%%%%%%%%%%%%%%%%%%%%%%%%%%%%%%%%%%
\section{Perturbative Theory Uncertainties}
\label{sec:pert_theory_unc}
%%%%%%%%%%%%%%%%%%%%%%%%%%%%%%%%%%%%%%%%%%%%%%%%%%%%%%%%%%%%%%%%%%%%%%%%%%%%%%%%

In \sec{philosophy} we elaborate on the criteria for meaningful theory
uncertainties. Readers who find them self-evident or are happy to accept them
can skip this subsection.
In \sec{tnp_method} we derive our basic approach to estimate theory uncertainties
as parametric uncertainties.
In \sec{scale_var} we discuss the limitations of scale variations and why
uncertainties derived from them cannot be regarded as parametric uncertainties.

%===============================================================================
\subsection{Philosophy}
\label{sec:philosophy}
%===============================================================================

As mentioned in the introduction, for a theory (or really any) uncertainty to be
meaningful, it must
\begin{enumerate}
   \item have a size that reflects our level of knowledge,
   \item provide correct correlations, and
   \item allow for some form of statistical interpretation.
\end{enumerate}
Before elaborating further on these criteria, we stress that despite the title
of this subsection, having meaningful theory uncertainties is not just a
philosophical or academic issue -- quite the opposite. As discussed in the
introduction, it has important implications for interpreting experimental
measurements.

%~~~~~~~~~~~~~~~~~~~~~~~~~~~~~~~~~~~~~~~~~~~~~~~~~~~~~~~~~~~~~~~~~~~~~~~~~~~~~~~
\subsubsection{Size and statistical interpretation}
\label{sec:philosophy_size}
%~~~~~~~~~~~~~~~~~~~~~~~~~~~~~~~~~~~~~~~~~~~~~~~~~~~~~~~~~~~~~~~~~~~~~~~~~~~~~~~

A theory uncertainty is a systematic uncertainty, and
as such will always require some element of human judgement. Nevertheless, like for any
systematic uncertainty, its size must reflect our level of knowledge or lack thereof.
With faithfully estimated theory uncertainties, the precision of a perturbative
prediction should be judged primarily
by its uncertainty and not so much by its formal perturbative accuracy.
Of course, for a given quantity, we expect a formally higher-order prediction to be more precise than a
formally lower-order one. The key point is that this should be
the \emph{outcome} of the uncertainty estimation procedure rather than an \emph{input} to it.
This essentially precludes estimating the theory uncertainty (solely) based on the size of
the last known perturbative correction.

To see this, consider the experimental analog of two measurements A and B of the same quantity,
where B is more precise than A due to increased statistics or
improved systematics or both. These ``formal'' improvements may make us more confident
in measurement B, but in the end what really counts is their respective uncertainty.
Assuming both have faithfully estimated uncertainties, we expect B's uncertainty to be
smaller than A's.
For simplicity, imagine that B's uncertainty is so much smaller than A's (and uncorrelated)
that only A's uncertainty matters for comparing A and B.
Consider the case that A does not agree with B: Since B is
deemed to be more reliable (formally ``better''), we would conclude that A's
uncertainty was underestimated, i.e., in this case we can invalidate A's
uncertainty. Crucially, the reverse
conclusion is not allowed: If A does agree with B within its uncertainty, this \emph{does not}
validate A's uncertainty, i.e., we \emph{cannot} conclude that A's uncertainty is not underestimated.
If that was allowed, then taken to its logical conclusion, if A's central value would agree
perfectly with B, we would have to conclude that A has a vanishingly small uncertainty, which
is clearly nonsense.

The above discussion applies identically when A is a lower-order and B a
higher-order calculation of the same quantity.
For A to agree with B within its uncertainty is only a necessary but
not sufficient condition for A's uncertainty to be correctly estimated.
In particular, we \emph{cannot} estimate the uncertainty of A by comparing its central
value to B. In other words, the difference in central values, i.e.\ the true
size of the higher-order correction, is at best a (rough) \emph{lower limit}
on A's uncertainty.

Unfortunately, this is exactly what happens frequently in perturbative predictions:
We are mistakenly led to think of the theory uncertainty as the
difference of our result to the all-order result (or a formally more accurate
higher-order one). This inevitably leads to the conclusion that we fundamentally
cannot know the theory uncertainty because we will never know the true all-order result.
Or perhaps less dramatically, that we will only really know the uncertainty at the
present perturbative order once we have calculated the next order(s).
The experimental analog would be to say that we can never know the uncertainty of a
measurement because we will never know the true value in nature.

To summarize the above discussion:
The theory uncertainty is not defined as the difference to the all-order (or the
next-order) result.
Instead, it must be a property of our present result reflecting
its intrinsic precision. When estimating it, we are meant to estimate a possible
range that contains the all-order result.
Of course, we cannot estimate this range with absolute
certainty. We can only hope to be able to estimate a range that contains the
all-order result with some probability or some level of confidence. This leads us to
the third criterium above, which basically means that we must have some way to quantify
this probability or level of confidence. Otherwise, we cannot actually utilize the prediction
for an interpretation, because to do so one must be able to interpret its
uncertainty in terms of some quantitative statistical meaning.

%~~~~~~~~~~~~~~~~~~~~~~~~~~~~~~~~~~~~~~~~~~~~~~~~~~~~~~~~~~~~~~~~~~~~~~~~~~~~~~~
\subsubsection{Correlations}
\label{sec:correlations}
%~~~~~~~~~~~~~~~~~~~~~~~~~~~~~~~~~~~~~~~~~~~~~~~~~~~~~~~~~~~~~~~~~~~~~~~~~~~~~~~

In practice theory predictions are
almost never utilized in isolation but almost always in combination with one another, at
which point the correlation in their uncertainties becomes relevant.
This is the case whenever one considers more than a single process or
phase-space region. Consider the following prototype of a data-driven method,
%%%
\begin{equation} \label{eq:ratio}
f = \bigl[g\bigr]_{\rm measured} \times \biggl[\frac{f}{g}\biggr]_{\rm theory}
   \,,\end{equation}
%%%
where a desired quantity $f$ (the ``signal'' region/process) is obtained from a
precisely measured quantity $g$ (the ``control'' region/process)
by multiplying it with their ratio predicted from theory.
Loosely speaking, if $f$ and $g$ are different but closely related, their
perturbative corrections should be very similar and largely cancel in the ratio,
such that \eq{ratio} yields an improved description of $f$ compared to its direct
prediction from theory alone.
More precisely, the theory uncertainties of $f$ and $g$ should be strongly correlated
in order to cancel in the ratio.
This cancellation is often implicitly assumed or relied on, but in reality it is very sensitive
to the exact correlation.

To appreciate this, consider $f$ and $g$ having relative uncertainty $\delta$
with correlation $\rho$. The relative uncertainty of their ratio, $\delta_{f/g}$, as a
function of $\rho$ is given by
%%%
\begin{equation}
\delta_{f/g} = \delta \sqrt{2(1-\rho)}
\,.\end{equation}
%%%
We are interested in the limit of strong correlation and large cancellation, i.e.,
$\rho$ close to 1. In this limit, $\delta_{f/g}$ is very sensitive to the precise
value of $\rho$, as illustrated in \tab{correlations}, because the square root
becomes infinitely steep for $\rho \to 1$.
For example, $\delta_{f/g}$ is 10 times smaller than $\delta$ for $\rho = 99.5\%$,
while already for $\rho = 98\%$ it doubles in size to only 5 times smaller than $\delta$.
The same correlation information is required whenever
one performs a simultaneous interpretation of two or more measurements. A prime
example is the interpretation of a differential spectrum, which requires
bin-by-bin (or point-by-point) theory correlations, as already discussed in
the introduction. The specific example of the transverse-momentum spectrum of $W$ and $Z$
bosons at the LHC is discussed in \sec{qT}. The importance of theory correlations in the context of modern machine learning methods was also stressed e.g.\ in \refcite{Ghosh:2021hrh}.

\begin{table}
\centering
\renewcommand{\arraystretch}{1.3}
\begin{tabular}{c|cccc}
\hline\hline
$\rho$ & 99.5\% & 98\% & 95.5\% & 87.5\%
\\
$\delta_{f/g}/\delta$ & $0.1$ & $0.2$ & $0.3$ & $0.5$
\\\hline\hline
\end{tabular}
\caption{Reduction of the relative uncertainty in the ratio $f/g$ for different
correlations $\rho$, see text for details.}
\label{tab:correlations}
\end{table}

It is important to keep in mind that different quantities we want to predict, such as cross sections
for different processes or at different points in phase space,
do not by themselves have a notion of being correlated to each other. A priori,
they are only more or less related to each other by the extent to which their theory
descriptions depend on common ingredients. What is correlated is the uncertainty
in their prediction due to the limited knowledge of those common ingredients. If two
quantities share a common source of uncertainty, the impact of that uncertainty
on both is 100\% correlated between them, and this is fundamentally the only way a
correlation can arise.

The simplest example is a common input parameter.
Its imprecise knowledge represents a common source of uncertainty and its resulting
uncertainty in all quantities that depend on it is 100\% correlated. When expressed
as a covariance matrix, it yields a 100\% correlated covariance matrix for all quantities.
A standard way to evaluate the correlated impact on all quantities
is to use a common nuisance parameter, which can be explicitly varied or floated
in a fit and whose variation is equivalent to varying the input parameter itself.

When several quantities depend on multiple independent sources of uncertainty,
the final correlation depends on the relative impact of the various 100\%
correlated contributions from each source. Expressed with covariance matrices, the total
covariance matrix is given by a sum of several 100\% correlated ones,
which is in general not 100\% correlated any longer.
Of course, different (nuisance) parameters can themselves have (partially) correlated
uncertainties, which can be propagated using standard error propagation.

More generally, the standard procedure to treat experimental systematic
uncertainties is to cast them into parametric uncertainties
by parameterizing the underlying source or effect in terms of one or
more nuisance parameters, which have true but a priori unknown values. Their
values are then constrained by auxiliary (real or imagined) control measurements.
The resulting best-estimate but imprecise values of the nuisance parameters
are then propagated to the nominal measurement and its interpretation.
Without any auxiliary constraint on a nuisance parameter, its uncertainty
is a priori infinite and its value will only be constrained by the
nominal measurement itself, reducing the power of the measurement
for constraining the parameters of interest.

To correctly quantify and account for theory correlations we simply follow the
same procedure: We identify and parameterize the common and mutually independent
sources of theory uncertainty and treat them respectively as 100\% correlated
and uncorrelated among all quantities of interest. This is exactly what the
theory nuisance parameters will do.

%===============================================================================
\subsection{Parametric perturbative uncertainties}
\label{sec:tnp_method}
%===============================================================================

Consider the formal perturbative expansion of a quantity $f$ in a small
parameter $\alpha$,
%%%
\begin{equation} \label{eq:f_series}
f(\alpha) = f_0 + f_1\,\alpha + f_2\,\alpha^2 + f_3\,\alpha^3 + \ord{\alpha^4}
\,.\end{equation}
%%%
By calculating the values of the first few coefficients of the series, we obtain a prediction
for $f$ at leading order (LO), next-to-leading order (NLO), next-to-next-to-leading
order (NNLO),
%%%
\begin{alignat}{9} \label{eq:f_pred}
\text{LO:}\qquad && f(\alpha) &= \hat f_0
\,,\nn\\
\text{NLO:}\qquad && f(\alpha) &= \hat f_0 + \hat f_1\,\alpha
\,,\nn\\
\text{NNLO:}\qquad && f(\alpha) &= \hat f_0 + \hat f_1\,\alpha + \hat f_2\,\alpha^2
\,,\end{alignat}
%%%
and so on.
We always denote the true value of a quantity by a hat, so $\hat f_n$ are the true
values of the series coefficient $f_n$.
When applying perturbation theory in this way, we always work under the general
assumption that the series in \eq{f_series} converges.%
\footnote{%
When $\alpha$ is a coupling constant, it is well know that
the series coefficients $f_n$ can grow
factorially, $f_n \sim n!$, which for sufficiently high $n$ overcomes the power suppression
by $\alpha^n$, so the series is only asymptotic. In practice, by using perturbation theory
to obtain predictions we implicitly assume (and confirm empirically) that the series is still
converging at the orders we are working, i.e., that the asymptotic behaviour
only becomes relevant at much higher orders than we are working at.
This can fail when the series is affected by (leading) renormalons, which
essentially spoil the convergence of the series already at low orders. This
can be remedied by working in an appropriate renormalon-free scheme in which
the nonconverging pieces of the series are absorbed into the definitions of suitable
(nonperturbative) parameters. Therefore, our general assumption is that $f$ is expanded
in a suitable perturbative scheme that is free of (leading) renormalons, such that
the factorial growth of the coefficients does not yet affect the convergence of the
series.}

The predictions for $f(\alpha)$ in \eq{f_pred} are not exact but approximations
of its all-order result. The theory uncertainty we consider here is
due to this intrinsic inexactness.%
\footnote{To be crystal clear, it is not the uncertainty due to the imprecisely
known value of $\alpha$ or any other input parameter.}
Its fundamental sources are the higher-order terms in \eq{f_series} that are
missing in \eq{f_pred}. Our assumption of convergence implies that the predictions get
increasingly better, i.e.\ more precise, by including more and more terms in the series.
This is equivalent to saying that the dominant source of theory uncertainty for the prediction
at a given order is the next missing term, i.e, that the sum of all missing higher orders is dominated
by the first missing one. (One might then consider treating the second missing one as the
``error on the error''~\cite{Cowan:2018lhq, Canonero:2023sua}.)

Strictly speaking, the actual source of uncertainty is not so much
the missing term as a whole; it is rather the unknown \emph{series coefficient} $f_n$,
as we do know the exact power $\alpha^n$ it comes with. At NNLO, if we knew
$\hat f_3$, we could add the next term $\hat f_3\,\alpha^3$ to increase the precision.
Hence, very strictly speaking, what is unknown is not the series coefficient per se
but rather its \emph{true value} $\hat f_n$.
We do know $f_n$ in the sense that we know its exact definition, we know it has a well-defined
true value, and we know how to calculate it in principle (even if we may not
have the means to compute it in practice). Importantly, these distinctions are not just
semantics, but are relevant in what follows.%
\footnote{%
We can draw the following contrast for illustration: It could be the
case that we do not know the structure of the series itself
but only how to obtain $f$ in some well-defined limit $\alpha\to 0$.
In this case, the missing higher-order terms as a whole are the source of the theory uncertainty,
which clearly makes it more difficult to estimate.
An example would be a theory where we only know the free theory but not the
interacting one.
A phenomenologically important example is the kinematic limit in which
parton showers are formulated, where we do not even in principle know the structure
of the expansion around this limit. Similarly, resummed predictions are performed
in a kinematic power expansion for which we know the leading-power limit, but we do not
necessarily know the structure of the associated power series (although there has
been a lot of progress in recent years to better understand it).
}

Let us stress another important logical
distinction: The unknown $\hat f_n$ is \emph{not} the theory uncertainty itself
(as discussed in \sec{philosophy_size}); it is only the \emph{source} of the
uncertainty. The theory uncertainty is the impact on the prediction of not
knowing $\hat f_n$, which also depends for example on the size of $\alpha^n$.
Therefore, to estimate the theory uncertainty we do not need a precise estimate of
$\hat f_n$. We rather need to quantify our limited (or lack of) knowledge of $f_n$.
We will discuss further how to do so in \sec{tnps}.
For now, it is sufficient to think of
$f_n$ as an unknown (or imprecisely known) parameter (not necessarily a scalar) which
is going to be varied in some way.
To estimate the theory uncertainty we have to propagate this variation into the
prediction. For this purpose, $f_n$ has to actually appear in our prediction,
which means we have to include the next term that contains the dependence on $f_n$.
For example, the NLO and NNLO predictions in \eq{f_pred} turn into
%%%
\begin{alignat}{9} \label{eq:f_pred_fn}
\text{N$^{1+1}$LO:}\qquad &&
f(\alpha, f_2) &= \hat f_0 + \hat f_1\,\alpha + f_2\, \alpha^2
\,,\nn\\
\text{N$^{1+2}$LO:}\qquad &&
f(\alpha, f_2, f_3) &= \hat f_0 + \hat f_1\,\alpha + f_2\, \alpha^2 + f_3\, \alpha^3
\,,\nn\\
\text{N$^{2+1}$LO:}\qquad &&
f(\alpha, f_3) &= \hat f_0 + \hat f_1\,\alpha + \hat f_2\, \alpha^2 + f_3\, \alpha^3
\,.\end{alignat}
%%%
The notation N$^{m+k}$LO is meant to indicate that in addition to the first $m$ fully known terms
we include $k$ further terms with unknown coefficients for estimating the theory uncertainty.%
\footnote{This notation implies a small departure from conventional wisdom in
that $1+1\neq 2$ and $1+2\neq 2+1$.}

We have now derived the essence of our approach: The missing series coefficients
are the sources of the theory uncertainty. They are well-defined parameters of
the perturbative series with a true but unknown value, and we simply treat them
accordingly: We include them in the prediction and vary them to account for the
theory uncertainty they cause. In this way, the theory uncertainty becomes
a truly parametric uncertainty, which is the basis for making it meaningful.
Note also that its source is actually different at each order, which also
implies that the theory correlations depend on the order of the prediction.

As discussed further in \sec{tnps}, in reality, the series coefficients have
internal structure (e.g.\ color,
partonic channels, etc.). They can also be functions of additional parameters
(e.g.\ quark masses) as well as kinematic variables.
Hence, instead of varying them directly, it will be more convenient to
parameterize them as $f_n(\theta_n)$ in terms of one or more
theory nuisance parameters $\theta_n$ that are the unknown parameters to be varied.

The actual range of variation for $f_n$ (really the $\theta_n$) is something
we have to decide, which we also discuss further in \sec{tnps}.
By default it will be a sufficiently large
range covering the generic, natural size of $f_n$ without knowing the true
value $\hat f_n$ or \emph{as if} we had no
knowledge of $\hat f_n$. In addition (or instead) we can also constrain $f_n$
(really the $\theta_n$) from experimental measurements.

If we are able to obtain a more precise estimate of $\hat f_n$, that is of course
even better. We can include this information to reduce the theory uncertainty due
to $f_n$. At this point, however, we have to remember the uncertainty due to $f_{n+1}$,
which so far we were only able to neglect because it was subdominant compared to $f_n$.
It has to be included now as soon as
it becomes relevant compared to the reduced uncertainty from $f_n$.
In this way, a lower-order prediction can gradually turn into a higher-order one.
For example, when $f_2$ is still unknown, we would start at N$^{1+1}$LO. As $f_2$
becomes better known, e.g., due to partial or approximate calculations and/or
experimental constraints, we switch at some point to N$^{1+2}$LO, which eventually
turns into N$^{2+1}$LO when $\hat f_2$ has been fully calculated.
Our approach thus allows continuously improving theory predictions
instead of being tied to large discrete steps from demanding complete formal
orders.

%===============================================================================
\subsection{Limitations of scale variations}
\label{sec:scale_var}
%===============================================================================

A well-known limitation of scale variations is that they only
have information from the known lower-order terms but no
information about the genuine higher-order terms or structure, which makes them
not very reliable and prone to underestimation due to accidentally small
lower-order terms or due to important new structures appearing at higher order
(e.g.\ new partonic channels or new functional dependences on kinematic invariants).
Since the amount of variation is largely arbitrary, one also runs the risk
of overestimating the uncertainties, which is of course also undesirable.

Even if with sufficient experience and appropriate care one is able to mitigate these dangers
of misestimation, scale variations suffer from an even more severe and fundamental
limitation: The scales that are being varied are unphysical: They are not actual
parameters of the calculation that have a true but
only imprecisely known value. There can easily be no value of the scale that actually
captures the higher-order result. This immediately tells us that it makes very
little sense to try and constrain them from data.
Since the scales have no notion of a true value or an uncertainty that is
being propagated, their variation also cannot be interpreted as such.
This implies that they are fundamentally incapable of correctly
determining theory correlations.

There is of course a more fundamental reason for the scales to appear, i.e.\ the
renormalization of the theory, which however has a priori nothing to do with
theory uncertainties. To all orders, the calculation does not depend on the
scales. By truncating the perturbative series at a finite order, a residual
scale dependence remains, i.e., it is an artifact of the calculation. Since this
residual dependence must be cancelled by the truncated higher-order terms, it
can be exploited to get a sense for the potential size of those missing higher-order
terms, but no more than that.

We can capture the essence of the scale-variation approach and expose its
limitations already at the level
of the generic expansion in \eq{f_series}. The key point is that the series
coefficients $f_n$ depend on the perturbative scheme by which we mean the exact
way of performing the perturbative expansion.
In our case here, it corresponds to the exact choice of the expansion parameter
$\alpha$. We can define a new scheme by introducing a new expansion
parameter $\talpha$ that differs from $\alpha$ by higher-order terms,
%%%
\begin{equation} \label{eq:talpha}
\talpha(\alpha)
= \alpha \bigl[1 + b_0\,\alpha + b_1\,\alpha^2 + b_2\,\alpha^3 + \ord{\alpha^4} \bigr]
\,.\end{equation}
%%%
The new scheme is uniquely determined by the coefficients $b_k$ appearing in \eq{talpha}.
Since $\alpha$ and $\talpha$ are the same at
lowest order, $\talpha = \alpha + \ord{\alpha^2}$, they are equally
good expansion parameters as long as we choose $b_k \sim \ord{1}$.
Apart from this condition, we can choose the $b_k$ freely, so \eq{talpha}
actually represents infinitely many possible expansion parameters.

To be concrete, for QCD scale variations we have
$\alpha \equiv \alpha_s(\mu_0)$ and $\talpha \equiv \alpha_s(\mu)$,
where $\mu_0$ is the central scale and
$\mu$ is the varied scale. Expanding
$\alpha_s(\mu)$ in terms of $\alpha_s(\mu_0)$, we can easily determine the
explicit $b_k$ in \eq{talpha} that are implied by scale variations,
%%%
\begin{alignat}{9} \label{eq:bk_L}
b_0 &= \frac{\beta_0}{2\pi} \ln\frac{\mu_0}{\mu}
&&= 0.85\, L
\,,\nn\\
b_1
&= \frac{\beta_0^2}{4\pi^2} \ln^2\frac{\mu_0}{\mu}
   + \frac{\beta_1}{8\pi^2} \ln\frac{\mu_0}{\mu}
&&= 0.72\,L^2 + 0.34\,L
\,,\nn\\
b_2
&= \frac{\beta_0^3}{8\pi^3} \ln^3\frac{\mu_0}{\mu}
   + \frac{5\beta_0\beta_1}{32\pi^2} \ln^2\frac{\mu_0}{\mu}
   + \frac{\beta_2}{32\pi^3} \ln\frac{\mu_0}{\mu}
&&= 0.61\,L^3 + 0.72\,L^2 + 0.13\,L
\,.\end{alignat}
%%%
They are ($k+1$)th-order polynomials in $\ln(\mu_0/\mu)$, and
$\beta_k$ are the ($k+1$)-loop QCD $\beta$ function coefficients
governing the $\mu$ dependence of $\alpha_s(\mu)$. In the second expressions
we used $n_f = 5$ and defined $L \equiv \ln(\mu_0/\mu)/\ln 2$ to give
explicit numerical results for illustration.
By convention, we vary $\mu$ by a factor of two around $\mu_0$ so
$L$ varies between $\pm 1$.
Note that scale variations do not actually provide us with the freedom
to choose all $b_k$ freely. Instead, they are all determined by choosing a
single value for $\mu$ or equivalently $L$.

Continuing our discussion,
we now have two (or really infinitely many) equally good ways to perform the
perturbative expansion for $f$, using either $\alpha$ or $\talpha$,
%%%
\begin{align} \label{eq:f_series_scheme}
f(\alpha)
&= f_0 + f_1\,\alpha + f_2\,\alpha^2 + f_3\,\alpha^3 + \ord{\alpha^4}
\,,\nn \\
\tf(\talpha) &= \tf_0 + \tf_1\,\talpha + \tf_2\,\talpha^2 + \tf_3\,\talpha^3 + \ord{\talpha^4}
\,.\end{align}
%%%
Since they are expansions of the same $f$, to all orders they are
identical: $f(\alpha) = \tf(\talpha) = f$.
Plugging \eq{talpha} back into $\tf(\talpha)$ and demanding that
$f(\alpha) = \tf(\talpha(\alpha))$ at each
order in $\alpha$, we can easily derive the scheme translation that relates
the $\tf_n$ to the original $f_n$,
%%%
\begin{equation} \label{eq:tf}
\tf_0 = f_0
\,,\qquad
\tf_1 = f_1
\,,\qquad
\tf_2 = f_2 - b_0 f_1
\,,\qquad
\tf_3
= f_3 - 2b_0 (f_2 - b_0 f_1) - b_1 f_1
\,.\end{equation}
%%%
Hence, the scheme choice essentially amounts to shuffling
around terms between orders in the series.

Although $f(\alpha) = \tf(\talpha)$ to all orders, when we truncate $\tf(\talpha)$
at a finite order to obtain predictions in our new scheme, they will differ
by higher-order terms from our original predictions truncating $f(\alpha)$ in
\eq{f_pred}. For example, up to NNLO we have
%%%
\begin{alignat}{9} \label{eq:tf_pred}
\text{LO:}\quad && \tf(\talpha)
&= \hat \tf_0 &&= \hat f_0
\,,\nn\\
\text{NLO:}\quad && \tf(\talpha)
&= \hat \tf_0 + \hat \tf_1\,\talpha
&&= \hat f_0 + \hat f_1\, \alpha + \textcolor{red}{b_0 \hat f_1\,\alpha^2 + b_1 \hat f_1\,\alpha^3 + \ord{\alpha^4}}
\,,\\\nn
\text{NNLO:}\quad && \tf(\talpha)
&= \hat \tf_0 + \hat \tf_1\,\talpha + \hat \tf_2\, \talpha^2\!
% \nn\\
&&= \hat f_0 + \hat f_1\,\alpha + \hat f_2\,\alpha^2
+\! \textcolor{red}{\bigl[2b_0 (\hat f_2 - b_0 \hat f_1) + b_1 \hat f_1\bigr]\alpha^3
\!+ \ord{\alpha^4}}
.\end{alignat}
%%%
In the second expressions we used \eqs{talpha}{tf} to rewrite
the prediction in terms of the original $\hat f_n$ and $\alpha$ to explicitly expose
the differences highlighted in red. In general, the N$^n$LO predictions
in the two schemes agree up to $\ord{\alpha^n}$ but differ by $\ord{\alpha^{n+1}}$ and
higher terms (except for the LO predictions, which happen to agree exactly because
there is no scheme dependence yet at this order).

In the scale-variation approach, this higher-order scheme dependence is now
exploited by taking the difference between the two schemes as an estimate $\Delta f$
of the theory uncertainty,
%%%
\begin{alignat}{9} \label{eq:Deltaf}
\text{LO:}\qquad && \Delta f(\alpha) &= 0
\,,\nn\\
\text{NLO:}\qquad &&
\Delta f(\alpha) &= b_0 \hat f_1\,\alpha^2 + b_1 \hat f_1\,\alpha^3 + \ord{\alpha^4}
\,,\nn\\
\text{NNLO:}\qquad &&
\Delta f(\alpha) &= \bigl[2b_0 (\hat f_2 - b_0 \hat f_1) + b_1 \hat f_1\bigr]\alpha^3 + \ord{\alpha^4}
\,.\end{alignat}
%%%

The limitations of the scale-variation approach should be clear from
the above discussion. They are fundamentally caused by the fact that the scalar
parameter $L$ (or $\mu$) that is being varied is not a true parameter of the prediction, i.e.,
it has no notion of having a true value $\hat L$ that reproduces the true
missing $\hat f_n$. This is because the coefficients of $\alpha^n$
in \eq{Deltaf} are in general not a valid parameterization of the missing
higher-order coefficients $f_n$. As soon as the $f_n$ have some nontrivial internal
structure, they will not just be given by fixed linear combinations of lower-order coefficients.%
\footnote{%
The attentive reader might have noticed that in the special case where the $f_n$
are pure scalars, we would in principle have enough degrees of freedom to correctly reproduce
each $\hat f_n$ if we were to choose each $b_k$ separately. However, apart from
the fact that scale variations do not actually provide this freedom, as we will
see in \secs{tnps}{parameterization}, the cases where we could
parameterize $f_n$ correctly as a single scalar are rare. Furthermore, this
essentially precludes accounting for any correlations.
}
The fact that $\Delta f(\alpha)$ is proportional to the
true values of the lower-order coefficients causes the common pitfall of underestimation
already mentioned at the beginning of this subsection. For example at NLO, if
$\hat f_1$ happens to be smaller than its natural size, or lacks relevant internal structures of $f_2$,
$b_0\hat f_1$ will underestimate the natural size of $f_2$ and thus the uncertainty
due to it. This is made even more severe by the fact that we practically always
use the same conventional value for $b_0$ regardless of the actual properties of
$\hat f_1$ and $f_2$.

Besides these dangers of misestimation, as $L$ is not a true parameter of the
prediction, its variation fundamentally cannot yield a meaningful theory
uncertainty to begin with. That is, it cannot imply
correlations or be constrained by measurements, and the resulting
uncertainty estimates do not admit a meaningful statistical interpretation.

These limitations of scale variations have been known for
a long time. A common method to alleviate the possible underestimation is to
perform a variety of scale variations. The individual differences are then combined
into a total uncertainty estimate $\Delta f$ by taking their envelope because
the different variations just probe the potential size of the same missing
higher-order terms in different ways and are not individually meaningful.
To mitigate the lack of correlations, the best we can do is to impose a
context-specific correlation model on the total $\Delta f$.
Dedicated correlation models have been discussed in the context of both
fixed-order predictions (see e.g.~\refscite{Stewart:2011cf, Banfi:2012yh, Gangal:2013nxa,
LHCHiggsCrossSectionWorkingGroup:2016ypw, Proceedings:2018jsb, Lindert:2017olm, Harland-Lang:2018bxd, NNPDF:2019ubu}) as well as resummed predictions (see e.g.~\refscite{Berger:2010xi,
Stewart:2013faa, Banfi:2012yh, Bizon:2019zgf, Ebert:2020dfc, Billis:2021ecs, Billis:2024dqq}).
Deciding whether or how to correlate or uncorrelate scale variations
for different predictions also just amounts to choosing some ad hoc correlation model.
While such correlation models can be theoretically motivated, they are
still ad hoc assumptions, so they are only bandaids and do not cure the
underlying problem.

In practice, the scale-variation based uncertainties are often propagated
by introducing ad hoc nuisance parameters $\theta_f$ by writing the predictions
at a given order as $f + \theta_f \Delta f$ with $\theta_f = 0\pm1$. Although this may
be done to simplify the technical implementation, we cannot stress enough that
doing so obviously does not turn $\Delta f$ automagically into an actual parametric
uncertainty. Such ad hoc nuisance parameters are not genuine nuisance parameters
and must not be treated or misinterpreted as such. In particular, despite the fact that this has
become a common mispractice, they \emph{may not} be profiled.

%%%%%%%%%%%%%%%%%%%%%%%%%%%%%%%%%%%%%%%%%%%%%%%%%%%%%%%%%%%%%%%%%%%%%%%%%%%%%%%%
\section{Theory Nuisance Parameters}
\label{sec:tnps}
%%%%%%%%%%%%%%%%%%%%%%%%%%%%%%%%%%%%%%%%%%%%%%%%%%%%%%%%%%%%%%%%%%%%%%%%%%%%%%%%

This section gives a general and largely self-contained discussion of theory
nuisance parameters (TNPs) and TNP-based theory predictions and uncertainties.
It is intended for all audiences. Readers should have read \sec{tnp_method}, but
not necessarily other subsections of \sec{pert_theory_unc}.

\Sec{tnp_method_continued} gives an introduction and general overview of the
TNP approach picking up where we left off in \sec{tnp_method}.
It is a prerequisite for the subsequent subsections and the rest of the paper.
In the subsequent subsections we further discuss several aspects of the TNP approach.
They are largely independent of each other, so readers not concerned with any one
of these aspects can freely skip the respective subsection:
\Sec{approx_implementation} discusses some implementation aspects.
\Sec{constraining} discusses constraining the TNPs based on theory
considerations and measurements.
\Sec{scheme_dependence} discusses how scale
and perturbative scheme choices figure into our approach.

%===============================================================================
\subsection{General overview}
\label{sec:tnp_method_continued}
%===============================================================================

We consider the expansion of a quantity $f$ in the small parameter $\alpha$,
%%%
\begin{equation} \label{eq:f_series2}
f(\alpha) = f_0 + f_1\,\alpha + f_2\,\alpha^2 + f_3\,\alpha^3 + \ord{\alpha^4}
\,.\end{equation}
%%%
As before, we use a hat to distinguish a parameter ($f_n$, $\theta_n$, ...) from
its true value ($\hat f_n$, $\hat\theta_n$, ...).
To obtain a perturbative prediction for $f$ at order
N$^{m+k}$LO in our approach, we include the true values of the first $m$ series
coefficients and in addition the next $k$ terms
whose coefficients are (considered to be) unknown parameters, which account for
the (dominant) theory uncertainty. For example,
%%%
\begin{alignat}{9} \label{eq:f_pred_tnps}
\text{N$^{1+1}$LO:}\qquad &&
f(\alpha, \theta_2) &= \hat f_0 + \hat f_1\,\alpha + f_2(\theta_2)\, \alpha^2
\,,\nn\\
\text{N$^{1+2}$LO:}\qquad &&
f(\alpha, \theta_2, \theta_3) &= \hat f_0 + \hat f_1\,\alpha + f_2(\theta_2)\, \alpha^2 + f_3(\theta_3)\, \alpha^3
\,,\nn\\
\text{N$^{2+1}$LO:}\qquad &&
f(\alpha, \theta_3) &= \hat f_0 + \hat f_1\,\alpha + \hat f_2\, \alpha^2 + f_3(\theta_3)\, \alpha^3
\,.\end{alignat}
%%%
In addition to \eq{f_pred_fn}, we have now
parameterized the unknown series coefficients $f_n(\theta_n)$
in terms of theory nuisance parameters $\theta_n$ for $n = m+1,\ldots,m+k$.

In general, $f_n$ has a nontrivial internal structure involving
various discrete and continuous variables.
In principle, some of this structure needs to be accounted
for in the ``TNP parameterization'' $f_n(\theta_n)$, which therefore
depends in general on multiple TNPs $\thetani$.
For example, when $f_n$ depends on different
flavor or color channels, we might need a $\thetani$ for each. When $f_n$ depends on a
continuous kinematic variable, we might need to parameterize this dependence in
terms of several $\thetani$.
The required number of TNPs thus depends on how $f_n$'s internal structure is
parameterized.
For notational simplicity we always let $\theta_n \equiv \{\thetani\}$ stand for
the full set of $\thetani$.

Different quantities can depend on a common $\thetani$ when their
coefficients internally depend on the same perturbative ingredient parameterized by
$\thetani$.
Some obvious examples are universal objects in QCD which appear
in many places, such as the beta function, splitting functions, or the cusp
anomalous dimension. In this case, a given $\thetani$ is always varied
simultaneously everywhere it appears and the resulting uncertainty is
treated as 100\% correlated.
This is how theory correlations among different quantities are correctly accounted for.
In fact, as we will discuss in \sec{parameterization}, which parts of the internal structure we need to
parameterize is directly determined by the theory correlations we need to account for.
On the other hand, different $\thetani$ should a priori be mutually independent and
correspond to independent sources of uncertainties. They can then be varied
independently and their resulting uncertainties can be treated as a priori uncorrelated.

An essential requirement on the TNP parameterization is that it must be able to
reproduce the coefficient's true value $\hat f_n$. That is, the TNPs must have
true values $\hat\theta_n\equiv\{\hat\theta_{n,i}\}$ corresponding to $\hat f_n$,
%%%
\begin{equation} \label{eq:hat_thetan}
\hat f_n = f_n(\hat \theta_n)
\,.\end{equation}
%%%
This is what makes the TNPs themselves true parameters of the perturbative series,
and what allows us to obtain meaningful constraints on their (a priori unknown) values.
That is, as for any physical parameter whose true value
is unknown, we can obtain a best estimate for the $\theta_n$ with some
uncertainty, which we denote as
%%%
\begin{equation} \label{eq:theta_estimate}
\theta_n = u_n \pm \Delta u_n
\,.\end{equation}
%%%
This estimate could come from theory considerations or experimental measurements
or both. It should be accompanied with a quantitative statistical interpretation
of the uncertainty, which may be more or less rigorous depending on where it comes from.
Statistically speaking, we want to treat \eq{theta_estimate} as coming from a real
or imagined auxiliary measurement, as for any other systematic uncertainty.
Normally, \eq{theta_estimate} will only provide a loose constraint for the $\theta_n$
to have their natural size but not a precise estimate of $\hat\theta_n$.
To emphasize this point, we mostly talk about \emph{constraints} on the $\theta_n$
rather than \emph{estimates} of them.
When we have several constraints for the same $\thetani$, we combine them
appropriately.
The central theory prediction is then obtained by setting the $\theta_n$ to their
central value $u_n$, while the theory uncertainty is evaluated
by appropriately propagating or incorporating the uncertainties
$\Delta u_n$, including their statistical interpretation, into the final results.
In this way, TNPs provide us with parametric, meaningful theory uncertainties.
They can (and should) always be propagated, combined,
and interpreted like standard parameter uncertainties.

To summarize, there are two main steps to obtain a theory prediction with TNP-based
uncertainties:
%%%
\begin{enumerate}
\item Derive an appropriate TNP parameterization $f_n(\theta_n)$ that
satisfies all requirements for all quantities of interest and implement it into the
predictions.
\item Obtain suitable auxiliary constraints on all relevant TNPs $\theta_n$
and propagate them into the final results.
\end{enumerate}
%%%
It is important to separate these two steps both logically and practically,
because they depend on different levels of approximations and assumptions.

The TNP parameterization in step 1 is determined by the internal structure
of $f_n$, which is what it is and not really debatable. As we will see in
\sec{parameterization}, all choices we can make here are based on external
requirements and can be framed as making approximations that are systematically
improvable if needed. Hence, the theory uncertainty and correlation
structure encoded by a given TNP parameterization is always
correct to some formal accuracy. Deriving it requires expert domain knowledge.
It must thus be provided as part of the prediction and cannot be left to users.
This also means that we cannot provide a generic
parameterization that is going to work in all cases.
Instead, in \sec{parameterization} we discuss the general principles and
strategies for constructing suitable parameterizations, and in \sec{qT} we
discuss a full-fledged example application.

On the other hand, in step 2 we can debate to
what extent a specific constraint (theoretical or experimental) is deemed
sufficiently reliable or not and informed users can choose to include it or not
based on their preferences or requirements.
We will see in \sec{tnp_scalar} that it is indeed possible to obtain robust theory constraints.
Furthermore, users can choose their
preferred method of propagating the TNP uncertainties.
One could either vary the TNPs explicitly or
derive a theory covariance matrix for all predictions
by performing a standard Gaussian error propagation.
When fitting to data one could repeat the fit for each variation (sometimes
called scanning or offset method), or use the derived theory covariance matrix,
or profile the TNPs as genuine nuisance parameters
with \eq{theta_estimate} imposed as an auxiliary constraint.
The option to profile the TNPs is of course a key advantage, and where their name comes from, as
it directly constrains the TNPs by the data. We will come back to this in \sec{constraining}.

%===============================================================================
\subsection{Approximate implementation}
\label{sec:approx_implementation}
%===============================================================================

In practice, to upgrade an existing N$^m$LO prediction to a full-fledged
N$^{m+k}$LO prediction with TNP uncertainties, one has to implement the correct
structure of the next $k$ orders in terms of the parameterized $f_{n}(\theta_{n})$.
Depending on the complexity of the prediction and parameterization this can
be a challenging task in itself. Therefore, as an approximation to the N$^{m+1}$LO
implementation one can also consider using the structure of the existing N$^m$LO
prediction and absorb the uncertainty term into the highest known order, for example,
%%%
\begin{alignat}{9}
\text{N$^{1+0}$LO:}\qquad &&
f(\alpha, \theta_2) &= \hat f_0 + \bigl[\hat f_1 + \alpha_0 f_2(\theta_2) \bigr]\,\alpha
\,,\nn\\
\text{N$^{2+0}$LO:}\qquad &&
f(\alpha, \theta_3) &= \hat f_0 + \hat f_1\,\alpha + \bigl[\hat f_2 + \alpha_0 f_3(\theta_3) \bigr]\, \alpha^2
\,.\end{alignat}
%%%
Here, $\alpha_0$ denotes a fixed value of $\alpha$, which is not part of
the formal series structure, e.g., it is does not participate in counting orders
of $\alpha$.
In extension to our notation, we denote this as N$^{m+0}$LO.%
\footnote{This approximation thus comes at the minor cost of further breaking
basic arithmetic: $m+0\neq m$.}

This approximation makes little difference for our simple
illustrative series, but it can make more of a difference for real-life series.
For example, it might require approximating or dropping parts of the internal structure
of $f_n(\theta_n)$ that cannot be absorbed into the existing structure of
$f_{n-1}$. Furthermore, when the full series
involves a product of several individual series (as e.g.\ in resummed
predictions), one correctly accounts for all
$\ord{\alpha^{m+1}}$ cross terms of lower-order pieces at N$^{m+1}$LO, while they
are neglected at N$^{m+0}$LO. So whilst this approximation still provides parametric
uncertainties, the impact of the parameters is only
approximately correct because one effectively uses the $\ord{\alpha^{m+1}}$ uncertainty
parameters with the lower $\ord{\alpha^m}$ uncertainty structure. We
might expect this to have only a limited effect on the overall size
of the theory uncertainty, while it might have a bigger effect on the theory correlations.
We generally recommend using the N$^{m+1}$LO prediction. If this is unfeasible for
practical reasons, one can still resort to N$^{m+0}$LO, but one should ideally
check with available orders how much of a difference this approximation makes.

As discussed at the end of \sec{tnp_method}, when the $\thetani$ become strongly constrained,
we have to include at some point the $\theta_{n+1,i}$, which means upgrading the prediction
from N$^{m+1}$LO to N$^{m+2}$LO. A convenient way to test if this is already warranted
or not is to include the $\theta_{n+1,i}$ in this approximate fashion, i.e, approximate
N$^{m+2}$LO by N$^{m+1+0}$LO. Another possible scenario is a mixed case where some $\thetani$ are
well estimated or exactly known such that their corresponding $\theta_{n+1,i}$ should
be included, while most others are still largely unconstrained.
In this case, it would be premature to upgrade to N$^{m+2}$LO but one can
already include the few required $\theta_{n+1,i}$ approximately.

%===============================================================================
\subsection{Constraining the TNPs}
\label{sec:constraining}
%===============================================================================

As already mentioned, since the TNPs are proper parameters with true values, it is perfectly
consistent to profile them in fits to data, in stark contrast to scale-variation based
approaches. We discuss several aspects of this in \sec{measurement_constraints} below.

Nevertheless, we still need a theory uncertainty estimate for the ``pre-fit'' theory
predictions, i.e., before confronting them with experimental measurements.
This is obviously necessary for any theoretical studies where we do not (yet) fit
to data. Even when fitting to data, it might be unfeasible or undesirable to always constrain all TNPs
entirely from data alone. Another reason is to be able to judge or test
whether the data constrains some $\thetani$ too strongly.
Therefore, we need some constraint on the TNPs based on theory considerations,
which we briefly discuss next and in much more detail in \sec{tnp_scalar}.

%~~~~~~~~~~~~~~~~~~~~~~~~~~~~~~~~~~~~~~~~~~~~~~~~~~~~~~~~~~~~~~~~~~~~~~~~~~~~~~~
\subsubsection{Theory constraints}
\label{sec:constraining_theory}
%~~~~~~~~~~~~~~~~~~~~~~~~~~~~~~~~~~~~~~~~~~~~~~~~~~~~~~~~~~~~~~~~~~~~~~~~~~~~~~~

As our default theory constraint, without any further information, we will have
$u_n = 0$ and $\Delta u_n$ given by the ``natural size'' of $\theta_n$, by which we mean
we would generically expect $\abs{\hat\theta_n}\lesssim \Delta u_n$.
To be more concrete, if we knew with 68\% confidence level that
$\abs{\hat\theta_n} \leq N_n$ we would take $\Delta u_n = N_n$.
The default theory constraint thus requires us to estimate the natural size
of $\theta_n$ and then allows $\theta_n$ to vary within it.
Without loss of generality, we assume that $\theta_n$ is normalized to have a
natural size of $\ord{1}$, i.e., such that we generically expect
$\abs{\hat\theta_n}\lesssim 1$ and thus $\Delta u_n \simeq 1$.

Estimating the natural size of $\theta_n$ is basically equivalent to estimating
the natural size of $f_n$, which is usually possible by considering its known higher-order
structure. For example, just pulling out the known leading color and loop factors
is usually sufficient to normalize $\theta_n$ to have $\ord{1}$ natural size.
As this estimate directly determines the eventual size of the theory uncertainty
we would of course like to narrow it down better than just a generic $\ord{1}$ factor,
ideally to within a factor of 2 or better. This can then be tested extensively
on many known series coefficients.
As we will see in \sec{tnp_scalar}, doing so we are able to obtain a
(almost surprisingly) robust estimate for $\Delta u_n$, well within a factor of 2,
and with a well-defined statistical interpretation.

In some cases we have further theoretical information that relates a priori
independent structures in $f_n$ (or different coefficients or quantities),
which have to satisfy certain relations.
An example are consistency relations between different anomalous
dimensions, which they must satisfy exactly. We have different options
to incorporate such information. If the relations are exact and simple enough,
one option is to solve them explicitly and eliminate some $\thetani$ by expressing them
in terms of others. This amounts to incorporating the constraint directly at the
level of the parameterization.
Otherwise, especially in cases with inexact relations, we
can keep all a priori independent $\thetani$ and account for each relation
by imposing a corresponding auxiliary theory constraint, which can
then lead to nontrivial a posteriori correlations between some $\theta_{n,i}$.

%~~~~~~~~~~~~~~~~~~~~~~~~~~~~~~~~~~~~~~~~~~~~~~~~~~~~~~~~~~~~~~~~~~~~~~~~~~~~~~~
\subsubsection{Measurement constraints}
\label{sec:measurement_constraints}
%~~~~~~~~~~~~~~~~~~~~~~~~~~~~~~~~~~~~~~~~~~~~~~~~~~~~~~~~~~~~~~~~~~~~~~~~~~~~~~~

Theory-based constraints, unless they are exact constraints, inevitably involve some
theoretical prejudice in the size of the uncertainty. (They can also
induce a potential bias due to scheme and parameterization dependences, as discussed
in \secs{scheme_dependence}{param_dependence}.)
However, when the theory predictions are used to interpret experimental
measurements, which is when the theory uncertainties arguably matter the most,
the TNPs can be constrained by the measurements themselves by including them in
the fit as actual nuisance parameters. Hence, we have the choice to avoid
(or at least minimize) the dependence on some undesired
theoretical prejudice by not imposing (or reducing) some theory-based constraint
and thereby rely more on the measurements. Of course, this comes at the expense
of some experimental sensitivity. A key advantage of our approach is that it
actually gives us this choice.
Thus, profiling the TNPs in fits to data has many important benefits:
\begin{itemize}
\item It allows constraining the theory uncertainties by data.
\item It avoids or reduces the susceptibility to possible theory prejudice or biases.
\item It allows taking into account possibly important correlations between the
TNPs and the parameters of interest.
\end{itemize}
The last point is because by profiling the TNPs we let the fit decide between moving
a parameter of interest vs.\ moving the theory predictions.

One might be worried that when the TNPs are constrained by the data, they also
absorb the effects of all yet higher-order corrections that have not been
included in the theory uncertainty estimate, or more generally, the effect
of any type of missing contribution or deficiency in our description. However, this problem is
always there: Any such effect is \emph{always} collectively
absorbed into all fitted parameters (both nuisance parameters and parameters of interest). The inclusion of TNPs in the fit does not make this any worse. In fact, it is likely
to \emph{reduce} this problem as far as missing theory contributions are concerned,
because it is not unlikely that they are structurally
similar to the theory uncertainty terms we now include. This means, they get
absorbed more likely into the fitted TNPs than into the parameters of
interest, thus reducing the contamination of the parameters of interest, which
is exactly what we want.

We should of course not blindly let the TNPs get misused for unintended purposes.
Formally, any unaccounted theory effect is really just an unaccounted source of
theory uncertainty. By neglecting it we assume that it is small enough to be
neglected against other uncertainties, which is equivalent to accepting that it
will be effectively absorbed somewhere (hopefully mostly into the TNPs).
However, this is exactly equivalent to the conditions
under which we are allowed to neglect $f_{n+1}$ compared to $f_n$
as discussed in \sec{tnp_method}.
The same discussion obviously applies to any other source of theory uncertainty
as well.
In particular, if with sufficiently precise data we want
to actually determine $\theta_n$, then we effectively elevate it to a parameter of
interest. We then have to at least include $\theta_{n+1}$ to account for the remaining leading theory uncertainty, as discussed at the end of \sec{tnp_method}, and
more generally also any other source of theory uncertainty of similar size.

%===============================================================================
\subsection{Scheme dependence}
\label{sec:scheme_dependence}
%===============================================================================

In our approach, we still have to choose a specific scale or scheme to perform
the perturbative expansion. For our purposes, the perturbative scheme includes
all choices of renormalization and factorization schemes as well as the choices
of all associated scales we have to make. One might wonder how the dependence on
this scheme figures into our approach now. In general, the scheme dependence is
not a problem. The scheme just has to be well defined so we can translate from
one scheme to another, and the scheme dependence has to be treated consistently.

We already discussed the scheme dependence at the level of our example series in
\sec{scale_var}. To briefly recap, by choosing different expansion parameters
$\alpha$ or $\talpha$, we have different ways to perform the perturbative
expansion for the same quantity $f$,
%%%
\begin{align} \label{eq:f_series_scheme_2}
f(\alpha)
&= f_0 + f_1\,\alpha + f_2\,\alpha^2 + f_3\,\alpha^3 + \ord{\alpha^4}
\,, \nn \\
\tf(\talpha) &= \tf_0 + \tf_1\,\talpha + \tf_2\,\talpha^2 + \tf_3\,\talpha^3 + \ord{\talpha^4}
\,.\end{align}
%%%
To all orders, the two series give identical results, $f(\alpha) = \tf(\talpha) = f$, but
at any truncated order they differ by higher-order terms [see \eq{tf_pred}]. The two schemes are
uniquely defined relative to each other by the relation between $\alpha$ and $\talpha$,
%%%
\begin{equation} \label{eq:talpha_2}
\talpha(\alpha)
= \alpha \bigl[1 + b_0\,\alpha + b_1\,\alpha^2 + b_2\,\alpha^3 + \ord{\alpha^4} \bigr]
\,,\end{equation}
%%%
from which the relation between the series coefficients $f_n$ and $\tf_n$
follows [see \eq{tf}],
%%%
\begin{equation} \label{eq:tf_2}
\tf_0 = f_0
\,,\qquad
\tf_1 = f_1
\,,\qquad
\tf_2 = f_2 - b_0 f_1
\,,\qquad
\tf_3
= f_3 - 2b_0 (f_2 - b_0 f_1) - b_1 f_1
\,.\end{equation}
%%%

To discuss the scheme dependence or ambiguity in the context of our TNP-based
predictions, it is important to distinguish two places where the scheme choice enters:
First, the scheme dependence of $f_n$ is inherited by $\theta_n$. We thus
pick a common ``reference scheme'' in which the $\theta_n$ are defined via the TNP
parameterization $f_n(\theta_n)$. We will come back to the question of how
to pick the reference scheme below. For notational simplicity, we continue
using $f_n$, $\theta_n$, $\alpha$ to denote the parameters in the reference
scheme, while we add tildes, $\tf_n$, $\tilde\theta_n$, $\tilde\alpha$ for the
parameters in some other scheme.

Second, as always we need to pick a scheme in which to evaluate the prediction itself.
An obvious and natural choice is to use the same scheme for both, i.e., we would
just use the reference scheme $f(\alpha)$, but in principle they could also be different.
To obtain the prediction in a different scheme, $\tf(\talpha)$, in terms of the
reference parameters $f_n(\theta_n)$, we simply translate
from the reference scheme by using \eq{tf_2} for the series coefficients of $\tf(\talpha)$.
For example, translating the predictions in \eq{f_pred_tnps}, we get
%%%
\begin{alignat}{9} \label{eq:tf_pred_tnps}
\text{N$^{1+1}$LO:}\quad &&
\tf(\talpha, \theta_2) &= \hat f_0 + \hat f_1\,\talpha
+ \bigl[f_2(\theta_2) - b_0\hat f_1\bigr]\, \talpha^2
\,,\\\nn
\text{N$^{2+1}$LO:}\quad &&
\tf(\talpha, \theta_3) &= \hat f_0 + \hat f_1\,\talpha
+ [\hat f_2 - b_0\hat f_1]\, \talpha^2
+ [f_3(\theta_3) - 2b_0(\hat f_2 - b_0\hat f_1) - b_1 \hat f_1]\, \talpha^3
\,.\end{alignat}
%%%
This makes it clear that the $\theta_n$ are always the same parameters and are
independent of the scheme of the prediction.

To discuss the residual scheme dependence of the prediction, first
consider the N$^{1+1}$LO prediction in \eq{tf_pred_tnps}:
Its residual scheme dependence is of $\ord{\talpha^3}$, because the $\ord{\talpha^2}$ term
includes by construction the correct scheme-dependent term $-b_0\hat f_1\,\talpha^2$ that
 cancels the scheme dependence of $\talpha$ in the previous term.
Similarly, at N$^{2+1}$LO (and also N$^{1+2}$LO which is not shown), the $\ord{\talpha^3}$ term
includes all necessary terms to cancel the $\ord{\talpha^3}$ scheme dependence of the
lower-order terms, so the residual scheme dependence is pushed to $\ord{\talpha^4}$.
In general, the residual scheme dependence of the N$^{m+k}$LO prediction
is by construction of $\ord{\alpha^{m+k+1}}$ and formally beyond the smallest included
theory uncertainty of $\ord{\alpha^{m+k}}$.
We can thus ignore it for the same reason we can drop the $\ord{\alpha^{m+k+1}}$
theory uncertainty caused by $f_{m+k+1}$.%
\footnote{%
More precisely, it causes a bias in the central value of our predictions, which
is small compared to the nominal theory uncertainty.}

We now come back to the question of how to pick the reference scheme for $\theta_n$.
Since $\theta_n$ plays the role of an input parameter, defining it in a different
scheme merely defines a different (but related) input parameter $\tilde\theta_n$
with a different but related true value $\hat{\tilde\theta}_n$. Their relation
follows from the relation between $f_n$ and $\tf_n$ in \eq{tf_2}.%
\footnote{Depending on the complexity of $f_n(\theta_n)$,
the exact relation between the individual $\thetani$ and $\tilde\theta_{n,i}$
can be more nontrivial than suggested by \eq{tf_2}, as it may not be immediately
obvious how to distribute the scheme difference between them.
There can also be some $\thetani$ that are scheme independent, namely those that
parameterize new structures in $f_n$ that cannot be generated by the scheme change
and are not captured by scale variations.}
Since this relation is exact, it a priori does not matter which parameter we use;
we can always translate exactly from one to the other.
Hence, choosing a common reference scheme for $\theta_n$ is akin to our conventional
choice of $\alpha_s(m_Z)$ (defined in a certain reference scheme namely \MSbar\
at $\mu = m_Z$) as the common input parameter for $\alpha_s$. We could have just as well chosen
$\alpha_s(m_W)$ or $\alpha_s(42\GeV)$. Since the relation between them is known
very precisely, it makes practically
no difference which one we decide to extract from data.%
\footnote{%
In contrast, for quark masses the scheme translation can induce a sizeable
uncertainty, so the optimal reference scheme for the mass parameter is the
scheme of the prediction that is used for its extraction.
}

The key difference to a purely data-determined parameter like $\alpha_s(m_Z)$
is that for $\theta_n$ we also
want to be able to obtain constraints based on theory considerations. For this
purpose, some reference schemes are better than others. A good
reference scheme is one where the $f_n$ are bounded by their natural size, i.e.,
they don't contain large scheme-induced artifacts, such that the corresponding
$\theta_n$ are of natural size.
We stress that this does not mean that the best reference scheme is necessarily the one
where $\hat f_n$ is the smallest, as this might just be accidental.
Instead, the best reference scheme is the one for which we are \emph{most confident} that
the $f_n$, and thereby the $\theta_n$, are of natural size, because this maximizes
the confidence we can ascribe to theory constraints that estimate the natural
size of $\theta_n$.

For the scale dependence, this is basically how we would usually choose
(or at least should be choosing) the central scale. We choose one for which we are most
confident that the $f_n$ do not contain large logarithms of the scale.
We usually do not (or at least should not) choose the central scale by minimizing
the highest-order $\hat f_n$. Hence, by default we can just recycle our
``best'' conventional or canonical central scales as reference scales.

To discuss the effect of choosing different reference schemes in more detail,
let us compare for concreteness the N$^{1+1}$LO predictions with TNPs defined in
different reference schemes,
%%%
\begin{alignat}{9} \label{eq:tf_pred_tnps_tilde}
\text{N$^{1+1}$LO:}\quad &&
f(\alpha, \theta_2) &= \hat f_0 + \hat f_1\,\alpha
+ f_2(\theta_2)\, \alpha^2
\,,\nn\\
\text{N$^{1+1}$LO:}\quad &&
f(\alpha, \tilde\theta_2) &= \hat f_0 + \hat f_1\,\alpha
+ \bigl[\tilde f_2(\tilde\theta_2) + b_0\hat f_1\bigr]\, \alpha^2
\,,\end{alignat}
%%%
where the two TNP parameterizations have to satisfy the scheme relation from \eq{tf_2},
%%%
\begin{equation} \label{eq:tilde_theta}
\tilde f_2(\tilde\theta_2) = f_2(\theta_2) - b_0\hat f_1
\,,\end{equation}
%%%
which determines the exact relation between the two parameters $\theta_2$ and
$\tilde\theta_2$. As long as they satisfy \eq{tilde_theta},
the predictions in \eq{tf_pred_tnps_tilde} are literally identical and it does
not matter at all which parameter we use.

A dependence on the reference scheme enters when the constraints we put on
$\theta_n$ vs.\ $\tilde\theta_n$ violate the scheme relation between them.
As already discussed above, real measurement constraints always respect this relation,
simply because they always constrain the quantity $f$ itself, which is scheme
independent. We can see this immediately from \eq{tf_pred_tnps_tilde}:
Any constraint we get on $f$, from data or elsewhere, yields exactly the same constraint on
either $f_2(\theta_2)$ or $\tf_2(\tilde\theta_2) + b_0\hat f_1$, and thus respects \eq{tilde_theta}.

On the other hand, our default theory constraints are scheme dependent because
we constrain $\theta_n$ directly. It clearly makes a difference
whether we decide to constrain $\theta_n = 0 \pm 1$ or $\tilde\theta_n = 0 \pm 1$.
They yield the same constraint for $f_n(\theta_n)$ or $\tf_n(\tilde\theta_n)$,
which means the (absolute) uncertainty on $f(\alpha, \theta_n)$ and $f(\alpha, \tilde\theta_n)$
is the same but their central value is shifted by the scheme-dependent terms.
Thus, the choice of reference scheme causes a bias (or prior) in the central value
of our theory constraint, but also nothing more.

At this point, we need to take a slight detour, as this type of bias
is actually not specific to
the theory uncertainty but can be the case for any systematic uncertainty. It amounts
to the inherent ambiguity that is always present when we have an unknown parameter that lacks
any constraints and for which we are therefore forced to pick a reasonable value.
In the absence of any external information, there is simply no unbiased way of doing so.

This is where the difference between a ``real'' vs.\ ``imagined'' auxiliary measurement
comes in. More precisely, this is how we can \emph{define} this distinction:
A real measurement or constraint imposes an unambiguous
central value. An imagined one, while also imposing a central value, leaves
open the choice on which parameter to impose it. Note that not all theory-based
constraints are necessarily of the latter type, e.g., an actual approximate
calculation of $\theta_n$ will usually apply in a specific scheme and thus resolve the
scheme ambiguity. The equivalent constraint for $\tilde\theta_n$ would then follow
from their scheme relation.

There are standard ways to deal with such biases in practice:
First, the bias from choosing a parameter's central value is \emph{not} an additional
source of uncertainty. It is a bias that may or may not be covered by the parameter's
uncertainty. If it is not, we might decide to enlarge the uncertainty or
explicitly state the choice that causes the bias as a precondition or both.
We then have several options for treating the bias:
%%%
\begin{enumerate}
\item If the parameter's bias is small compared to the parameter's uncertainty,
we can formally neglect it and move on.
\item Otherwise, if the final analysis or interpretation is insensitive to the bias, i.e.,
the resulting bias induced in the final result is small compared to its other
uncertainties, we can ignore it for practical purposes and move on.
\item Otherwise, the final result is sensitive to the bias. In this case,
   \begin{enumerate}
   \item if possible, we leave the parameter unconstrained and let the data itself
   constrain it.
   This removes any bias at the potential cost of reducing the power of the data for
   determining other parameters.
   \item Otherwise, if possible and still useful, we quote the final result explicitly
   stating the preconditions under which it is valid.
   \item Otherwise, we have to accept the fact that the analysis or interpretation
   is not possible or useful with current knowledge.
   \end{enumerate}
\end{enumerate}
%%%
In cases 1 and 2, we always have the option to further
constrain the parameter's uncertainty by the data.
Note that these cases require us to be able to quantify the bias, otherwise
we are automatically in case 3.

We now return  to our discussion at hand.
First, the exact scheme choice for the prediction itself is actually an example of case 1.
It also causes a small bias in the prediction's central
value, but as discussed above, this ambiguity is formally at least one order higher
than the theory uncertainty.

The bias caused by the choice of reference scheme of
$\theta_n$ in our default theory constraint should be covered by its
uncertainty on $\theta_n$ as long as we are comparing two equally good schemes
and the uncertainty is not underestimated. In other words, if the bias is not
covered by the uncertainty, the scheme difference $\abs{\theta_n -
\tilde\theta_n}$ exceeds what we estimated to be $\theta_n$'s natural size. This
means one of the schemes is not a good one. If we cannot figure out which one,
then the uncertainty estimate, i.e., our estimate of $\theta_n$'s natural size,
is too small. Often however, we really do have a theoretically preferred
reference scheme, for instance when there is an obvious canonical scale choice,
which effectively reduces the bias to be smaller than the uncertainty.%
\footnote{
Evaluating the bias is of course subjective, but we should only consider
alternative schemes for which we really are as confident as for our reference
scheme that the $\tf_n$ are of natural size.
In other words, the bias is not automatically given now by varying the scale by
a factor of 2. When we (used to) do scale variations by a factor of 2, it was not to
estimate the scheme bias, we just exploited the scheme dependence to
guess the theory uncertainty.
}

We should also stress that at the end of the day this bias is not a major issue.
For the pre-fit theory predictions we are effectively in case 3b, unless we can argue for case 1.
However, at this stage the exact central value is not actually that useful or
interesting, what matters more are the uncertainties. We just have to keep in mind
when discussing pre-fit predictions that we had to make an explicit choice for
the exact central value and we could have made a slightly different one within the uncertainties.
The central value actually becomes relevant when the predictions are confronted
with data, but at this point the bias can be reduced or even eliminated by the data itself.
To be prudent, we can in addition weaken any biased theory constraint,
e.g.\ by taking twice its uncertainty, to reduce its constraining power and put
more emphasis on the data as much as desired.

Finally, the above discussion provides another way to highlight the key limitation of scale
variations. They can at best provide an estimate of the scheme-induced bias but not of
the actual uncertainty, because they cannot actually probe the underlying unknown parameter
(the missing $f_n$) whose central value is implicitly chosen to be zero.

%%%%%%%%%%%%%%%%%%%%%%%%%%%%%%%%%%%%%%%%%%%%%%%%%%%%%%%%%%%%%%%%%%%%%%%%%%%%%%%%
\section{Parameterization Guide}
\label{sec:parameterization}
%%%%%%%%%%%%%%%%%%%%%%%%%%%%%%%%%%%%%%%%%%%%%%%%%%%%%%%%%%%%%%%%%%%%%%%%%%%%%%%%

This section discusses how the series coefficients $f_n$ can be parameterized in
terms of theory nuisance parameters $\theta_n$. It is intended for readers
wishing to implement TNPs into their predictions as well as curious readers
wishing to use such predictions. It assumes readers are familiar with
\sec{tnp_method_continued}.

As already mentioned in \sec{tnp_method_continued}, deriving an optimal TNP
parameterization $f_n(\theta_n)$ amounts to
deriving the correct and relevant theory uncertainty and
correlation structure for the prediction at hand. It must thus be regarded as
an integral part of performing and providing the prediction itself.%
\footnote{%
To put it more bluntly: It must not be left to users to figure out for themselves
what to do as it happens too often right now with scale variations.}
This is in general a nontrivial task and requires expert knowledge on the structure
of the underlying perturbative series.
It is clearly not as easy or convenient as performing scale variations; there
is no free lunch.

The particular parameterization strategy or combination of strategies to follow
will depend on the case at hand. We hope our discussion here will serve as a useful
guideline and starting point for future investigations.

In the next subsection we setup the basic problem and along the way give an
outline of the rest of this section. We also provide a brief executive summary
for the impatient reader at the beginning of each subsequent subsection.

%===============================================================================
\subsection{Overview and outline}
\label{sec:param_overview}
%===============================================================================

The internal structure of $f_n$ is determined by various dependences on both
discrete and continuous parameters, variables, or labels. Typical discrete dependencies
are partonic channels, color channels, or any type of discrete quantum numbers.
Examples of continuous dependencies are kinematic variables or particle masses.
In some cases, $f_n$ is mathematically a continuous function
of a parameter which in practice only takes discrete (typically integer) values.
Examples are the number of fermions, $n_f$, or the number of colors, $N_c$.
In \sec{param_examples}, we will discuss these and other examples to illustrate
our general discussion.

For now, let us denote any one of these variables (discrete or continuous) by $x$.
For the sake of simplicity and without much loss of
generality we focus on the case of $f_n(x)$ being a function of a
single variable $x$ at a time.
The true value of $f_n(x)$ is again denoted by $\hat f_n(x)$.
The dependence on multiple variables can be treated as a direct
generalization as discussed in \sec{param_multiple}.

Our goal is
to construct a TNP parameterization $f_n(x, \theta_n)$ that satisfies the
key requirement in \eq{hat_thetan}, which now reads
%%%
\begin{equation} \label{eq:hat_thetan_x}
\hat f_n(x) = f_n(x, \hat\theta_n)
\,.\end{equation}
%%%
Another goal is that the
$\thetani$ should be mutually independent, i.e., they should correspond to
mutually independent sources of theory uncertainties.
As a minimal (but not sufficient) requirement, they must parameterize $f_n(x)$ in a mathematically
independent way, so all $\hat\theta_{n,i}$ are uniquely determined by \eq{hat_thetan_x}.
We will come back to this distinction in \sec{param_dependence}, where we discuss
the parameterization dependence.
In \sec{param_strategies} we discuss various strategies for deriving suitable
parameterizations satisfying these requirements. Before doing so, we first
discuss which dependencies we actually need to parameterize in
\sec{param_requirements} next.

%===============================================================================
\subsection{Correlation requirements}
\label{sec:param_requirements}
%===============================================================================

The first question we have to ask ourselves is which parts of the internal
structure of $f_n$ we actually need to account for, i.e., which of its internal
$x$ dependencies we need to explicitly parameterize.
The answer is that it depends on our usage requirements: The dependencies we have
to parameterize are in one-to-one correspondence with the correlations we
are required to take into account.
To see this, we will discuss three different cases:
\begin{enumerate}
\item Predictions not requiring $x$ dependence
\item Predictions requiring $x$ dependence without correlations
\item Predictions requiring $x$ dependence with correlations
\end{enumerate}

%~~~~~~~~~~~~~~~~~~~~~~~~~~~~~~~~~~~~~~~~~~~~~~~~~~~~~~~~~~~~~~~~~~~~~~~~~~~~~~~
\subsubsection{Predictions not requiring \texorpdfstring{$x$}{x} dependence}
%~~~~~~~~~~~~~~~~~~~~~~~~~~~~~~~~~~~~~~~~~~~~~~~~~~~~~~~~~~~~~~~~~~~~~~~~~~~~~~~

In this case 1), we only require predictions for which the $x$ dependence is effectively not
needed. There are two basic scenarios for this:
%%%
\begin{enumerate}
\item[(a)] We only require predictions at a given fixed value $x_0$. For example, we
always work in QCD at fixed $N_c = 3$ or fixed $n_f = 5$. Or we only require
cross sections at a fixed center-of-mass energy.

In this case, we can consider $f_n(x_0)$ as a
scalar coefficient parameterized by a single TNP $\theta_n$,
%%%
\begin{equation} \label{eq:param_1a}
f_n(x_0, \theta_n) = N_n(x_0)\, \theta_n
\,,\end{equation}
%%%
where $N_n(x_0)$ is a normalization factor and the true value of $\theta_n$ is
given by
%%%
\begin{equation}
\hat\theta_n = \frac{\hat f_n(x_0)}{N_n(x_0)}
\,.\end{equation}
%%%

\item[(b)] We only require predictions summed or integrated over a fixed range $[x_a, x_b]$
in $x$.
For example, we only require a total cross section summed over all partonic channels
and integrated over phase space.

In this case, we can consider the integral of $f_n(x)$ (or the sum for discrete $x$),
%%%
\begin{equation}
F_n(x_a, x_b) = \int_{x_a}^{x_b}\!\df x\, f_n(x)
\,,\end{equation}
%%%
as a scalar coefficient parameterized by a single TNP $\theta_n$,
%%%
\begin{equation} \label{eq:param_1b}
F_n(x_a, x_b, \theta_n) = N_n(x_a, x_b)\, \theta_n
\,.\end{equation}
%%%
Here, $N_n(x_a, x_b)$ is again a normalization factor and the true value of
$\theta_n$ is given by
%%%
\begin{equation}
\hat\theta_n = \frac{\hat F_n(x_a, x_b)}{N_n(x_a, x_b)}
\,.\end{equation}
%%%
If one of the integration limits is always fixed, say $x_a = 0$,
this is the same as case (a) applied to the cumulant function
$F_n(x) = \int_0^x\df x\, f_n(x)$.
\end{enumerate}
%%%
In either case, we can choose the normalization factor $N_n$
such that $\theta_n$ has $\ord{1}$ natural size. It may or may not have to depend
on the value $x_0$ or the integration limits $(x_a, x_b)$, depending to what
extent the value of $x$ determines the natural size of $f_n(x)$.

%~~~~~~~~~~~~~~~~~~~~~~~~~~~~~~~~~~~~~~~~~~~~~~~~~~~~~~~~~~~~~~~~~~~~~~~~~~~~~~~
\subsubsection{Predictions requiring \texorpdfstring{$x$}{x} dependence without correlations}
\label{sec:param_case2}
%~~~~~~~~~~~~~~~~~~~~~~~~~~~~~~~~~~~~~~~~~~~~~~~~~~~~~~~~~~~~~~~~~~~~~~~~~~~~~~~

In this case 2), we require predictions at several discrete values of $x$
(e.g.\ at different $n_f$), or several bins in $x$
or as a function of $x$ (e.g.\ a binned or unbinned differential spectrum),
but we do not need to have correct correlations in $x$.
Even so, being differential in $x$ forces us to assume some correlations
in $x$, for which we have different options:
\begin{enumerate}
\item[(a)] We assume the uncertainties to be fully correlated for all $x$, which
means we are happy to neglect any shape uncertainties in $x$
and only care about some overall uncertainty in $f_n(x)$.
We can then parameterize $f_n(x)$ in terms of the same single $\theta_n$
appearing in case 1) above, so
%%%
\begin{align}
\text{1(a)}\qquad
f_n(x, \theta_n) &= N_n(x_0)\,\theta_n\, \phi_n(x)
&&\text{with}\qquad
\phi_n(x_0) = 1
\,,\\\nn
\text{1(b)}\qquad
f_n(x, \theta_n) &= N_n(x_a, x_b)\,\theta_n\, \phi_n(x)
&&\text{with}\qquad
\int_{x_a}^{x_b}\!\df x\,\phi_n(x) = 1
\,.\end{align}
%%%
The normalization factors $N_n$ are the same as in \eqs{param_1a}{param_1b}.
The function $\phi_n(x)$ determines how the uncertainty is distributed over $x$.
Its normalization condition is chosen such that $\theta_n$ parameterizes the exact
same uncertainty as in cases 1(a) or 1(b) and we
assume there are no shape uncertainties in $x$, i.e., we assume to know
the shape perfectly given by $\phi_n(x)$.
This also means we are explicitly giving up that
\eq{hat_thetan_x} holds point-by-point in $x$. Instead, we only
require that it holds as in case 1) either at $x_0$ or integrated over $[x_a, x_b]$.

There are various choices we might consider for $\phi_n(x)$. For example,
a constant absolute uncertainty in $x$ is achieved by taking $\phi_n(x)$ to
be a constant, $\phi_n(x) = a$. A constant relative uncertainty in $x$
is achieved by taking $\phi_n(x) = a [\hat f_0(x) + \dotsb + \hat f_{n-1}(x)\alpha^{n-1}]$.
Another typical choice would be to take $\phi_n(x)$ proportional to the lowest-order
$x$ dependence, $\phi_n(x) = a\, \hat f_0(x)$.
In either case, the proportionality constant $a$ is fixed by the normalization
condition for $\phi_n(x)$.

\item[(b)] We assume the uncertainties for some set of $x$ values $\{x_i\}$
are fully uncorrelated. This amounts to using case 1) for each $x_i$ with its own
independent TNP $\theta_{n,i}$,
%%%
\begin{equation}
f_n(x_i, \theta_n) = N_n(x_i)\, \thetani
\,,\end{equation}
%%%
where $N_n(x_i)$ are again normalization factors, and
$\theta_n \equiv \{\thetani\}$ now stands for the set of $\thetani$.
The true values of the $\thetani$ are
%%%
\begin{equation}
\hat\theta_{n,i} = \frac{\hat f_n(x_i)}{N_n(x_i)}
\,,\end{equation}
%%%
and \eq{hat_thetan_x} is now satisfied at each $x_i$.

We can now extend this to all $x$ by generalizing case (a) above as follows,
%%%
\begin{equation}
f_n(x, \theta_n) = \sum_i N_n(x_i)\, \thetani\, \phi_{n,i}(x)
\qquad\text{with}\qquad
\phi_{n,i}(x_j) = \delta_{ij}
\,.\end{equation}
%%%
The functions $\phi_{n,i}(x)$ now determine how the uncertainty due to $\thetani$ is distributed
away from $x_i$. Their form is more complicated now due to the additional requirement
that they must vanish at all but one $x_i$. An analogous construction can be used
for a set of bins instead of $x$ values.
\end{enumerate}

We stress again, that the above options do not provide correct correlations in $x$.
They should only be used if it is known that correlations in $x$ do not matter or in order
to test whether or not this is the case.

%~~~~~~~~~~~~~~~~~~~~~~~~~~~~~~~~~~~~~~~~~~~~~~~~~~~~~~~~~~~~~~~~~~~~~~~~~~~~~~~
\subsubsection{Predictions requiring \texorpdfstring{$x$}{x} dependence with correlations}
%~~~~~~~~~~~~~~~~~~~~~~~~~~~~~~~~~~~~~~~~~~~~~~~~~~~~~~~~~~~~~~~~~~~~~~~~~~~~~~~

In this case 3), we require $x$-dependent predictions as in case 2) but now with correct
correlations in the uncertainties at different $x$.
In other words, we require predictions with correct shape uncertainties in $x$.

In this case, we have to explicitly parameterize the correct $x$
dependence of $f_n(x)$. In other words, $f_n(x,\theta_n)$ must parameterize the
true functional form of $\hat f_n(x)$ such that there are true values
$\hat\theta_n$ for which it reproduces $\hat f_n(x)$ exactly, i.e.,
\eq{hat_thetan_x} is satisfied at \emph{any} $x$.
Clearly, this requires us to have some
knowledge of the true functional form of $f_n(x)$.

When $x$ is a discrete label,
knowing the functional form in $x$ simply means knowing the complete set of
possible values $\{x_i\}$, which is basically always the case. We can then
assign an independent $\theta_{n,i}$ for each $f_n(x_i)$ as in case 2(b) above.

It gets more complicated when $f_n(x)$ is a continuous function of $x$, which
has in principle infinitely many degrees of freedom. We will discuss several
strategies to deal with this situation in \sec{param_strategies} next. For the
sake of our discussion here, let us consider a simple example: Say we know on
theoretical grounds that $f_n(x)$ is a polynomial of degree $k$ (as is the case
e.g.\ for $x = n_f$),
%%%
\begin{equation}
f_n(x) = f_{n,0} + f_{n,1}\,x + \dotsb + f_{n,k}\,x^k
\,.\end{equation}
%%%
The scalar coefficients $f_{n,i}$ are parameters of $f_n(x)$
with true but possibly unknown values $\hat f_{n,i}$. Without further information,
we have to treat them as independent unknown parameters and thus
parameterize each with its own TNP, $f_{n,i} = N_{n,i}\,\thetani$, such that
%%%
\begin{equation}
f_n(x, \theta_n) = \sum_{i = 0}^k N_{n,i}\, \thetani\, x^i
\,,\end{equation}
%%%
where $N_{n,i}$ are normalization factors of our choice, and the true
values of the $\thetani$ are
%%%
\begin{equation}
\hat\theta_{n,i} = \frac{\hat f_{n,i}}{N_{n,i}}
\,.\end{equation}
%%%
The key point is
that in contrast to case 2), the $\thetani$ are now defined to be the actual
parameters of the true functional form of $f_n(x)$ and therefore encode the
correct correlation structure in $x$.

%===============================================================================
\subsection{Parameterization strategies}
\label{sec:param_strategies}
%===============================================================================

As we have seen, to account for the correct correlations in $x$ we have to
parameterize the series coefficient $f_n(x)$ in terms of its correct underlying
$x$ dependence. There are different basic strategies for doing so, depending on
how much or little we know about the true functional form of $f_n(x)$:
%%%
\begin{enumerate}
\item We know it well enough to be able to parameterize it explicitly
in terms of a small number of parameters.
\item We know it well enough to apply strategy 1) in some well-defined limit
and can perform a systematic expansion around that limit.
\item Having insufficient information for strategies 1) or 2), we can
still perform an expansion in a suitable complete functional basis.
\end{enumerate}
%%%
We now discuss each of these in turn.

%~~~~~~~~~~~~~~~~~~~~~~~~~~~~~~~~~~~~~~~~~~~~~~~~~~~~~~~~~~~~~~~~~~~~~~~~~~~~~~~
\subsubsection{Known functional form}
%~~~~~~~~~~~~~~~~~~~~~~~~~~~~~~~~~~~~~~~~~~~~~~~~~~~~~~~~~~~~~~~~~~~~~~~~~~~~~~~

If we know the true functional form of $f_n(x)$ well enough, we can parameterize it
explicitly.
In general, we can imagine $\hat f_n(x)$ to be some functional $\hat\phi_n$ of
$x$-dependent building blocks $\hat\phi_{n,i}(x)$ and scalar coefficients $f_{n,i}$,
%%%
\begin{equation} \label{eq:hatphi_n}
\hat f_n(x)
= \hat\phi_n\bigl[\{\hat\phi_{n,i}(x)\}, \{\hat f_{n,i}\}\bigr]
\,.\end{equation}
%%%
Knowing the true functional form of $f_n(x)$ but not the true $\hat f_n(x)$ itself means
we know the true $\hat\phi_n$ and $\hat\phi_{n,i}(x)$ but we do not know the
true values $\hat f_{n,i}$.
We can then parameterize each coefficient $f_{n,i} = N_{n,i}\,\thetani$ in terms
of its own $\thetani$ to obtain the TNP parameterization
%%%
\begin{equation}
f_n(x, \theta_n) = \hat\phi_n\bigl[\{\hat\phi_{n,i}(x)\}, \{N_{n,i}\, \thetani\}\bigr]
\,,\end{equation}
%%%
where as before $N_{n,i}$ are normalization factors of our choice, and the true
values are $\hat\theta_{n,i} = \hat f_{n,i}/N_{n,i}$.
A common case is that $\hat\phi_n$ is a linear functional, such that
%%%
\begin{equation} \label{eq:hatphi_n_linear}
f_n(x, \theta_n) = \sum_{i = 0}^k N_{n,i}\,\thetani\, \hat\phi_{n,i}(x)
\,.\end{equation}
%%%
Common special cases are $\hat\phi_{n,i}(x) = x^i$ or $\hat\phi_{n,i}(x) = \ln^i x$
corresponding to polynomials in $x$ or $\ln x$ of degree $k$.

The key point point here is that we have to know the true functional form well enough to
be able to write \eq{hatphi_n} with a \emph{finite} number of unknown parameters $f_{n,i}$.
This is where we need expert knowledge on the structure of the perturbative series of
the quantity $f$.
Furthermore, we want to have a ``minimal parameterization'' in the following sense:
We want to choose $\hat\phi_n$ and $\hat\phi_{n,i}(x)$ such that we
have the minimal possible number of a priori unknown parameters $f_{n,i}$.
Firstly, this means we should not introduce additional a priori known
parameters. For example, we should not needlessly split the $\hat\phi_{n,i}(x)$ into smaller
pieces at the expense (or for the purpose) of introducing additional fake parameters
that would effectively always be known and which we then pretend to be unknown.
Vice versa, we also should not eliminate parameters
whose true values we happen to know already but which could a
priori be unknown. Instead, we should leave the
decision for later whether to use our knowledge of the true value to reduce the uncertainty
or not.%
\footnote{Even with this definition of ``minimal'' there might be corner cases
where we might debate whether a parameter is a priori known or unknown. To
decide these, we just have to remember that the uncertainties
not only reflect our knowledge but also our common sense.}
Note that even if we know in principle the allowed $\hat\phi_{n,i}(x)$, this strategy
can fail because the number of $\hat\phi_{n,i}(x)$
might simply be too large to be practical.

Even a minimal parameterization is not unique. This is easily seen from the linear
example in \eq{hatphi_n_linear}. We can always choose a different independent combination
of the $\hat\phi_{n,i}(x)$ and correspondingly use a different combination of the
$\thetani$ as the independent parameters. This parameterization ambiguity is conceptually
analogous to the scheme dependence of the perturbative series discussed in
\sec{scheme_dependence} and we will come back to it in \sec{param_dependence}.

Finally, let us point out that there are cases for which the functional form in $x$
is simple and known, particularly when $x$ is a discrete label (e.g.\ the
partonic channel), and for which it can be of advantage to parameterize the
$x$ dependence explicitly even if correlations in $x$ are not required. For example, when the
$x$ dependence strongly affects the size of $f_n(x)$, it can be much easier to
figure out the natural size of the individual $x$-independent $\theta_{n,i}$
than of some single overall $\theta_n$.

%~~~~~~~~~~~~~~~~~~~~~~~~~~~~~~~~~~~~~~~~~~~~~~~~~~~~~~~~~~~~~~~~~~~~~~~~~~~~~~~
\subsubsection{Supplementary power expansion}
%~~~~~~~~~~~~~~~~~~~~~~~~~~~~~~~~~~~~~~~~~~~~~~~~~~~~~~~~~~~~~~~~~~~~~~~~~~~~~~~

If we can identify a suitable small parameter $\eps$, we can perform a
supplementary power expansion of $f_n(x)$ in $\eps$,
%%%
\begin{equation} \label{eq:fn_eps}
f_n(x) = f_{n0}(x) + f_{n1}(x)\, \eps  + f_{n2}(x)\, \eps^2 + \ord{\eps^3}
\,.\end{equation}
%%%
Whilst we might not know the functional form of $f_n(x)$ well enough to apply
strategy 1), we might know the functional form of its
$f_{nl}(x)$ series coefficients well enough to apply strategy 1) for each of them.
This is clear when the expansion parameter $\eps$ is related to $x$ itself, e.g.,
$\eps = x$ or $\eps = 1-x$. If $f_{nl}$ is independent of $x$ then
this is simply the Taylor expansion of $f_n(x)$ around $x = 0$ or $x = 1$, but it can also be more general. A primary example is the small-$p_T$ expansion we will employ in \sec{qT}
in which case $f_{nl}(x)$ are known to be polynomials in $\ln x$.
Expanding around some point in $x$ of course only helps us when we are actually close to that point.
However, $\eps$ does not necessarily have to be $x$ itself in order to simplify the $x$ dependence. It may also be related to another variable $y$. When $f_n(x, y)$ has
a nontrivial two-dimensional structure, expanding in $y$ can simplify not just the $y$
dependence but also the $x$ dependence significantly, and expanding in $y$ can
be justified even when expanding in $x$ is not.

The $\eps$ series in \eq{fn_eps} is conceptually completely equivalent to our
perturbative series in $\alpha$ and we can apply exactly the same logic for
treating its uncertainties.
The coefficients $f_{nl}$ are parameters of the series with true but possibly unknown values.
As long as the expansion converges, we can keep the first $m$ known coefficients, include the next $k$ terms to parameterize the dominant uncertainties, and truncate the remaining terms since their uncertainties are formally small compared to the ones we keep. For example,
%%%
\begin{alignat}{9} \label{eq:fn_pred_eps}
\text{N$^{0+1}$LP$_\eps$:}\qquad &&
f_n(x, \theta_{n0}) &= f_{n0}(x, \theta_{n0})
\,,\nn\\
\text{N$^{0+2}$LP$_\eps$:}\qquad &&
f_n(x, \theta_{n0}, \theta_{n1}) &= f_{n0}(x, \theta_{n0}) + f_{n1}(x, \theta_{n1})\, \eps
\,,\nn\\
\text{N$^{1+1}$LP$_\eps$:}\qquad &&
f_n(x, \theta_{n1}) &= \hat f_{n0}(x) + f_{n1}(x, \theta_{n1})\, \eps
\,.\end{alignat}
%%%
where the notation refers to the next-$(m+k)$-leading-power expansion in $\eps$.

We stress that the primary reason for using this expansion is to provide us
the formal justification and practical ability to only parameterize the
leading-in-$\eps$ dependence on $x$ for the purpose of
correctly parameterizing the uncertainties in $x$. Doing so does not
force us in any way to perform this expansion also for the known series coefficients,
for which we may not want to do so.

In case we happen to know the true value of $\hat f_{n0}(x)$ we can include it exactly
as in the N$^{1+1}$LP$_\eps$ result in \eq{fn_pred_eps}.
In this case, the $\eps$ expansion even allows us to include further information and thus reduce
the uncertainties, which we would not be able to do otherwise. In fact, this is exactly
a case where thanks to our approach we are able to reduce the uncertainty due to $f_n(x)$
by including partial higher-order information. Consequently, we then also have to
consider whether or not to include the uncertainty due to $f_{(n+1)0}(x)$.

%~~~~~~~~~~~~~~~~~~~~~~~~~~~~~~~~~~~~~~~~~~~~~~~~~~~~~~~~~~~~~~~~~~~~~~~~~~~~~~~
\subsubsection{Generic basis expansion}
%~~~~~~~~~~~~~~~~~~~~~~~~~~~~~~~~~~~~~~~~~~~~~~~~~~~~~~~~~~~~~~~~~~~~~~~~~~~~~~~

When we do not have sufficient information to parameterize $f_n(x)$ directly or
in some limit, we face the basic mathematical problem of how to best parameterize an unknown
function such that we can guarantee that the $\thetani$ have true
values. We start by expanding $f_n(x)$ in a suitable complete basis
$\phi_{n,i}(x)$,
%%%
\begin{equation}
f_n(x) = \sum_{i = 0}^\infty f_{n,i}\, \phi_{n,i}(x)
\,.\end{equation}
%%%
Thanks to the Weierstrass approximation theorem this expansion converges for polynomial
bases on any bounded interval in $x$ as long as $f_n(x)$ is continuous.
This means that formally the $f_{n,i}$ are proper parameters with true values
$\hat f_{n,i}$.
This expansion is not particularly useful yet, because it has infinitely
many parameters. To make it useful for our purposes, we have to truncate
the series after a few terms to limit the number of parameters.

The key question then is how to justify truncating the series. In principle, we
like to apply the same argument as for our perturbative series in $\alpha$ or
the $\eps$ expansion in strategy 2,
namely that the uncertainties due to the truncated terms can be neglected as small
compared to the uncertainties from the terms we keep.
However, this argument is harder to make now because we
lack a parameter like $\alpha$ or $\eps$ which would allow us to control the size
of the truncated terms and decide where to truncate it without knowing $\hat f_n(x)$.
Instead, we have to rely more on experience and being able to test it on known coefficients.

Therefore, the most suitable basis we can choose is not necessarily the one which yields the best
approximation for $\hat f_n(x)$ for a given number of terms, but rather the one for which we are \emph{most
confident} that it yields a sufficient approximation for a given number of terms.
In other words, we want a basis for which we are confident that it convergences quickly with the first couple of terms to a point where we can safely neglect the remainder. Beyond that point, the remaining series may converge as slowly as it likes. The region of quick convergence should include some
safety margin to allow including additional terms in case the first few terms
we would keep by default get constrained too strongly.

For simplicity and without loss of generality, let us assume the relevant $x$ range to be $x\in[-1, 1]$.
Standard polynomial bases on this interval which are known to converge very fast for
sufficiently smooth functions are Legendre and Chebyshev polynomials.%
\footnote{Roughly speaking, for differentiable functions that have $p-1$ continuous derivatives
and a $p$th derivative of bounded variation, Legendre, Chebyshev, and similar polynomial
expansions converge algebraically $\sim 1/k^p$. For analytic functions they converge geometrically
$\sim 1/\rho^k$ where the constant $\rho$ depends on how far the function can be analytically continued
into the complex plane. The corresponding precise mathematically theorems can be found e.g.\ in \refcite{Trefethen}.}
An advantage of Legendre polynomials is that they are orthogonal with respect to the unit
weight function. An advantage of Chebyshev polynomials is their equioscillation
property, which means that all their minima and maxima in the interval $[-1,1]$
are at $\pm 1$. Even if we do not know the full functional form of $f_n(x)$ we
rarely know nothing about it. We can improve the
convergence of the series by starting from some ansatz $\phi_n(x)$ and
expanding the ratio to $f_n(x)$ to obtain the TNP parameterization,
%%%
\begin{align}
\frac{f_n(x)}{\phi_n(x)} &= \sum_{i = 0}^\infty f_{n,i} \, \phi_{n,i}(x)
\,,\nn\\
f_n(x, \theta_n) &= \phi_n(x) \sum_{i = 0}^k N_{n,i}\, \theta_{n,i}\, \phi_{n,i}(x)
\,.\end{align}
%%%
Alternatively, we can expand the difference to obtain
%%%
\begin{align}
f_n(x) - \phi_n(x) &= \sum_{i = 0}^\infty f_{n,i} \, \phi_{n,i}(x)
\,,\nn\\
f_n(x, \theta_n) &= \phi_n(x) + \sum_{i = 0}^k N_{n,i}\, \theta_{n,i}\, \phi_{n,i}(x)
\,.\end{align}
%%%
Note that $\phi_n(x)$ and $\phi_{n,i}(x)$ should be normalized suitably such that
the overall size of the uncertainty and the natural size of $\theta_{n,i}$ is
determined by their normalization factors $N_{n,i}$.

Another general method to accelerate the convergence is to use a variable transformation
to account for some known general behaviour of $f_n(x)$. For example, if $f_n(x)$
is known to have poles or branch cuts in the complex plane, using a variable transformation
that maps these to infinity can significantly improve the rate of convergence.
One particular option if $\phi_n(x)$ is square-integrable and positive definite,
$\phi_n(x) = \abs{\phi_n(x)}$, is to construct a custom orthonormal basis on top of it, as
was done in \refcite{Ligeti:2008ac} in a different context. The idea is to use
the cumulant of $\phi_n(x)$ as a variable transformation as follows. Let us normalize
$\phi_n(x)$ such that $\int_{-1}^1 \df x\,[\phi_n(x)]^2 = 1$ and define
%%%
\begin{align}
y(x) &= -1 + 2 \int_{-1}^x\!\df x'\, [\phi_n(x')]^2
\,,\nn\\
\phi_{n,i}(x) &= \sqrt{y'(x)}\,p_i[y(x)]
\,,\qquad
p_i(y) = \sqrt{\frac{2n + 1}{2}} \frac{1}{2^n n!}\,\frac{\df^n}{\df y^n}(y^2 - 1)^n
\,,\end{align}
%%%
Here, $p_i(y)$ are normalized Legendre polynomials, which are
orthonormal on $y\in[-1,1]$, and thus $\phi_{n,i}(x)$ are
orthonormal on $x\in [-1,1]$. Since $\sqrt{y'(x)} = \sqrt{2} \phi_n(x)$,
the 0th basis function $\phi_{n,0}(x) = \phi_n(x)$ itself, while the higher basis
functions are orthogonal polynomial modulations on top of it.
As a result, if $\phi_n(x)$ captures the overall shape of $f_n(x)$, the series
is expected to converge much more rapidly than expanding in $p_i(x)$ directly,
at least for the first few terms until the detailed shape starts to matter,
which is all we really want.

Hence, the key to finding a suitable basis in the above sense is to start from a
suitable approximation $\phi_n(x)$. Importantly, the goodness of this
approximation is not a fundamental limitation, as it only serves in one or
another way as a starting point for a complete expansion. There are various ways
we can imagine choosing $\phi_n(x)$:
%%%
\begin{itemize}
\item Pick $\phi_n(x) = \hat\phi_n(x)$, or more
generally $\phi_n(x) = \hat\phi_n(x, \theta_n)$, where $\hat\phi_n(x)$ encodes
some known aspect of the true functional form of $f_n(x)$, e.g., it has (or parameterizes)
the correct asymptotic behaviour or the correct poles. This essentially supplements
strategy 1) in case the information we have is not sufficient for using it standalone.

\item Pick $\phi_n(x) = f_{n0}(x,\theta_{n0})$ or $\phi_n(x) = \hat f_{n0}(x)$,
where $f_{n0}(x)$ is the leading-power limit from strategy 2) which either has a
simpler known $x$ dependence or is fully known. This essentially supplements
strategy 2) in case it cannot be used standalone, e.g., when the $\eps\to0$ limit is known
but the expansion itself is not or does not apply to all $x$.

\item Use the known lower-order shape $\phi_n(x) = a \hat f_0(x)$ or
$\phi_n(x) = a[\hat f_0(x) + \dotsb + \hat f_{n-1}(x)\,\alpha^{n-1}]$, with
$a$ determined by the appropriate normalization condition on $\phi_n(x)$.
This essentially supplements case 2) in \sec{param_case2}. For the multiplicative
case it effectively expands the $K$ factor, which makes sense whenever we are
confident that $f_n(x)/\hat f_0(x)$ is much flatter in $x$ than $f_n(x)$ itself.

\item Use some approximation $\phi_n(x) \approx \hat f_n(x)$,
which is known to work in similar cases, e.g.\ a Pad{\'e} approximation. This
supplements any ad hoc approximation method, extending it into a formally complete parameterization.
\end{itemize}
%%%

Before concluding, let us comment that one might naively think that having
to know or parameterize the functional form
of $f_n(x)$ is a drawback of our approach. It is not. It is simply a necessity
for obtaining the correct correlation structure in $x$.
In practice, we almost always have some, even if limited, information about
the functional form and it is in fact a key advantage of our approach that all
information we have can be systematically incorporated.
In contrast, with uncertainties derived from scale variations $f_n(x)$ is
silently modelled by some linear combination of lower-order coefficients, see \eq{Deltaf}.
If this is indeed believed to be a sufficient correlation model, one can always
use the lower-order coefficients to construct the ansatz $\phi_n(x)$
as mentioned in the third bullet point above. This is still much better than scale variations,
because it provides explicit control over the assumptions made and
furthermore provides a systematic extension to a formally complete parameterization.

%===============================================================================
\subsection{Parameterization dependence}
\label{sec:param_dependence}
%===============================================================================

Regardless of the strategy used to derive the TNP parameterization,
a given parameterization is never unique. For example, we can always choose some
linearly independent combination of the $\theta_{n,i}$ as new independent
parameters.
The ambiguity in the choice of the parameterization is conceptually
analogous to the scheme dependence of the perturbative series, and our
discussion here will resemble much of the discussion in \sec{scheme_dependence}.

Let us therefore start with an executive summary:
It is important to distinguish the uncertainty and correlation
\emph{structure}, which is unique and correctly encoded by any valid parameterization,
from the actual \emph{values} of the uncertainties and correlations, which are determined
by whatever constraints we choose to impose on the parameters.
Before any constraints they are simply unknown, which means their uncertainties are infinite and their correlations do not matter, which is a parameterization independent statement.
When the parameters are solely constrained by data or other parameterization-independent constraints,
the uncertainties and correlations reflect the combined uncertainties
and correlations due to all constraints in a parameterization-independent way.
A parameterization-dependent bias is only induced when we impose parameterization-dependent
constraints.

To discuss the possible parameterization dependence in more detail,
let us denote by $f_n(x, \theta_n)$ our default parameterization
and by $f_n'(x, \theta_n')$ some alternative parameterization.
For the sake of discussion let us also consider them to still be exact, so before
truncating the $\eps$ series in strategy 2) or the basis expansion in strategy 3).
Different parameterizations must then be equal by definition,
%%%
\begin{equation} \label{eq:fnprime}
f_n(x, \theta_n) = f_n'(x, \theta_n')
\,,\end{equation}
%%%
as they both parameterize the same function $f_n(x)$ and reproduce the same true
value $\hat f_n(x)$. It follows that from \eq{fnprime} the $\theta_n'$ are uniquely
determined in terms of the $\theta_n$ (and vice versa) in exactly the same way
the true $\hat\theta_n$ and $\hat\theta_n'$ are uniquely determined by \eq{hat_thetan_x}.

Note that in principle there could be more $\theta_{n,i}'$ than $\theta_{n,i}$
parameters. If so, it would imply that $f_n(x, \theta_n)$ contains more
information on the true functional form in $x$ than $f_n'(x, \theta_n')$.
So from the point of view of $f_n(x, \theta_n)$ some of the $\theta_{n,i}'$ are
either known or not independent. Let us therefore assume that both
parameterizations are based on the same information and thus have the same
number of parameters.

Any valid parameterization, meaning it satisfies \eq{hat_thetan_x}, encodes the
correct theory uncertainty and correlation structure.
The $\theta_n$ and $\theta_n'$ play the role of different but related input parameters.
If we treat them as unknown parameters to be determined from data, it does not
matter at all which one we choose, since there is an exact
relation between them. Any constraints imposed by the data are
parameterization independent as they always constrain $f_n(x)$, so they respect
the relation between $\theta_n$ and $\theta_n'$ implied by \eq{fnprime}.

A dependence on the parameterization (only) appears when we make
parameterization dependent assumptions, i.e., by imposing theory constraints on the
$\theta_n$ of a specific parameterization. This also includes the assumption of
their mutual independence, which only enters when we choose to impose
independent (uncorrelated) theory constraints on them.
Doing so for the $\theta_{n,i}$ implies in general some nontrivial correlated
uncertainties for $\theta_{n,i}'$ (and vice versa). The point is that the condition
for some parameters to be \emph{mathematically} independent is only a necessary but not
sufficient condition for them to be \emph{conceptually} independent, i.e.\ to correspond to
independent sources of uncertainties, and thus to be a priori uncorrelated.
Whenever we talk about the $\theta_{n,i}$ being mutually independent we really
refer to their conceptual independence.

To illustrate this with a simple example, say we know $f_n(x)$ to be a
$k$th-order polynomial. Consider the two equivalent parameterizations
%%%
\begin{align} \label{eq:param_example}
f_n(x, \theta_n) &= \theta_{n,0} + \theta_{n,1}\,x + \dotsb + \theta_{n,k}\,x^k
\,,\nn\\
f_n'(x, \theta_n') &= \theta'_{n,0} + \theta'_{n,1}\,(1-x) + \dotsb + \theta'_{n,k}\,(1-x)^k
\,.\end{align}
%%%
By setting them equal, we can easily derive the exact relation between $\theta_n'$
and $\theta_n$, e.g.,
%%%
\begin{equation}
\theta_{n,0}' = \theta_{n,0} + \theta_{n,1} + \dotsb + \theta_{n,k}
\,,\qquad
\theta_{n,1}' =  - \theta_{n,1} - 2\, \theta_{n,2} - \dotsb - k\, \theta_{n,k}
\,,\end{equation}
%%%
and so on.
A fit to data always chooses the $k$th-order polynomial that best fits the data,
regardless of the specific parameterization, with the post-fit uncertainties and correlations
of the parameters reflecting the uncertainties and correlations of the fitted
measurements in a parameterization independent way.
On the other hand, imposing a theory
constraint that the $\theta_{n,i}$ have mutually uncorrelated uncertainties of
$\Delta u_{n,i} = 1$ yields an uncertainty
for $f_n(x = 0)$ of $1$ and for $f_n(x = 1)$ of $\sqrt{k}$. On the other hand,
imposing the same constraint on the $\theta_{n,i}'$ yields an uncertainty for
$f_n(x = 0)$ of $\sqrt{k}$ and for $f_n(x = 1)$ of $1$.

Hence, the choice of parameterization in principle induces a bias
in the uncertainties and correlations \emph{if} we let it determine
which parameters to impose independent theory constraints on.
Therefore, we should not choose the independent parameters based on the parameterization,
but rather the other way around. We should choose a parameterization for which we
are most confident that its $\theta_{n,i}$ can be considered to correspond to
independent sources of uncertainty. Furthermore, we can avoid
a parameterization bias by imposing theory constraints at the level of $f_n(x)$
itself. For example, we should always choose the central value directly for $f_n(x)$, which by
default can just be $f_n(x) = 0$. This is a parameterization-independent condition
on the central values of $\theta_n$ or $\theta_n'$ and thus avoids any
parameterization bias in the central value.
Similarly, we can impose for example an uncertainty based on the natural size of the
integral of $f_n(x)$ or its value at special points. Ultimately, however, this is
just another way of choosing what we consider to be the independent sources of uncertainty.
The TNPs can only help us to parameterize the independent sources
of uncertainty once we have identified them. They cannot decide for us what
they are. As soon as we want or need to impose theory constraints we cannot avoid making
this decision. This is another place where clearly domain knowledge is required.
What we can avoid though is to make an implicit or uninformed decision.
If we do not have enough information to decide, we have to limit
ourselves to unambiguous parameterization-independent constraints.

Our above discussion implies for strategy 3) that the choice between different
linearly related bases (polynomial or otherwise) is actually irrelevant
when fitting to data (apart from effects due to numerical stability etc.)
or imposing other parameterization-independent constraints.
It is only relevant for arguing where we are allowed to truncate
and perhaps for arguing which parameters we should consider to be independent.
The truncation itself does induce a parameterization-dependent bias, which
however can be phrased in terms of the above discussion: We can always think of
it as imposing a theory constraint on the parameters of the truncated
terms that their central value vanishes with an uncertainty whose net effect we
can neglect.

%===============================================================================
\subsection{Multiple dependencies}
\label{sec:param_multiple}
%===============================================================================

So far, we have assumed that the $x$-independent coefficients $f_{n,i}$ are
scalars so we can parameterize them by a scalar nuisance parameter. This is no
longer the case when $f_n$ depends on multiple internal variables whose
dependence we are required to parameterize, i.e., which fall into cases 2) or 3)
in \sec{param_requirements}. To generalize our discussion to this situation, it
is sufficient to discuss how to extend from the one-dimensional case $f_n(x)$ to
the two-dimensional case $f_n(x, y)$. The generalization to further variables
then proceeds in exactly the same way.

When we assemble the final prediction from
its ingredients there is typically a natural progression of the dependencies from the inner layers
to the outer layers, which we can also follow here. For example, the innermost
layer could be the color and $n_f$ dependence, then comes the kinematic dependence, next
we sum over partonic channels, and eventually at the outermost layer we sum or combine
different processes.

To be concrete, let us
denote the innermost relevant variable by $x$ and the next outer variable by $y$.
To parameterize the $y$ dependence, we can follow the same strategies
discussed in the previous subsections. The only difference is that once the $y$
dependence is stripped away, the $y$-independent coefficients $f_{n,j} \equiv f_{n,j}(x)$ are not
scalars but still functions of $x$. Each of them we can then parameterize in $x$
in terms of scalar parameters as we have discussed so far for $f_n(x)$.
For example, if $y$ is a discrete variable or only needed at fixed values
we simply have $f_{n,j}(x) = f_n(x, y_j)$.

The only more complicated case is when $x$ and $y$ are both continuous and appear
at the same layer, e.g., when considering a double-differential
spectrum in two kinematic variables.
If their dependence is separable, $f_n(x, y) = f_n(x) g_n(y)$, we can treat
each one-dimensional factor as before.
Finally, when we have a genuinely two-dimensional function
$f_n(x, y)$ and require correlations in both $x$ and $y$, we need to parameterize
the $x$ and $y$ dependencies simultaneously, for which we can follow the two-dimensional
generalization of the strategies in \sec{param_strategies}.

For strategy 1), we have to consider two-dimensional basic building blocks $\hat\phi_{n,ij}(x, y)$.
Clearly, finding a minimal parameterization of the true functional form is going to be more
difficult now, but it can still be possible if the $x$ and $y$ dependence is separable
or if it can be reduced to several one-dimensional functions which only depend on certain
combinations of $x$ and $y$.

For strategy 2), the $\eps$ expansion coefficients $f_{nl}(x, y)$ are in general
two-dimensional now. This strategy can be quite powerful to make the two-dimensional case
more tractable. By expanding in $\eps$, we might be able to simplify one or both dependencies or
make them separable or otherwise reduce the problem to the one-dimensional case.

For strategy 3), we have to consider two-dimensional functional bases $\phi_{n,ij}(x, y)$.
The approximation of multivariate functions is surprisingly more difficult than
the univariate case, and an active area of mathematical research.
Finding suitable multivariate parameterizations for an unknown multivariate
function is a similarly difficult problem. However, it is ultimately necessary to correctly
account for a genuinely multidimensional correlation structure if we lack the ability
to apply strategies 1) or 2).
The most straightforward is to consider a product basis $\phi_{n,ij}(x, y) = \phi_{n,i}(x)\phi_{n,j}(y)$,
which simply amounts to expanding $f_n(x, y)$ in $\phi_{n,j}(y)$ for fixed $x$ and then further
expanding the resulting $x$-dependent series coefficients in $\phi_{n,i}(x)$. Unfortunately,
the number of terms quickly proliferates -- the curse of dimensionality.
However, we can often
identify a primary variable $x$ and a secondary variable $y$, whose correlations
might matter less or which is going to be integrated over first. In this case
we can mitigate the curse of dimensionality by optimizing the basis in favor of $x$.

%===============================================================================
\subsection{Examples}
\label{sec:param_examples}
%===============================================================================

In this subsection, we discuss various dependencies to illustrate the general
discussion of the previous subsections.

%~~~~~~~~~~~~~~~~~~~~~~~~~~~~~~~~~~~~~~~~~~~~~~~~~~~~~~~~~~~~~~~~~~~~~~~~~~~~~~~
\subsubsection{\texorpdfstring{$N_c$}{Nc} dependence and color structure}
%~~~~~~~~~~~~~~~~~~~~~~~~~~~~~~~~~~~~~~~~~~~~~~~~~~~~~~~~~~~~~~~~~~~~~~~~~~~~~~~

When we are only interested in QCD corrections, the dependence on the number of
colors, $N_c$, is an example of case 1) in \sec{param_requirements}: We always
have fixed $N_c = 3$ and do not require correlations between different values of $N_c$.
This means we do not need separate $\theta_{n,i}$ for individual color
coefficients but only a single overall one for $f_n(N_c = 3)$.

Nevertheless, if we were to parameterize the $N_c$ dependence, it is a good example for
strategy 1) where the functional form is fully known, as we know exactly
which color coefficients composed of $C_A$, $C_F$, $T_F$, as well as higher invariants,
appear for a given $f_n$.

When considering QCD and QED corrections, we still do not need the full $N_c$-dependent
structure but effectively two pieces of it. The abelian parts of the QCD coefficients
are clearly correlated with the QED
coefficients. To correctly account for this correlation we have to separate the
abelian and nonabelian parts of the QCD color structure and parameterize each
with a separate TNP. The abelian one will then be shared by the QCD and QED
coefficients, whereas the nonabelian one only appears in the QCD coefficients.
In this way, the partial correlation between QCD and QED coefficients is correctly
accounted for.

An analogous discussion applies to QCD and electroweak corrections at
sufficiently high energies where the masses $m_V$ of the electroweak gauge bosons can
be neglected. When the gauge boson masses cannot be neglected it requires a
more detailed investigation to identify the possibly common parts. Effectively
one has to consider in addition the dependence on $m_V$ at two fixed points, namely
at the physical value of $m_V$ and in the $m_V\to 0$ limit.

%~~~~~~~~~~~~~~~~~~~~~~~~~~~~~~~~~~~~~~~~~~~~~~~~~~~~~~~~~~~~~~~~~~~~~~~~~~~~~~~
\subsubsection{\texorpdfstring{$n_f$}{nf} dependence}
%~~~~~~~~~~~~~~~~~~~~~~~~~~~~~~~~~~~~~~~~~~~~~~~~~~~~~~~~~~~~~~~~~~~~~~~~~~~~~~~

When the number of flavors, $n_f$, is the same in all considered predictions,
as is often the case with $n_f = 5$, we are in case 1) and do not require
correlations in $n_f$ and only a single TNP for $f_n(n_f = 5)$.
When we do cross flavor thresholds and require $f_n(n_f)$ at
different $n_f$ values, we do need to parameterize the $n_f$
dependence to account for the (de)correlation between say $n_f = 5$ and $n_f = 4$.

The $n_f$ dependence is another example where strategy 1)
is easily applicable, since $f_n(n_f)$ is a polynomial in $n_f$ of known degree.
This actually poses an interesting theoretical question, namely which parts
of the $n_f$ dependence are conceptually independent. Neither
the naive choice to consider the coefficients of $n_f^i$ as independent nor
rewriting $n_f$ in terms of $\beta_0(n_f)$ and considering the coefficients of $\beta_0(n_f)^i$
as independent seem to be supported by empirical evidence.
Instead, empirical evidence suggests that the coefficients of $(C_A - T_F n_f)^i$
are independent. This can likely
be attributed to the screening effect of quarks, see \sec{tnp_normalization}
for some further discussion.

%~~~~~~~~~~~~~~~~~~~~~~~~~~~~~~~~~~~~~~~~~~~~~~~~~~~~~~~~~~~~~~~~~~~~~~~~~~~~~~~
\subsubsection{Partonic channels}
%~~~~~~~~~~~~~~~~~~~~~~~~~~~~~~~~~~~~~~~~~~~~~~~~~~~~~~~~~~~~~~~~~~~~~~~~~~~~~~~

The dependence on different partonic channels is a primary example of a
genuinely discrete dependence. Here, strategy 1) is immediately applicable, as
we know exactly which partonic channels appear at a given order, and it
amounts to separately parameterizing each partonic channel.

One might ask the question when we are actually required to
separate the partonic channels. One reason is when we require hadron-collider
predictions at different center-of-mass energies, $\Ecm$, since the
$\Ecm$ dependence enters via the different parton luminosities for each channel,
which can have very different scaling with $\Ecm$. Another reason is to capture correlations
between different processes that share common partonic channels, see below.

Another important reason to separately parameterize partonic channels is
to anticipate new channels that only open up at higher orders but can
have sizeable contributions, which is a classic case where scale variations can
fail badly. This is an example where parameterizing the dependence, even if not
required for correlations, can be of advantage for figuring out the natural
size of the TNPs.

%~~~~~~~~~~~~~~~~~~~~~~~~~~~~~~~~~~~~~~~~~~~~~~~~~~~~~~~~~~~~~~~~~~~~~~~~~~~~~~~
\subsubsection{Process dependence}
%~~~~~~~~~~~~~~~~~~~~~~~~~~~~~~~~~~~~~~~~~~~~~~~~~~~~~~~~~~~~~~~~~~~~~~~~~~~~~~~

Another type of correlation is that between different processes. This tends to
be a more complicated dependence to take into account as it requires
detailed knowledge of the internal structure.
To correctly correlate the process dependence we basically have to map it into the
dependence on some internal variables $x$. Luckily, the most relevant cases, namely
closely related processes expected to be strongly correlated,
are also the most straightforward. For example, for $W$ vs.\ $Z$ production, the
process dependence essentially maps into the dependence on
partonic channels and electroweak gauge couplings and boson masses.
We will see an explicit example in \sec{qT}.

%~~~~~~~~~~~~~~~~~~~~~~~~~~~~~~~~~~~~~~~~~~~~~~~~~~~~~~~~~~~~~~~~~~~~~~~~~~~~~~~
\subsubsection{Continuous dependencies}
%~~~~~~~~~~~~~~~~~~~~~~~~~~~~~~~~~~~~~~~~~~~~~~~~~~~~~~~~~~~~~~~~~~~~~~~~~~~~~~~

A typical example of a genuinely continuous dependence is that of a differential
spectrum. We will discuss the example of the $q_T$ spectrum in detail in \sec{qT},
which is going to involve a repeated application of strategies 1 and 2.
Another generic example is the dependence on the partonic momentum fractions $z_{a,b}$
of partonic cross sections in hadronic collisions. Here, if we are only asking about
a total cross section, the $z_a$ and $z_b$ dependence is effectively projected onto a single
number. If we consider a kinematic distribution that effectively measures the total invariant
mass $Q$ of the hard process, then we need the one-dimensional dependence on $z = z_a z_b$.
Finally, if we are sensitive to both the total invariant mass and rapidity of the hard
process we need the full dependence on $z_a$ and $z_b$. Suitably parameterizing this dependence
is in general nontrivial. Often though, the cross section tends to be dominated
by the $z\to 1$ limit, which can be a good starting point by applying strategy 2
with $\varepsilon = 1-z$. This strategy has already proven very useful in other
cases where the dependence on partonic momentum fractions arises, namely to parameterize
the unknown parts of beam function matching kernels~\cite{Billis:2019vxg} or QCD splitting functions~\cite{McGowan:2022nag} in terms of TNPs.

%%%%%%%%%%%%%%%%%%%%%%%%%%%%%%%%%%%%%%%%%%%%%%%%%%%%%%%%%%%%%%%%%%%%%%%%%%%%%%%%
\section{Theory Constraints for Scalar Series}
\label{sec:tnp_scalar}
%%%%%%%%%%%%%%%%%%%%%%%%%%%%%%%%%%%%%%%%%%%%%%%%%%%%%%%%%%%%%%%%%%%%%%%%%%%%%%%%

In this section, we discuss TNPs for perturbative series with scalar coefficients $f_n$ and
how to obtain robust theory constraints on them, which belongs to the second step
of applying the TNP approach as discussed in \sec{tnp_method_continued}.

We assume that the parameterization of any relevant outer levels of $x$
dependences as discussed in \sec{parameterization} has happened and has reduced
the remaining perturbative series to have scalar coefficients $f_n$,
as will be relevant for the application in \sec{qT}. We limit ourselves to QCD corrections at
fixed $n_f = 5$. Further investigations beyond this case are of course
warranted but are well beyond our scope here and are left to future work.

Hence, the starting point for our discussion in this section is that we have
a QCD series in $\alpha_s$ with scalar coefficients $f_n$ that can be parameterized
by a single theory nuisance parameter,
%%%
\begin{equation}
f_n(n_f = 5, \theta_n) = N_n(n_f = 5)\, \theta_n
\,.\end{equation}
%%%
To simplify the notation, we will suppress the $n_f = 5$ argument from here on.
The normalization factor $N_n$ accounts for the expected natural size of
$f_n$, i.e., it should be chosen such that we generically
expect $\abs{\hat f_n} \lesssim N_n$. Consequently, the expected
natural size of $\theta_n$ is $\abs{\hat\theta_n}\lesssim 1$.

%===============================================================================
\subsection{Overview}
%===============================================================================

Not knowing the true value $\hat\theta_n$ of $\theta_n$, our goal is to obtain an estimate
as in \eq{theta_estimate},
%%%
\begin{equation} \label{eq:theta_estimate_2}
\theta_n = u_n \pm \Delta u_n
\,,\end{equation}
%%%
based on theoretical arguments.
This will be our baseline theory constraint on $\theta_n$, which we use to
evaluate the theory uncertainty in the absence of any additional constraint
from other sources of information.

Without additional information we will usually just take $u_n = 0$ as our
best-guess central value.
We then need to assign an uncertainty $\Delta u_n$ to this choice, which
determines the amount by which we vary $\theta_n$ and thus the size of the
resulting theory uncertainty.
When we need a statistical treatment of \eq{theta_estimate_2},
we also need the probability distribution $P(u_n|\theta_n)$ of $u_n$.
For this purpose, we treat $u_n$ as if it came from a measurement with
a Gaussian 1$\sigma$ uncertainty of $\Delta u_n$.
More precisely, we model our estimator $u_n$ for $\theta_n$ as a
Gaussian-distributed random variable with mean
$\mu = \theta_n$ and standard deviation $\sigma = \Delta u_n$.
This is a standard assumption also used for
nuisance parameters of experimental systematic uncertainties,
whose justification basically stems from the central-limit theorem.
In \sec{tnp_statistics}, we will find strong empirical evidence that
$u_n$ can indeed be considered as a Gaussian-distributed random
variable. We will thus refer to the theory uncertainties that result from
varying a theory constraint by $\pm\Delta u_n$ as one ``theory-$\sigma$'' uncertainty
or 68\% ``theory CL''. Similarly, 95\% theory CL refers to varying by $\pm2\Delta u_n$.

Following our discussion in \sec{philosophy}, $\Delta u_n$ is not given by the distance
$\abs{\hat\theta_n - u_n}$ of our estimate $u_n$ to the true value $\hat\theta_n$.
Thus, to estimate $\Delta u_n$ we \emph{do not} need to estimate a
precise value of $\hat\theta_n$. (Our best guess for $\hat\theta_n$ is already represented
by $u_n$). Rather, $\Delta u_n$ must reflect our limited knowledge.
With the above statistical interpretation this means we need to choose $\Delta u_n$ such that
$\abs{\hat\theta_n - u_n} \leq \Delta u_n$ with 68\% confidence.
For $u_n = 0$, $\Delta u_n$ is thus determined by the natural size
of $\theta_n$, $\abs{\hat\theta_n} \lesssim \Delta u_n$, and so with our choice of normalization
we have $\Delta u_n \simeq 1$.

If we believe to know absolutely nothing about $f_n$, it would imply to take
$\Delta u_n = \infty$ so $\theta_n$ would be left to vary unconstrained within
$[-\infty, \infty]$. In other words, we would treat $\theta_n$ as a truly
unknown parameter to be determined from data. In many cases, this is of course
too pessimistic as we do have some expectations and plenty of experience of the
typical size of higher-order corrections. Therefore, to choose an appropriate
$\Delta u_n$ we proceed in two steps: In the first step in \sec{tnp_normalization}, we use
theoretical arguments to estimate the expected natural size of $f_n$. That is,
we determine the normalization $N_n$ for which we expect $\abs{\hat f_n} \lesssim N_n$ so
$\abs{\hat\theta_n}\lesssim 1$ and $\Delta u_n \simeq 1$.
Based purely on theoretical expectations we can only
hope to narrow $\Delta u_n$ down to an $\ord{1}$ factor, perhaps a factor of two
at best. Therefore, in the second step in \sec{tnp_statistics} we study
the true values $\hat\theta_n$ of many known series
of a common category. This will provide us with the empirical evidence
to verify and further narrow down the value of $\Delta u_n$ and also
to confirm its statistical interpretation in terms of the probability distribution
$P(u_n|\theta_n)$.

%===============================================================================
\subsection{Normalization and estimate of natural size}
\label{sec:tnp_normalization}
%===============================================================================

We consider two general categories of perturbative quantities. The first are quantities
corresponding to the finite constant terms of matrix elements, which we refer to
as matrix-element ``constants'' and for which we continue to use the generic
notation $\f(\alpha_s)$. This includes total cross sections and decay rates as well as
the constant (nonlogarithmic) terms (RG boundary conditions) of
matching coefficients and matrix elements of renormalized operators.
The second type are anomalous dimensions, denoted generically as $\gamma(\alpha_s)$,
which correspond to the coefficients of $1/\eps$ poles in the bare perturbative series.
We distinguish these two categories because we expect and find their perturbative
series to behave somewhat differently.

%~~~~~~~~~~~~~~~~~~~~~~~~~~~~~~~~~~~~~~~~~~~~~~~~~~~~~~~~~~~~~~~~~~~~~~~~~~~~~~~
\subsubsection{Matrix-element constants}
\label{sec:tnp_constants}
%~~~~~~~~~~~~~~~~~~~~~~~~~~~~~~~~~~~~~~~~~~~~~~~~~~~~~~~~~~~~~~~~~~~~~~~~~~~~~~~

We write the perturbative series for matrix-element constants as
%%%
\begin{equation} \label{eq:const_series}
\f(\alpha_s) = 1 + \sum_{n = 1} \f_n\,\Bigl(\frac{\alpha_s}{4\pi}\Bigr)^n
\,,\end{equation}
%%%
which defines their perturbative coefficients $\f_n$.
We normalize all quantities such that their leading-order result is $\f_0 = 1$,
since it only contains overall couplings and prefactors, which are always known,
and so does not yet contain nontrivial information about the perturbative series.
We choose the normalization $N_n^\f$ to parameterize $\f_n$ in terms of $\theta_n^\f$ as
%%%
\begin{equation} \label{eq:NnF}
\f_n(\theta_n^\f) = N_n^\f\,\theta_n^\f
\qquad\text{with}\qquad
N_n^\f = 4^n C_n (n-1)!
\,.\end{equation}
%%%
Here, $C_n  = C_r C_A^{n-1}$ is the leading color factor of $\f_n$ with $C_r$ the one-loop
color factor, which depends on the color representation of the external particles,
i.e., $C_r = C_F$ for external quarks and $C_r = C_A$ for external gluons.
Note that we merely use the leading-color limit to determine the normalization.
We do not make a leading-color approximation anywhere. As discussed in \sec{param_examples},
we do not need to parameterize the full color structure of the coefficients
because here we are only interested in QCD and fixed values of $N_c$. We will
explain the other factors in a moment.

The above discussion applies to tree-level quantities. Considering quantities that are
inherently loop induced, their overall normalization is defined to be consistent
with that of an associated tree-level quantity. Typical examples would be an off-diagonal partonic
channel that has an associated diagonal partonic channel, or
a singlet coefficient that has an associated nonsinglet coefficient.
Their leading $n$-loop color factor is then given by the color factor of the first
appearing loop order times one power of $C_A$ for each additional loop order.

As an instructive example to understand this choice of $N_n^\f$, let us consider
the $q\bar q$ vector, $q\bar q$ scalar, and
$gg$ matching coefficients, corresponding to the infrared-finite
parts of the respective QCD form factors, which are known to four loops~\cite{Lee:2022nhh, Chakraborty:2022yan}.
The perturbative series of their respective constant terms are denoted as $c_{q\bar qV}(\alpha_s)$,
$c_{q\bar qS}(\alpha_s)$, and $c_{gg}(\alpha_s)$. (The $c_{q\bar qV}(\alpha_s)$
coefficient is defined in more detail in \sec{qT_TNPs}.)
In \tab{wilson_coefficients}, we show the true values $\hat \f_n/N_n$ for all three
matching coefficients successively dividing out the normalization factor $N_n^\f$:
%%%
\begin{itemize}
\item The first line in each block shows the raw values for $\hat \f_n$, which
grow very large for increasing $n$. Naively, there would be little hope to
directly estimate the correct expected size of these numbers.

\item In the second lines, we divide out a factor of $4^n$, which
basically removes the $1/4^n$ in \eq{const_series}. It is clear that the conventional
$1/(4\pi)^n$ loop factor is artificial in this regard and a main reason for the
quickly increasing magnitude of the coefficients.
We could of course have directly expanded \eq{const_series} in terms of $\alpha_s/\pi$,
which is actually known to be a more appropriate expansion parameter.
The reason we did not do so is for the
sake of illustration here and because defining the series coefficients with respect
to $\alpha_s/(4\pi)$ is the most commonly used convention.
Nevertheless, the resulting numbers are still far from $\ord{1}$.

\item In the third lines, we further divide out the leading color factor $C_r C_A^n$
appearing at $n$-loop order, which brings the numbers to $\ord{1}$ as we might expect.

\item Finally, in the fourth and last line, we further divide out a factor of
$(n-1)!$, which amounts to $\{1, 1, 2, 6\}$ for $n = \{1, 2, 3, 4\}$ and which is
clearly still present in $\hat \f_3/N_3$ and $\hat \f_4/N_4$ in the previous line.
The appearance of this factor also matches our expectation of the factorial
growth of the series coefficients.
\end{itemize}
%%%
The last line in each block in \tab{wilson_coefficients} corresponds to the
nominal $N_n^\f$ in \eq{NnF} with the numbers in
bold corresponding to the true values $\hat\theta_n^\f$, which indeed
satisfy $\abs{\hat\theta_n^\f} \lesssim 1$ to well within a factor of two as
desired.
The above arguments leading to this choice of $N_n^\f$ are generic and not specific to
the given examples. Hence, we consider it as a very plausible generic expectation for the
natural size of $\f_n$, and consequently we can consider $\Delta u_n \simeq 1$
as a plausible uncertainty.

%-------------------------------------------------------------------------------
\begin{table}[t]
\centering
\renewcommand{\arraystretch}{1.1}
\begin{tabular}{cc|rrrr}
\hline\hline
$\f(\alpha_s)$ & $N_n$ & $\hat \f_1/N_1$ & $\hat \f_2/N_2$ & $\hat \f_3/N_3$ & $\hat \f_4/N_4$
\\\hline
$c_{q\bar qV}(\alpha_s)$
& 1                          & $-8.47$ & $-48.6$ & $-1387$ & $-42015$
\\
& $4^n$                      & $-2.12$ & $-3.04$ & $-21.7$ & $-164$
\\
& $4^n C_F C_A^{n-1}$        & $-1.59$ & $-0.76$ & $-1.81$ & $-4.56$
\\
& $4^n C_F C_A^{n-1} (n-1)!$ & \bf $-1.59$ & \bf $-0.76$ & \bf $-0.90$ & \bf $-0.76$
\\\hline
$c_{q\bar qS}(\alpha_s)$
& 1                          & $-0.47$ & $+87.1$ & $+2309$ & $+76100$
\\
& $4^n$                      & $-0.12$ & $+5.44$ & $+36.1$ & $+297$
\\
& $4^n C_F C_A^{n-1}$        & $-0.09$ & $+1.36$ & $+3.01$ & $+8.26$
\\
& $4^n C_F C_A^{n-1} (n-1)!$ & \bf $-0.09$ & \bf $+1.36$ & \bf $+1.50$ & \bf $+1.38$
\\\hline
$c_{gg}(\alpha_s)$
& 1                          & $+4.93$ & $-24.0$ & $-4066$ & $-123979$
\\
& $4^n$                      & $+1.23$ & $-1.50$ & $-63.5$ & $-484$
\\
& $4^n C_A C_A^{n-1}$        & $+0.41$ & $-0.17$ & $-2.35$ & $-5.98$
\\
& $4^n C_A C_A^{n-1} (n-1)!$ & \bf $+0.41$ & \bf $-0.17$ & \bf $-1.18$ & \bf $-1.00$
\\\hline\hline
\end{tabular}
\caption{True values of the series coefficients $\hat \f_n$ divided by various normalization
factors $N_n$ for the quark vector (top block), quark scalar (middle block), and gluon (bottom block)
matching coefficients. The numbers in bold in the last line of each block are the $\hat\theta_n^\f$.}
\label{tab:wilson_coefficients}
\end{table}
%-------------------------------------------------------------------------------

This exercise already teaches us several interesting things and dispels
some common lore. First, gluonic quantities do not necessarily have genuinely larger perturbative corrections than quark ones. Once the different overall color factor of $C_r = C_A$ vs.\ $C_r = C_F$ is accounted for, the remaining normalized coefficients for $C_{gg}$ have the same generic size as those for
$C_{q\bar q V}$ and $C_{q\bar qS}$. In fact, one of the latter is always larger than $C_{gg}$ at each order.
Secondly, once normalized to their natural size, the specific size of the coefficient(s) of previous order(s)
is not a good indicator for the size of the coefficient(s) at the following order(s).
In other words, one should not look at this table from left to right but only from
top to bottom.
Thirdly, the coefficients are not always or mostly positive and may change sign
at different orders.

The convention to have $\f_0 = 1$ does not yet uniquely fix
the overall convention for $\f(\alpha_s)$, as we could still raise
$\f(\alpha_s)$ to some power, which keeps $\f_0 = 1$, but changes the $\f_n$. For example,
by squaring a series with all $\f_n = 1$ we get
%%%
\begin{equation}
\biggl[1 + \sum_{n = 1} \alpha^n \biggr]^2 = 1 + \sum_{n=1} (n+1)\,\alpha^n
\,.\end{equation}
%%%
Therefore, by taking the square or square root of $\f(\alpha_s)$, the natural size
of $\f_n$ can change by an $\ord{n}$ factor, which we clearly have to account for
if we aim for an estimate to within a factor of two or better.
We find the normalization in \eq{NnF} to be appropriate for the
convention that $\f(\alpha_s)$ is raised to an appropriate power such that it
effectively scales as a matrix element with two (resolved or Born-level) external
QCD partons, as is the
case for the matching coefficients considered above, or equivalently a squared matrix
element with a single (resolved or Born-level) external QCD parton. This means we
consider jet and beam functions as they are, since they can be regarded either as
forward $1\to 1$ matrix elements or $1$-parton squared matrix elements.
On the other hand, we consider the square root of $0\to2$ cross sections and
decay rates and also of soft functions with two Wilson lines.
We might argue that this is also natural from the point of view of identifying the conceptually
independent perturbative corrections, since they fundamentally
appear for the matrix element and not its square. A typical example is a large
correction to the NLO matrix element whose square then also causes a large NNLO correction
to the cross section. By considering the square root of the cross section, this
is effectively accounted for.%
\footnote{In the future, instead of just taking the square root for cross-section-like
quantities, it might be worth to investigate the option of directly parameterizing
and estimating the real and imaginary parts of the underlying complex amplitude.
This is clearly more challenging due to the presence of IR divergences
and also because in the literature perturbative results are often provided only
for the cross section.}

Finally, there is one more subtlety to consider. The attentive reader might have
wondered that the factorial factor in \eq{NnF} is $(n-1)!$ and not just $n!$.
As it turns out, an $n!$ factor is indeed appropriate in the pure gauge theory
with $n_f = 0$. The factorial growth can be attributed to bubble chains inserted
into gluon propagators. Including fermions, the leading
$n_f$ dependence comes from replacing a gluon bubble by a fermion loop and
generically appears as $C_A - T_F n_f$, which vanishes for $n_f = 6$ (for QCD
with $C_A = N_c = 3$ and $T_F = 1/2$).
Empirically, we find that this quark screening effect reduces the size
of the corrections by $1/n$ turning the $n!$ behaviour for $n_f = 0$ into $(n-1)!$
for $n_f \simeq 6$. This applies when the $n_f$ dependence
starts at $n = 2$. In some (but not all) cases with external gluons, the $n_f$
dependence starts at $n = 1$, in which case $n$ must be increased by one in
the factorial factor, so we would use $n!$ in \eq{NnF} for $n_f = 5$.

%~~~~~~~~~~~~~~~~~~~~~~~~~~~~~~~~~~~~~~~~~~~~~~~~~~~~~~~~~~~~~~~~~~~~~~~~~~~~~~~
\subsubsection{Anomalous dimensions}
\label{sec:tnp_anom_dims}
%~~~~~~~~~~~~~~~~~~~~~~~~~~~~~~~~~~~~~~~~~~~~~~~~~~~~~~~~~~~~~~~~~~~~~~~~~~~~~~~

We write the perturbative series for anomalous dimensions as
%%%
\begin{equation} \label{eq:gamma_series}
\gamma(\alpha_s) = \sum_{n = 0} \gamma_n\, \Bigl(\frac{\alpha_s}{4\pi}\Bigr)^{n+1}
\,,\end{equation}
%%%
which defines their coefficients $\gamma_n$.
Their overall normalization is less obvious than for $\f(\alpha_s)$ since they
start at loop-level, so $\gamma_0$ appears at the
same order as $\f_1$ and already contains nontrivial perturbative information.
The anomalous dimensions correspond to logarithmic $\mu$ derivatives of matrix elements so
the ambiguity of raising $\f(\alpha_s)$ to some power corresponds to multiplying
$\gamma(\alpha_s)$ by some overall factor. We therefore decide to fix the normalization
convention for $\gamma(\alpha_s)$, including its overall
sign, analogous to that of $\f(\alpha_s)$ so it corresponds to
the anomalous dimension of some $\f(\alpha_s)$, i.e., the derivative with respect to
$\ln\mu$ of a matrix element with two external QCD partons.

%-------------------------------------------------------------------------------
\begin{table}[t]
\centering
\renewcommand{\arraystretch}{1.1}
\begin{tabular}{cc|rrrrr}
\hline\hline
$\gamma(\alpha_s)$ & $N_n$ & $\hat\gamma_0/N_0$ & $\hat \gamma_1/N_1$ & $\hat \gamma_2/N_2$ & $\hat \gamma_3/N_3$ & $\hat \gamma_4/N_4$
\\\hline
$\beta$
& 1                          & $-15.3$ & $-77.3$ & $-362$ & $-9652$ & $-30941$
\\
& $4^{n+1}$                  & $-3.83$ & $-4.83$ & $-5.65$ & $-37.7$ & $-30.2$
\\
& $4^{n+1} C_F C_A^n$        & \bf $-1.28$ & \bf $-0.54$ & \bf $-0.21$ & \bf $-0.47$ & \bf $-0.12$
\\\hline
$\gamma_m$
& 1                          & $-8.00$ & $-112$ & $-950$ & $-5650$ & $-85648$
\\
& $4^{n+1}$                  & $-2.00$ & $-7.028$ & $-14.8$ & $-22.1$ & $-83.6$
\\
& $4^{n+1} C_F C_A^n$        & \bf $-1.50$ & \bf $-1.76$ & \bf $-1.24$ & \bf $-0.61$ & \bf $-0.77$
\\\hline
$2\Gamma_{\rm cusp}^q$
& 1                          & $+10.7$ & $+73.7$ & $+478$ & $+282$ & $(+140000)$
\\
& $4^{n+1}$                  & $+2.67$ & $+4.61$ & $+7.48$ & $+1.10$ & $(+137)$
\\
& $4^{n+1} C_F C_A^n$        & \bf $+2.00$ & \bf $+1.15$ & \bf $+0.62$ & \bf $+0.03$ & \bf $(+1.27)$
\\\hline\hline
\end{tabular}
\caption{True values of the series coefficients $\hat \gamma_n$ divided by various normalization
factors $N_n$ for the QCD $\beta$ function~\cite{Tarasov:1980au, Larin:1993tp, vanRitbergen:1997va, Czakon:2004bu, Baikov:2016tgj, Herzog:2017ohr, Luthe:2017ttg}, the quark-mass anomalous dimension~\cite{Tarasov:1982plg, Larin:1993tq, Chetyrkin:1997dh, Vermaseren:1997fq, Baikov:2014qja, Luthe:2016xec}, and the quark cusp
anomalous dimension~\cite{Moch:2004pa, Vogt:2004mw, Henn:2019swt, vonManteuffel:2020vjv}. The numbers in bold in the last line of each block are the $\hat\theta_n^\gamma$.
The 5-loop result for the quark cusp anomalous dimension~\cite{Herzog:2018kwj} in brackets is only known approximately.
}
\label{tab:anom_dims}
\end{table}
%-------------------------------------------------------------------------------

We then choose the normalization $N_n^\gamma$ to parameterize $\gamma_n$ in terms of $\theta_n^\gamma$ as
%%%
\begin{equation}
\gamma_n(\theta_n^\gamma) = N_n^\gamma\,\theta_n^\gamma
\qquad\text{with}\qquad
N_n^\gamma = 4^{n+1} C_{n+1}
\,,\end{equation}
%%%
where $C_{n+1}$ is again the leading $(n+1)$-loop color factor, typically given by
$C_{n+1} = C_r C_A^n$ with $C_r$ the one-loop color factor determined by the color
representation of the external legs.
To motivate this normalization, we show the known true values for a few anomalous
dimensions in \tab{anom_dims} successively dividing out the normalization factor $N_n^\gamma$.
We find a quite similar pattern as before for the constants $\f_n$:
\begin{itemize}
\item The first line in each block shows the raw values for $\hat \gamma_n$, which
quickly grow large as $n$ increases. There would again be little hope to
directly estimate the size of these numbers.

\item In the second lines, we divide out a factor of $4^{n+1}$, which
removes the $1/4^{n+1}$ in \eq{gamma_series}. We see again that the conventional
$1/(4\pi)^{n+1}$ loop factor artificially enlarges the size of the coefficients.

\item In the third lines, we further divide out the leading $n$-loop color factor,
which yields numbers that are $\lesssim 1$ within a factor of two.
\end{itemize}
%%%
In contrast to the matrix-element constants, no factorial factor appears for the
anomalous dimensions, which is not entirely unexpected. However, we still find
that for $n_f = 0$ the coefficients are enhanced by a factor of $n$ due to the
absence of the quark screening compared to $n_f \simeq 6$. Also, the sign of the
higher-order coefficients now tends to be determined by the sign of $\gamma_0$
for $n_f \leq 5$, while for $n_f = 6$ the coefficients do change sign at
different orders. We leave a more detailed investigation and parameterization of
the $n_f$ dependence to the future.

%===============================================================================
\subsection{Validation and statistical interpretation}
\label{sec:tnp_statistics}
%===============================================================================

%~~~~~~~~~~~~~~~~~~~~~~~~~~~~~~~~~~~~~~~~~~~~~~~~~~~~~~~~~~~~~~~~~~~~~~~~~~~~~~~
\subsubsection{Statistical model and interpretation}
%~~~~~~~~~~~~~~~~~~~~~~~~~~~~~~~~~~~~~~~~~~~~~~~~~~~~~~~~~~~~~~~~~~~~~~~~~~~~~~~

For a real measurement, performing a single measurement corresponds to drawing
a value $u_n$ from $P(u_n|\hat\theta_n)$ with the measurement's uncertainty
$\Delta u_n$ corresponding to the standard deviation of $P(u_n|\hat\theta_n)$.
To verify the assigned $\Delta u_n$ and shape of $P(u_n|\hat\theta_n)$
we would repeat the measurement many times, i.e., we would draw a sample of many
$u_n$ values from $P(u_n|\hat\theta_n)$ for fixed $\hat\theta_n$ and study
its sample distribution.

With our idealized measurement we do not have the option to repeat
the measurement, so we cannot sample $P(u_n|\hat\theta_n)$
in $u_n$ for fixed $\hat\theta_n$. However, $P(u_n|\theta_n)$ models
our entire estimation procedure, which we can apply to all perturbative
series $f$ that we consider to belong to a common category. We can thus sample
$P(u_n|\hat\theta_n)$ over $\hat\theta_n$ by applying our
estimation procedure to many different parameters $\theta_n^f$ of the same category
whose true values $\hat\theta_n^f$ are known.

Given our estimator $u_n$ for the parameter $\theta_n$ with estimated uncertainty
$\Delta u_n$, we can consider the pull
%%%
\begin{equation}
t_n = \frac{\hat\theta_n - u_n}{\Delta u_n}
\,,\end{equation}
%%%
which is invariant under a linear transformation
$\theta_n\to a\theta_n+b$, $u_n\to a u_n+b$, $\Delta u_n\to a \Delta u_n$.
Since $P(u_n|\theta_n)$ should be invariant under such a rescaling,
we can consider it to be a function of the pull only,
%%%
\begin{equation} \label{eq:P_un_thetan}
P(u_n|\theta_n) = p\Bigl(\frac{\theta_n - u_n}{\Delta u_n}\Bigr)
\,.\end{equation}
%%%
In particular, if we model $u_n$ as a Gaussian random variable with
mean $\theta_n$ and standard deviation $\Delta u_n$, then $t_n$ is normally distributed,
i.e., $p(t_n)$ is a Gaussian with zero mean and unit variance.

As long as our estimation procedure is deterministic and involves choosing
specific values for $u_n$ and $\Delta u_n$, we can always perform a linear rescaling
and redefine $\theta_n$ to have $u_n = 0$ and $\Delta u_n = 1$ so $t_n = \hat\theta_n$.
In fact, we already did so by choosing our common normalization conventions as discussed in
\sec{tnp_normalization}, which are thus
an integral part of the estimation procedure. The likelihood
for $\theta_n$ is then given by
%%%
\begin{equation} \label{eq:L_thetan}
L(\theta_n) = P(u_n=0|\theta_n) = p(\theta_n)
\,,\end{equation}
%%%
and so we like to learn about the distribution $p(\theta_n)$.

Let us denote the collection of perturbative series $f$ of a given category by $F$
and the corresponding collection of their series coefficients $f_n$ as $F_n$.
In principle, the distribution $p(\theta_n)$ could be specific to
each parameter $\theta_n^f$, so to be clear for the moment let us label it and use a generic
argument, $p_{\theta_n^f}(x)$.
However, since it is primarily a property of our
estimation procedure, which is common to all $f_n\in F_n$,
we can assume it to be the same for all of their respective $\theta_n^f$.
Furthermore, we can naturally identify this common distribution with
the distribution of true values $\hat\theta_n^f$ of all $f_n \in F_n$, which we
denote as $\hat p_{F_n}(x)$, so
%%%
\begin{equation} \label{eq:hatp_Fn}
p_{\theta_n^f}(x) \equiv \hat p_{F_n}(x) \qquad \forall \theta_n^f \text{ whose } f_n \in F_n
\,.\end{equation}
%%%
Although this identification comes natural it is an assumption we
make. Intuitively, we can think of it as follows:
The collection of series coefficients in $F_n$ is a QCD bag of balls. Each ball
has a visible label $f_n$ on it and a not visible number $\hat\theta_n^f$
inside it.
We now consider a specific coefficient $f_n$ of interest for which we need
an estimate. With the identification in \eq{hatp_Fn},
we think of this situation as having just taken the ball labelled $f_n$
out of the bag, which is not random. But we are not allowed (or able) to look at its
number inside it, so we have effectively drawn a random member from the population of
hidden $\hat\theta_n^f$ numbers in the bag.
Knowing that it came out of this bag (and nothing else about it), our best estimate of
its value is simply the population mean and its uncertainty the population variance.%
\footnote{More precisely, we can obtain an estimate for $\theta_n$ based on the
likelihood $L(\theta_n) = \hat p_{F_n}(\theta_n)$. For a Gaussian (or similar)
distribution the maximum likelihood estimate coincides with the mean of the
distribution.}

For the rest of our discussion, we will work under the premise of this
identification. Without it, we would have to live with a stronger assumption of assuming
a certain shape for $p_{\theta_n^f}(x)$. We could also be somewhere in the middle and
consider the form of $\hat p_{F_n}(x)$ only as a motivation for the assumed shape of $p_{\theta_n^f}(x)$
but without making the explicit identification. Ultimately, the precise interpretation
is a choice the user of our theory constraint can make.

%~~~~~~~~~~~~~~~~~~~~~~~~~~~~~~~~~~~~~~~~~~~~~~~~~~~~~~~~~~~~~~~~~~~~~~~~~~~~~~~
\subsubsection{Distribution of known perturbative series}
\label{sec:tnp_distribution}
%~~~~~~~~~~~~~~~~~~~~~~~~~~~~~~~~~~~~~~~~~~~~~~~~~~~~~~~~~~~~~~~~~~~~~~~~~~~~~~~

We now discuss the distribution $\hat p_{F_n}(x)$ of the population of
true values $\hat\theta_n^f$ of all $f_n \in F_n$.
We consider two collections of perturbative series belonging to the two broad categories of
matrix-element constants ($F_\f$) and anomalous dimensions ($F_\gamma$) defined
at the beginning of \sec{tnp_normalization}.

The true distribution $\hat p_{F_n}(x)$ is obviously not known to us, as we would
have to know all possible $\hat\theta_n^f$. Instead, we can follow the standard
procedure of estimating an unknown population
distribution by drawing a random sample from the population and using the resulting
sample distribution as an approximation of the true population distribution.
In our case, we can use the sample $\{\hat\theta_n^f\}$ belonging to a
subset of known series coefficients $\{\hat f_n \} \subset F_n$, which
is indeed random because we choose it without having a prior look at the
actual values $\hat\theta_n^f$.
In terms of our QCD bag of balls, by default all balls
are locked and we cannot look inside them. While taking out a specific (not random)
set of balls which someone has graciously unlocked for us, we are not yet
looking at their numbers inside. Hence, just like when we are asking about a specific $f_n$,
this amounts to drawing a random sample of $\hat\theta_n^f$ from the population
inside the bag.

We might still worry that the sample distribution could be biased by the
fact that the perturbative series that are known to high order are
naturally simpler to calculate than the ones we do not yet know. Whilst this makes
the quantities themselves special in some sense, the only relevant question is whether
this also makes their values $\hat\theta_n^f$ somehow special and not representative
of the full population, which is not necessarily the case.
The sample distribution not being representative (yet) can indeed be a valid concern
when only a handful and perhaps even closely related series are available.
To alleviate this concern and ensure a sample as representative as possible, we
have made an effort to include a large variety of different QCD quantities.
Furthermore, from our repeated experience of adding new results to the existing
samples over time, we do not believe this to be a concern any longer.

A detailed list of the quantities included in our sample is given in \app{series_sample}.
We have included all four-loop results for matrix-element constants and all
four-loop and five-loop results for anomalous dimensions we are aware of
(without any claim of completeness), as well as all known three-loop matrix-element constants
relevant for $q_T$ and thrust resummation to N$^4$LL.
To include a quantity in our sample, we have to be sure that it actually
belongs to one of our common categories and roughly obeys our natural size estimate.
For this reason
we focus on series that are known to at least third order including their
$n_f$ dependence, which allows for sufficient sanity checks.

%-------------------------------------------------------------------------------
\begin{figure}
\includegraphics[width=\WidthTwoSubfigs]{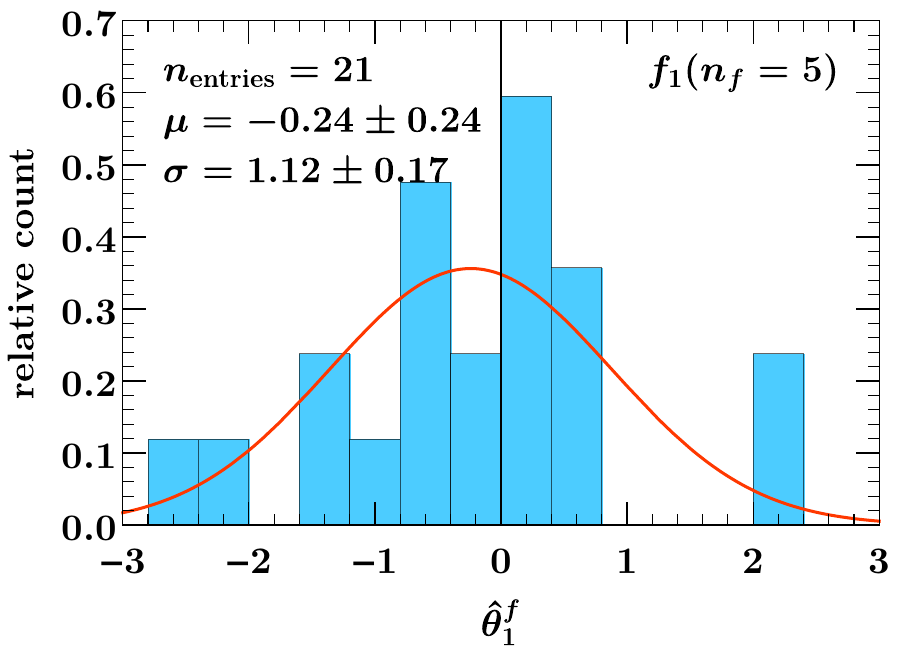}%
\hfill%
\includegraphics[width=\WidthTwoSubfigs]{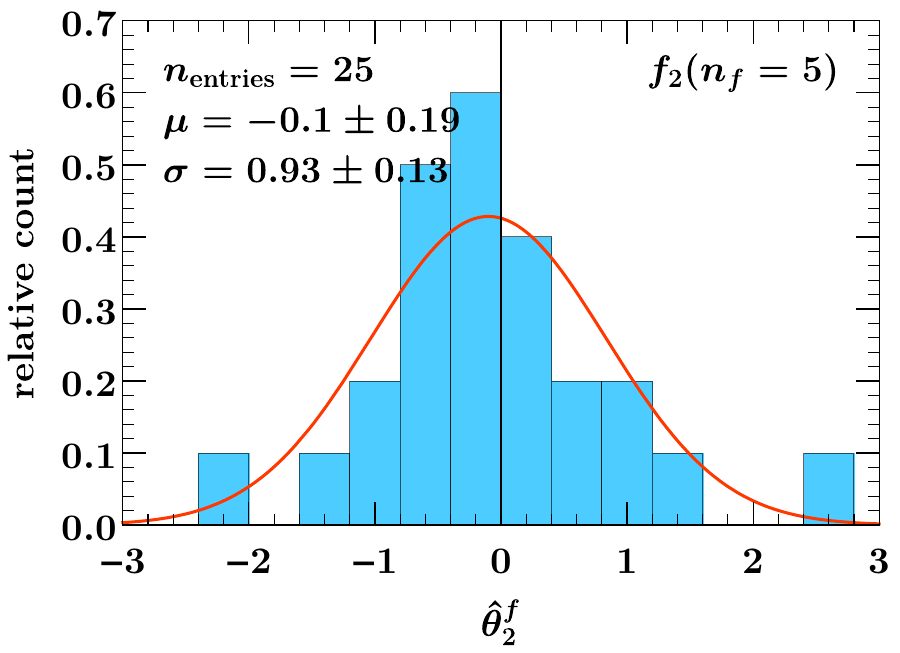}%
\\%
\includegraphics[width=\WidthTwoSubfigs]{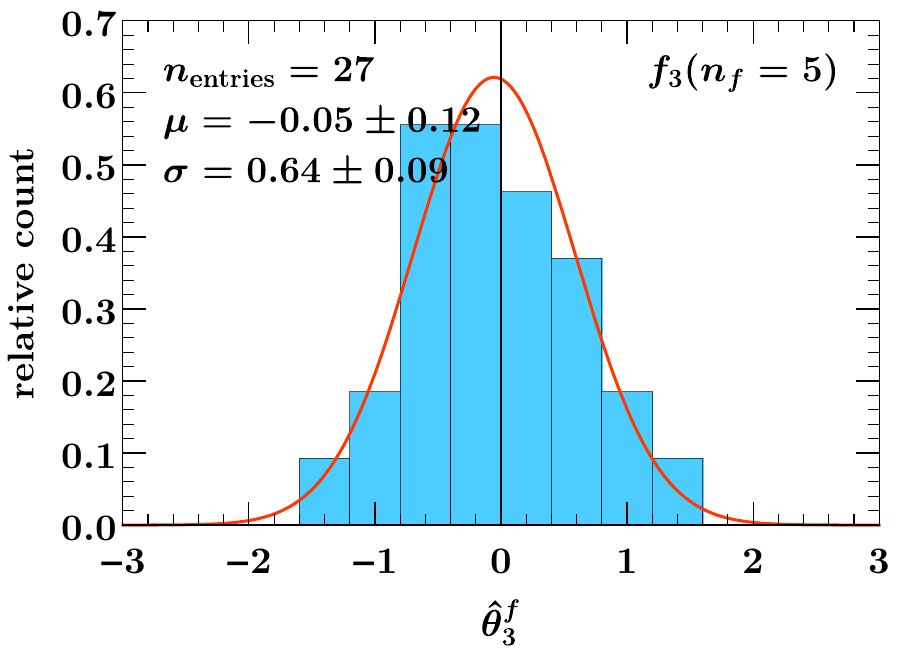}%
\hfill%
\includegraphics[width=\WidthTwoSubfigs]{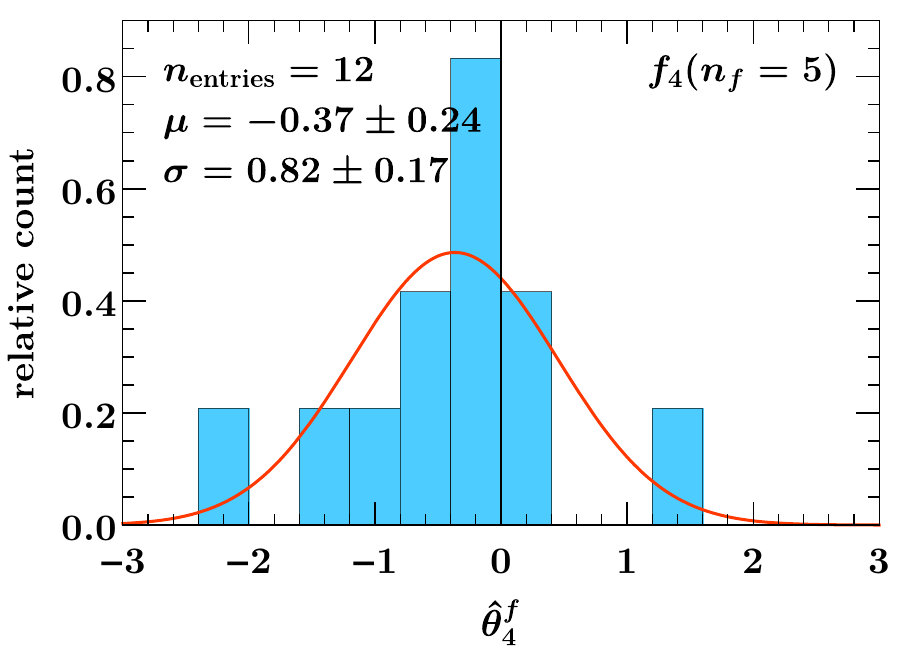}%
\caption{Distribution of true values of theory nuisance parameters
for QCD matrix-element constants for $n_f = 5$ at $n$-loop order for $n = 1$ to $4$.}
\label{fig:matrix_elements}
\end{figure}
%-------------------------------------------------------------------------------

%-------------------------------------------------------------------------------
\begin{figure}
\includegraphics[width=\WidthTwoSubfigs]{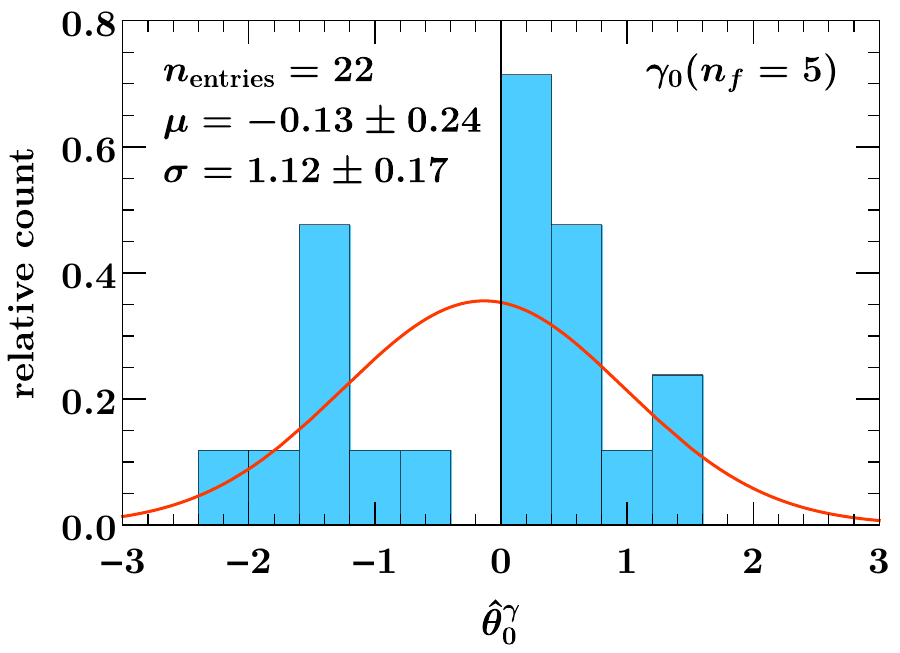}%
\hfill%
\includegraphics[width=\WidthTwoSubfigs]{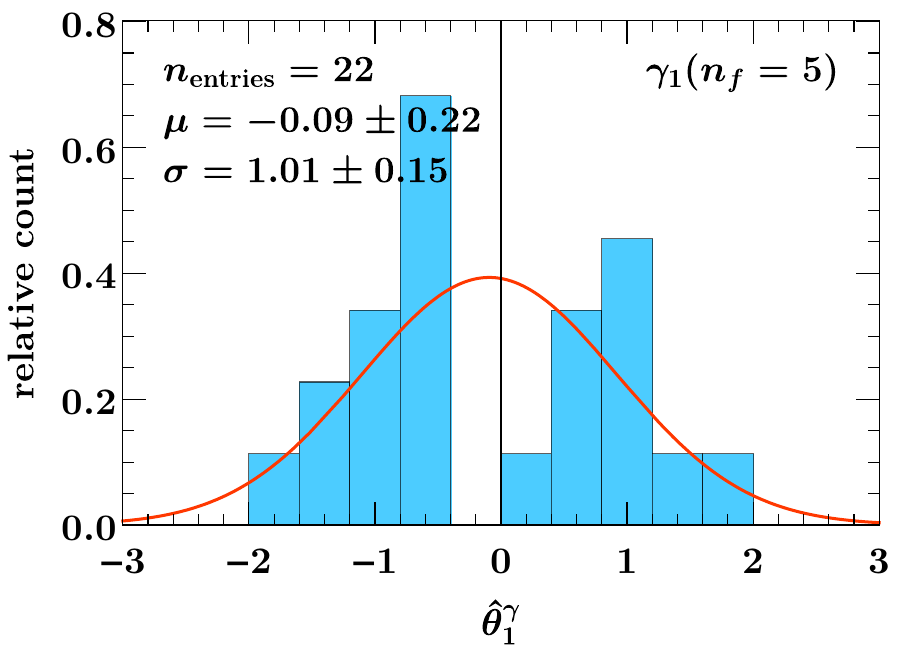}%
\\%
\includegraphics[width=\WidthTwoSubfigs]{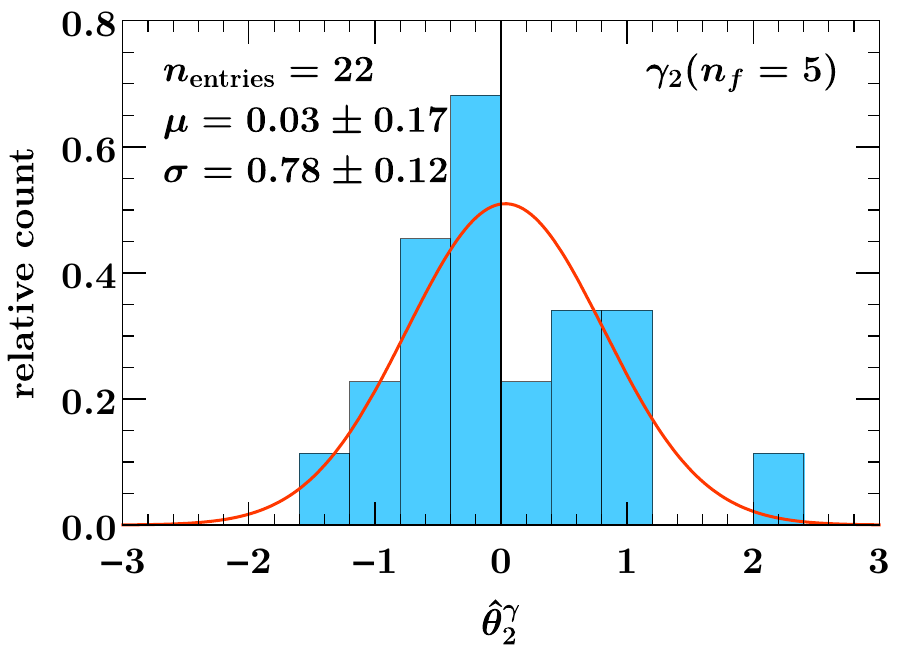}%
\hfill%
\includegraphics[width=\WidthTwoSubfigs]{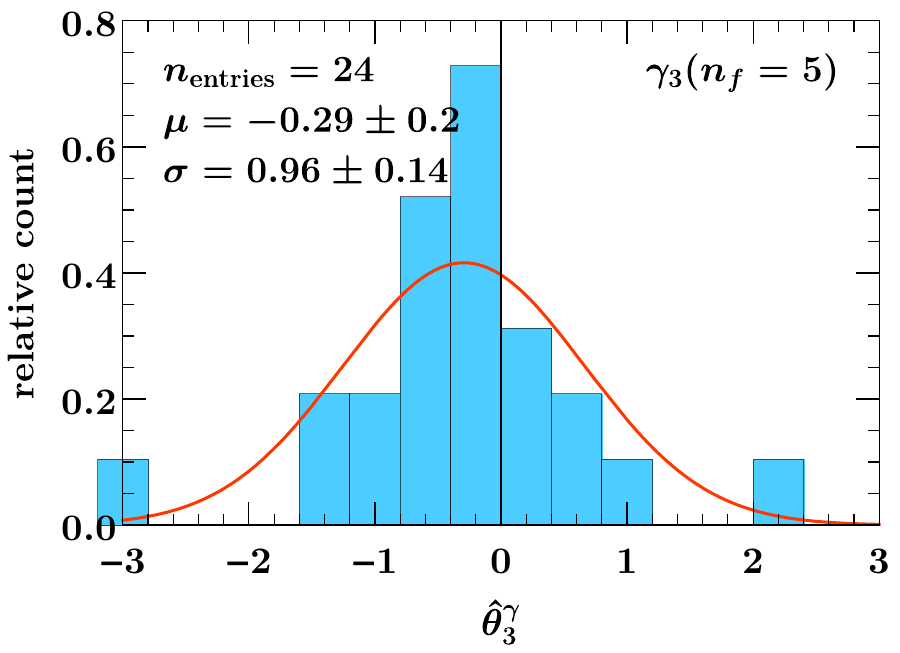}%
\\%
\hspace*{\fill}%
\includegraphics[width=\WidthTwoSubfigs]{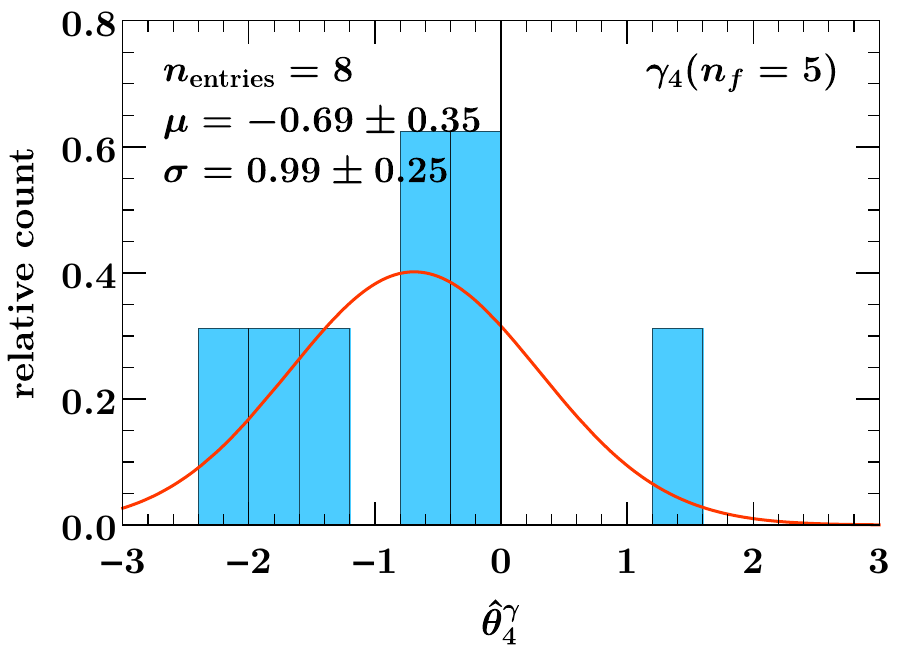}%
\hspace*{\fill}%
\caption{Distribution of true values of theory nuisance parameters for QCD anomalous
dimensions for $n_f = 5$ at $(n+1)$-loop order for $n+1 = 1$ to $5$.}
\label{fig:anom_dim}
\end{figure}
%-------------------------------------------------------------------------------

%-------------------------------------------------------------------------------
\begin{figure}
\includegraphics[width=\WidthTwoSubfigs]{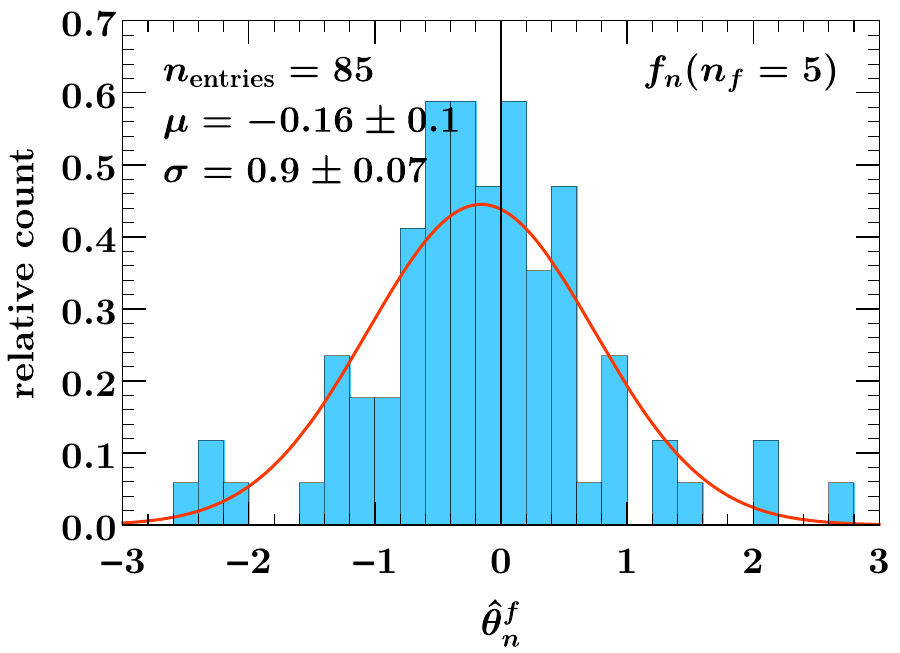}%
\hfill%
\includegraphics[width=\WidthTwoSubfigs]{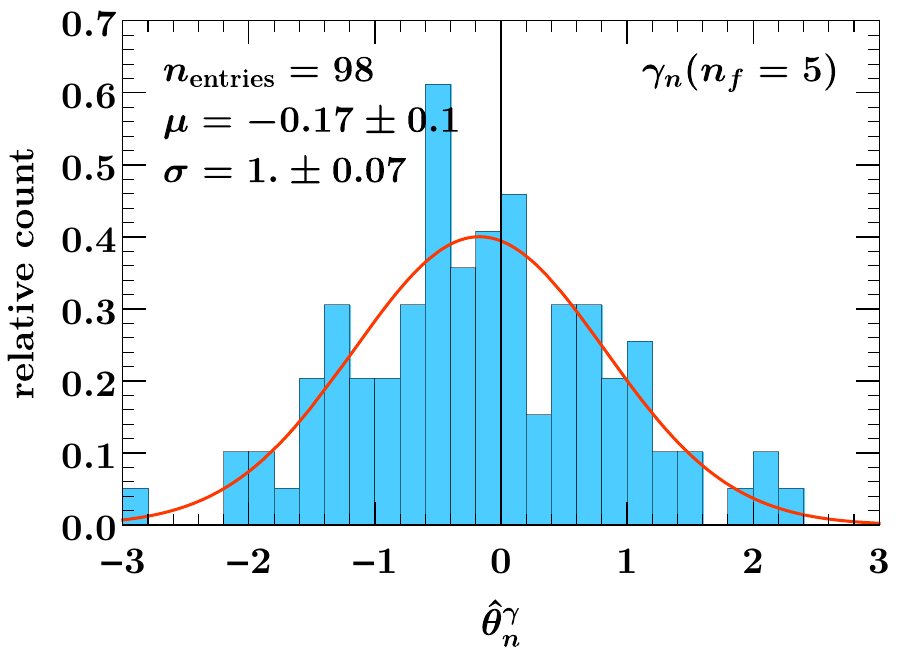}%
\caption{Distribution of true values of theory nuisance parameters
for QCD matrix-element constants (left) and anomalous dimensions (right)
for $n_f = 5$ combining all available orders.}
\label{fig:all}
\end{figure}
%-------------------------------------------------------------------------------

The lower-order coefficients of some quantities are directly related to each other
by naive Casimir scaling.
Some anomalous dimensions are equivalent due to trivial consistency relations
of the form $\gamma_a + \gamma_b = 0$.
In these cases, we only include the coefficients once.
On the other hand, some anomalous dimensions are related by consistency relations
of the form $\gamma_a + \gamma_b + \gamma_c = 0$. For these cases we do include
all three series for several reasons. First, each of the $\gamma_i$ is an actual anomalous
dimension of some quantity and should in principle obey our estimate. Second,
there is no obvious choice which one of the three to eliminate and we rather
introduce a minor correlation into the sample by keeping all three
than making an arbitrary selection which might cause some bias.

The sample distributions are shown for the matrix-element constants in \fig{matrix_elements}
and for the anomalous dimensions in \fig{anom_dim}.
The data is shown by the light blue histograms. It is binned only for visualization
purposes. For each sample, we perform an unbinned fit to a Gaussian
distribution with mean $\mu$ and variance $\sigma^2$ as free parameters.
The result is shown by the orange line and the fitted values for $\mu$ and $\sigma$
are quoted in each plot. The number of entries ($n_{\rm entries}$) for each sample
is also given.
The computed sample variance agrees with the fitted Gaussian variance to within
a few percent in all cases except at the highest order with few entries where
it differs by at most 8\%.

We first observe that the standard deviation $\sigma$ for all samples is
consistent with unity, which provides a clear validation of our natural-size
estimate in \sec{tnp_normalization}. (For the three-loop constants and
anomalous dimensions the variance
is somewhat smaller than one, which is not a concern.) The mean $\mu$ for both
matrix-element constants and anomalous dimensions is consistent with zero.
The shape of the sample distributions is generally well described
by the Gaussian fit for the given number of fitted data points.
The one-loop and two-loop anomalous dimensions show some noticeable clusterings
away from zero, which we can likely attribute to the fact that the their
coefficients do not yet contain sufficient entropy.

Given that the samples for different $n$ all show similar distributions,
with in particular the same mean and standard deviation within uncertainties,
we can take a step further and assume that the populations for different $n$
can be described by a common distribution,
%%%
\begin{equation}
\hat p_{F_n}(x) \equiv \hat p_F(x) \qquad \forall n
\,,\end{equation}
%%%
where $\hat p_F(x)$ is the distribution of all $\hat\theta_n$ in $F$ for any $n$.
This allows us to combine the samples for different $n$ in each category. The resulting
distributions of the combined samples for $F_f$ and $F_\gamma$ are shown in \fig{all}.
Their approximately Gaussian shape is clearly evident.
Most importantly, their fitted $\sigma = 0.90\pm0.07$ and $\sigma=1.00\pm 0.07$
are perfectly consistent with unity, and also agree with the computed sample
variances. The means of the distributions $\mu = -0.16\pm 0.10$
and $\mu = -0.17 \pm 0.10$ show a small shift away from zero, but they are still
consistent with zero. We could in principle account for this shift by adjusting
our estimated central value. However, given that the effect is only marginal,
we do not see a good reason for doing so in practice at this point.

In summary, we find very strong and convincing evidence for the robustness of
our estimation procedure. Based on a large sample of known series coefficients,
and with the identification in \eq{hatp_Fn}, we find the distribution
$p(\theta_n)$ in \eqs{P_un_thetan}{L_thetan} to have approximately zero mean and
unit variance, confirming our original estimate of $u_n = 0$ with $\Delta u_n = 1$.
We further find it to be well approximated by a Gaussian in agreement with our original
assumption.

%~~~~~~~~~~~~~~~~~~~~~~~~~~~~~~~~~~~~~~~~~~~~~~~~~~~~~~~~~~~~~~~~~~~~~~~~~~~~~~~
\subsubsection{Further discussion}
%~~~~~~~~~~~~~~~~~~~~~~~~~~~~~~~~~~~~~~~~~~~~~~~~~~~~~~~~~~~~~~~~~~~~~~~~~~~~~~~

The fact that our estimation procedure yields distributions closely resembling Gaussians
as in \fig{all} speaks for itself. This can be contrasted with the very long-tailed
distributions obtained from an analogous exercise using scale variations in \refcite{Ghosh:2022lrf}
or the typically very non-Gaussian distributions produced by the methods of \refscite{Cacciari:2011ze, Bagnaschi:2014wea, Bonvini:2020xeo, Duhr:2021mfd}.

It is also undeniable that the distributions are closer
to a Gaussian than a flat box, in contrast to what one might
have expected, as such a box-like distribution is sometimes advocated to be more
appropriate than a Gaussian for perturbative theory uncertainties (albeit typically
in the context of scale variations).
We might also ask why to expect the distributions $\hat p_{F_n}(\hat\theta_n)$ and $\hat p_F(\hat\theta_n)$
to be sensible or useful in the first place.
In fact, even though the $\hat f_n \in F_n$ all belong to some common category
of perturbative series, we want this category to be as broad as possible to be
as useful as possible. This means the distribution of $\hat f_n$, which is
solely a property of the collection $F_n$ might very well be quite irregular and
not very useful by itself.
However, as we have stressed already, the distribution $\hat p_{F_n}(\hat\theta_n)$
is a property of our estimation procedure for $F_n$. Its goal is precisely
to strip the $\hat f_n$ of their individuality and reduce them to a generic
bunch of numbers of a common more-or-less random origin, namely arising as
a more-or-less random sum of Feynman diagrams.
It is then perhaps not surprising after all that the resulting population of $\hat\theta_n$
can be well described by a Gaussian distribution. We might think of it as the
central-limit theorem of Feynman diagrams.%
\footnote{The credit for coining this term goes to Glen Cowan.}

It is also instructive to think about how we would notice if there was
something going wrong in our estimation procedure.
If we find a Gaussian with different mean or variance,
then our estimation procedure itself is sound but our final choice of $u_n$
or $\Delta u_n$ is off, which we can easily adjust for if necessary.
If the resulting distribution is irregular, e.g., with large
tails or other undesired features, then this signals that our estimation
procedure is suboptimal or missing some important aspect. A long tail would be
indicative of underestimating the natural size for some coefficients in $F_n$.
To illustrate this with a simple example, imagine we had not accounted
for the overall $C_r$ color factor. As long as our collection $F_n$
contains only quark-like or only gluonic quantities, this would not be much of a
problem, we would simply find an uncertainty of $C_F$ or $C_A$ instead of one.
However, when $F_n$ contains both, we would end up with a superposition of
two Gaussians of different variance. We could still work with this
distribution, but it would be suboptimal because it would lead to overestimating
the uncertainty for quark-like quantities and underestimating it for gluonic quantities.

This is also why it is prudent to consider separate collections $F_n$ for each $n$
at first, as this allows us to test and identify the appropriate $n$-dependent
normalization. For example, without the $(n-1)!$ in $N_n^\f$ in \eq{NnF} we would find distributions
of correspondingly larger variance for $n \geq 3$, which is in fact how we became
aware of this factor during the course of our investigations.

We conclude this subsection with two more comments.
First, when applying our estimation procedure to a known $\theta_n$
we should not use our knowledge of its true value $\hat\theta_{n}$, but by including it
in the sample of known $\hat\theta_n$
we do indirectly use it. However, this is acceptable since the impact of any one
coefficient on the sample distribution is minor.
Second, when applying our estimation procedure to a new and still unknown $\theta_n$,
we still have to make a judgement whether or not it belongs to a particular
category. There is a priori no guarantee for that. It might well be the case that
there are genuinely different types of quantities
than those considered so far that cannot be reduced to fit into an existing category
but instead require defining and studying a new category.

%===============================================================================
\subsection{Designing theory estimators}
%===============================================================================

An interesting question to consider is whether it is possible to improve upon our estimator
or design alternative estimators, which could be tested using the same procedure
as above. We leave this for future investigation and only give some general remarks here.
An improved estimator should on average yield a tighter estimate of the
natural size but without underestimating it for some subset of series either.
In other words, it should yield a reduced variance and ideally still produce a
roughly Gaussian distribution.

For example, one could imagine devising an estimator based on the actual leading-color
approximation of a series coefficient, i.e., using the
large-$N_c$ expansion as a supplementary expansion following strategy 2 in \sec{param_strategies}.
This of course requires performing an actual
calculation, but typically the calculation in the large-$N_c$ limit is easier to
perform than the full calculation. Having a robust estimator based on this limit
would allow one to robustly use such approximate results.

Another natural question that arises is whether one could utilize the information from
the known lower orders $\hat f_{k < n}$ of a given quantity $f$ to
improve the estimate for $f_n$. This effectively amounts to devising alternative
parameterizations for $f_n$ involving the information of the known $\hat f_{k<n}$
in some way.
In our initial attempts we found that this overall leads to less reliable estimates.
For example, Pad{\'e} approximations can work extremely well in some cases and utterly
fail in others. The basic problem of relying on lower-order information is that this
introduces the same generic pitfall also present for scale variations: It
can lead to random significant underestimations when the lower-order coefficient(s)
happen to be randomly smaller than their own natural size. This is in fact not
unlikely to happen since as we have seen the mean of the distribution of true values
is around 0. Examples of this are already present in \tabs{wilson_coefficients}{anom_dims},
namely $\hat f_1$ of $c_{q\bar q S}$, $\hat f_2$ of $c_{gg}$, and notably $\hat\gamma_3$
of $\Gamma_\cusp^q$.

One might also consider applying the Bayesian inference models of
\refscite{Cacciari:2011ze, Bagnaschi:2014wea, Bonvini:2020xeo, Duhr:2021mfd} to
estimate a given $\theta_n$ based on its known lower-order $\hat\theta_{k<n}$,
as was mentioned already in \refcite{Bonvini:2020xeo}.
In this case, similar care has to be exercised to avoid the above pitfall.
Another general option would be to utilize series transformations or series
acceleration methods as in \refcite{David:2013gaa}. In fact, taking the square
root of a quantity can be considered a simple form of a series transformation.

From our experience so far, the most useful way to utilize the known lower-order information
is as an important cross check of the estimation procedure rather than as a direct input to it.
If a quantity consistently violates its estimated natural size at lower orders,
it might indicate that we are not estimating its natural size correctly, which
can have various reasons. We might be using the wrong normalization factor
or a suboptimal reference scheme, or we might be associating it incorrectly with
a given category. The latter can happen when not using the appropriate conventions,
e.g.\ we should be parameterizing $\sqrt{f}$ or $f^2$ instead of $f$.
It can also happen when we are using a suboptimal parameterization, for example
when the scalar series has
important internal structures (e.g.\ new color or partonic channels) which affect
its natural size but which we have not explicitly parameterized.

More generally, the question is to what extent the higher-order coefficients
are correlated with the lower-order ones and how to best exploit this correlation
to our advantage. Such correlations could arise for example
from cross terms of lower-order coefficients appearing as part of the higher-order
coefficient (an obvious example is again using $f^2$ instead of $f$), or we believe
the higher-order correction to be a genuinely multiplicative correction on top
of the lower-order result (in which case we would parameterize their ratio).
In general, such information is specific to a given quantity $f$. Therefore,
all genuine lower-order information that we believe to be relevant should be accounted for
explicitly by the specific parameterization of $f_n(\theta_n)$ itself. An optimal parameterization
would then be one for which the $\theta_n$ are uncorrelated for different $n$.
In this limit, no more information can be
gained from the lower-order $\hat\theta_{k<n}$ and the easiest and
safest approach is to estimate the natural size of $\theta_n$ without further
reference to $\hat\theta_{k<n}$. The
boundary between parameterization and estimation is of course somewhat blurry,
since as we have seen, the final step of parameterizing the remaining scalar
coefficients in terms of scalar TNPs is in fact an important part of the estimation
procedure itself.

%%%%%%%%%%%%%%%%%%%%%%%%%%%%%%%%%%%%%%%%%%%%%%%%%%%%%%%%%%%%%%%%%%%%%%%%%%%%%%%%
\section{Application to Transverse-Momentum Resummation}
\label{sec:qT}
%%%%%%%%%%%%%%%%%%%%%%%%%%%%%%%%%%%%%%%%%%%%%%%%%%%%%%%%%%%%%%%%%%%%%%%%%%%%%%%%

The $q_T$ spectrum of $Z$ and $W$ bosons produced in hadronic collisions, where
$q_T\equiv p_T^{Z,W}$ is the transverse momentum of the produced vector boson,
is a benchmark observable of the LHC precision physics program
and has been measured to incredible precision
by the ATLAS~\cite{Aad:2014xaa, Aad:2015auj, Aad:2019wmn, ATLAS:2023lsr},
CMS~\cite{Chatrchyan:2011wt, Khachatryan:2016nbe, Sirunyan:2019bzr},
and LHCb collaborations~\cite{LHCb:2015mad, LHCb:2016fbk}.
In this section, we discuss the application of our approach to precision
predictions for the $q_T$ spectrum using resummed perturbation theory.

Correlations within the $p_T^W$ and $p_T^\ell$ spectra, and depending
on the analysis strategy also between $p_T^W$ and $p_T^Z$, are critical for
a precise measurement of the $W$-boson mass at hadron
colliders~\cite{CDF:2022hxs, Aaboud:2017svj, ATLAS:2024erm, LHCb:2021bjt, CMS:2024lrd}.
As shown in \refcite{TNPalphas}, the theory correlations in $p_T^Z$ are also critical
if one wants to perform fits to the precisely measured small-$p_T^Z$
spectrum to extract nonperturbative parameters~\cite{Bacchetta:2022awv, Moos:2023yfa, Bacchetta:2024qre}
or the strong coupling constant~\cite{Camarda:2022qdg, ATLAS:2023lhg}.

In \sec{qT_resummation} we give a brief account of the aspects of
$q_T$ factorization and resummation that are relevant to our discussion.
In \sec{qT_TNPs} we identify and discuss
the necessary TNPs, and in \sec{qT_results} we present numerical results
that illustrate the power of the TNP approach to obtain predictions with proper
theory correlations. Finally in \sec{qT_subleading}, we briefly discuss the treatment
of subleading effects, which we neglect here for simplicity.

%===============================================================================
\subsection{Aspects of \texorpdfstring{$q_T$}{qT} resummation}
\label{sec:qT_resummation}
%===============================================================================

We denote the four-momentum of the vector boson by $q^\mu$, its invariant mass and
rapidity by $Q \equiv \sqrt{q^2}$ and $Y$, and its transverse momentum by $q_T = \abs{\qt}$.
The quantity of our interest is the cross section fully differential in $Q$, $Y$, and $q_T$,
which we write for brevity as $\df\sigma/\df^4 q$.

We start by applying strategy 2 of \sec{param_strategies} and expand the cross section
in a power series in $\varepsilon \equiv q_T^2/Q^2$,
%%%
\begin{equation} \label{eq:power_expansion}
\frac{\df\sigma}{\df^4 q} = \frac{\df\sigma^\zero}{\df^4 q} \Bigl[1 + \ORd{\frac{q_T^2}{Q^2}} \Bigr]
\,.\end{equation}
%%%
Compared to our discussion in \sec{param_strategies}, where we expanded the series
coefficient in $\varepsilon$, here it is much more useful to first perform the expansion in
$\varepsilon$ and only later the perturbative expansion in $\alpha_s$. The reason
is that we actually know the functional form in $q_T$ (and $Q$) of the leading-power
term $\df\sigma^\zero$ to all orders in $\alpha_s$, allowing us to apply strategy 1
and obtain the exact correlations in $q_T$ and $Q$.
Furthermore, we will also resum certain parts of the perturbative
series to all orders in $\alpha_s$, although the precise way of doing so is not of
immediate concern to us here, so we will not discuss it but refer the interested reader
to \refscite{Ebert:2020dfc, Billis:2024dqq}.

The power expansion in \eq{power_expansion} converges very well, even better than the $q_T^2/Q^2$
scaling suggests, such that the power corrections remain below $\lesssim 5\%$ even up to
moderately large $q_T \lesssim Q/3$ and even $Q/2$. As a result, the leading-power
term $\df\sigma^\zero$ dominates and effectively determines the spectrum over this
entire small-$q_T$ region, and thus also causes the dominant perturbative uncertainties.
We can therefore focus our discussion on $\df\sigma^\zero$. In particular, it will
serve us to demonstrate a nontrivial example application of the TNP approach.
We will comment further on the treatment of the
$\ord{q_T^2/Q^2}$ power corrections and other subleading effects in \sec{qT_subleading}.

The leading-power term $\df\sigma^\zero$ is the subject of the $q_T$ factorization
and resummation program. %~\cite{xxx}.
We do not intend to provide a detailed review of $q_T$ resummation here. Rather, our
focus is on the kinematic and process dependence, which we wish to break
down and parameterize in terms of theory nuisance parameters.
We use the SCET resummation framework of \refscite{Ebert:2020dfc, Billis:2024dqq}.
We closely follow the notation of those references and refer there for
more details and further references. The leading-power cross section can be written as
%%%
\begin{align} \label{eq:factorization}
\frac{\df \sigma^\zero}{\df^4 q}
&= \frac{1}{2\Ecm^2}\, L_{VV'}(q^2) \,
\sum_{a,b} H_{VV'\,ab}(q^2, \mu)
\\\nn & \quad\times
\int\!\!\frac{\df^2\bt}{(2\pi)^2} \, e^{\img \bt \cdot \qt} \,
   \tB_a(x_a, b_T, \mu, \nu/Q) \, \tB_b(x_b, b_T, \mu, \nu/Q)
   \tS(b_T, \mu, \nu)
\,.\end{align}
%%%
Here, $VV' = \{\gamma\gamma,\gamma Z,Z\gamma,ZZ,W^+W^+,W^-W^-\}$ labels the produced
vector boson including possible interferences.
The leptonic tensor $L_{VV'}(q^2)$ contains the vector-boson propagator
and decay and receives no QCD corrections, so its $q^2$ dependence is known.
The hard function $H_{VV'\,ab}(Q^2,\mu)$ encodes the production of the vector boson in the underlying
hard interaction $ab \to V$, with the sum over $a,b$ running over all relevant combinations of quark and antiquark flavors.
The functional form of its $q^2$ and process dependence is known to all orders.
The second line in \eq{factorization} contains all soft and collinear physics at the low scale $\mu \sim q_T$
encoded respectively in the soft function $\tS$ and beam function $\tB_{a,b}$.
The $q_T$ dependence arises entirely from the second line,
and its functional form is fully determined by the functional dependence on its
Fourier-conjugate variable $b_T = \abs{\bt}$. The functional form of the $b_T$ dependence
of the beam and soft functions is in turn known to all orders in $\alpha_s$.
The beam function also depends on the flavor of the (anti)quark participating in the hard interaction
and on $Q$. The functional form of these dependencies is also known to all orders.
Finally, the variables $x_{a,b} = (Q/\Ecm) e^{\pm Y}$ encode the
dependence on the rapidity $Y$ and center-of-mass energy $\Ecm$. The functional form
of the $x_{a,b}$ dependence of the beam function is not known to all orders but
depends on their perturbative order.

The factorization in \eq{factorization} is very powerful for our purposes as it predicts the
complete functional form in $q_T$ and also in $Q$ for given $x_{a,b}$. Furthermore, it fully
parameterizes the exact dependence on the process and partonic channels.
We are therefore able to apply strategy 1 and obtain exact correlations in all these dependencies.
Although
it does not predict the complete functional form in $x_{a,b}$ it still reduces it
from a generic two-dimensional dependence to a product of common, universal one-dimensional
beam functions.

For simplicity we have limited ourselves to the inclusive $q_T$ spectrum in \eq{factorization}.
Including the full kinematics of the vector-boson decay products is also possible.
Importantly, at leading power doing so only increases the complexity of the leptonic tensor but
does not induce any additional sources of QCD uncertainties~\cite{Ebert:2020dfc}.%
\footnote{%
More precisely, leptonic observables can give rise to enhanced power corrections,
which for azimuthally symmetric observables can be taken into account
in terms of the leading-power QCD contributions, and thus without inducing additional
sources of QCD uncertainties. Starting at $\ord{q_T^2/Q^2}$ also genuinely new QCD
structures can contribute, see \refcite{Ebert:2020dfc} for a detailed discussion.}
We can therefore also capture the correlations in leptonic kinematic variables,
most notably the lepton transverse momentum $p_T^\ell$, or between the $q_T$ spectrum
and the $q_T$-dependent forward-backward asymmetry.

In principle, \eq{factorization} could be applied to each coefficient of
the perturbative series of $\df\sigma^\zero$.
However, at each order in $\alpha_s$, (double) logarithms of $q_T/Q$ appear, which render
a fixed-order expansion of $\df\sigma^\zero$ unstable. Instead, \eq{factorization} also
provides the basis for systematically resumming the unstable logarithmic contributions to all orders
in $\alpha_s$, leading to precise and perturbatively stable predictions. We will not discuss
how the resummation is carried out in practice but refer to \refscite{Ebert:2020dfc, Billis:2024dqq} for details. The key point for us is that a given perturbative resummation order, N$^n$LL, is uniquely defined
by including all underlying scalar perturbative series discussed below to a specific order in $\alpha_s$.
We then define our generalized counting including TNPs, N$^{n+k}$LL, to include
the true values for all coefficients relevant for N$^n$LL and in addition for each
series the TNP parameterization of the next $k$ terms. In analogy to \sec{approx_implementation},
we also define the approximate N$^{n+0}$LL implementation by absorbing the TNPs appearing
at N$^{n+1}$LL as an additive correction to the respective highest coefficients appearing at N$^n$LL.

%===============================================================================
\subsection{TNPs for \texorpdfstring{$q_T$}{qT} resummation}
\label{sec:qT_TNPs}
%===============================================================================

The perturbative ingredients required in \eq{factorization} are the hard, beam,
and soft functions. Their functional dependence on the kinematic variables, except $x$,
is fully predicted to all orders in $\alpha_s$ by their renormalization group equations,
which we now discuss in turn. At the end we will be left with a set of (mostly) scalar
perturbative series that fully determine the (fixed-order and/or resummed) perturbative
series of $\df\sigma^\zero$. We will give a summary in \sec{qT_TNPs_summary}, so readers not
interested in the detailed definitions can directly skip there.

%~~~~~~~~~~~~~~~~~~~~~~~~~~~~~~~~~~~~~~~~~~~~~~~~~~~~~~~~~~~~~~~~~~~~~~~~~~~~~~~
\subsubsection{Hard function}
%~~~~~~~~~~~~~~~~~~~~~~~~~~~~~~~~~~~~~~~~~~~~~~~~~~~~~~~~~~~~~~~~~~~~~~~~~~~~~~~

The leptonic tensors for inclusive $Z\to\ell\ell$ and $W\to\ell\nu$ in \eq{factorization}
are given by
%%%
\begin{align}
L_{ZZ}(q^2)
&= \frac{2}{3} \frac{\aem}{q^2}\, (v_\ell^2 + a_\ell^2)\, \ABS{\frac{q^2}{q^2 - m_Z^2 + \img \Gamma_Z m_Z}}^2
\,,\nn\\
L_{W^+ W^+}(q^2)
&= \frac{1}{6} \frac{\aem}{q^2}\, \frac{1}{\sin^2 \theta_w}\, \ABS{\frac{q^2}{q^2 - m_W^2 + \img \Gamma_W m_Z}}^2
\,.\end{align}
%%%
Their $q^2$ dependence is known exactly in QCD.
The corresponding hard functions have the form
%%%
\begin{align} \label{eq:hard}
H_{ZZ \, q\bar q'}(q^2, \mu)
 &= \frac{8\pi \aem}{N_c} \delta_{qq'} \biggl\{ (v_q^2 + a_q^2) \abs{C_q(q^2, \mu)}^2
\nn\\ &\quad
 + 2 \Re \sum_f \Bigl[ v_q v_f C_q^*(q^2, \mu) C_{vf}(q^2, \mu)
 + a_q a_f C_q^*(q^2, \mu) C_{af}(q^2, \mu) \Bigr] + \dotsb \biggr\}
\,,\nn\\
H_{W^+ W^+ \, q\bar q'}(q^2, \mu)
&= \frac{2\pi \aem}{N_c} \frac{\abs{V_{qq'}}^2}{\sin^2 \theta_w} \abs{C_q(q^2, \mu)}^2
\,.\end{align}
%%%
The expressions for the remaining $VV'$ combinations can be found in
appendix A of \refcite{Ebert:2020dfc}. The $v_i$ and $a_i$ are the usual axial and
vector couplings of the $Z$ boson, $Q_q$ is the electromagnetic charge of quark $q$,
and $V_{qq'}$ are the CKM-matrix elements.

The $q^2$ dependence of the hard function is determined by that of the matching coefficients
$C_i(q^2, \mu)$, which correspond to the infrared-finite parts of the respective QCD form factors.
Here, $C_q = 1 + \ord{\alpha_s^2}$ is the dominant
vector nonsinglet coefficient corresponding to diagrams where the vector
boson couples to the external quark line.
The $C_{af}$ and $C_{vf}$ in \eq{hard} are
axial-singlet and vector-singlet coefficients corresponding to diagrams
where the vector boson couples to a closed fermion loop,
which only contribute to $Z$ production but not to $W$ production.
They have separate perturbative series starting at $\ord{\alpha_s^2}$ and $\ord{\alpha_s^3}$,
respectively, and have to be parameterized separately. In practice, their contributions
are very small
even at the order they contribute~\cite{Billis:2024dqq}, so
we can neglect them here for simplicity. In principle they would have to be included
(starting at N$^3$LL)
to fully account for the correct (de)correlation between $W$ and $Z$ production.
The ellipses in $H_{ZZ\,q\bar q'}$ denote terms
proportional to the square of $C_{af}$ and $C_{vf}$, which only contribute starting at
$\ord{\alpha_s^4}$.

The functional form of the $q^2$ dependence of $C_q(q^2, \mu)$ is known
because by dimensional analysis it can only depend on the ratio $q^2/\mu^2$. The
$q^2$ dependence is therefore fully predicted by the $\mu$ dependence, which in turn
is governed by $C_q$'s renormalization group evolution (RGE) equation,
%%%
\begin{align}\label{eq:Cq_RGE}
\mu\frac{\df}{\df\mu} \ln C_q(q^2,\mu)
&= \Gamma_\cusp^q[\alpha_s(\mu)] \ln\frac{-q^2 - \img 0}{\mu^2} + 2\gamma_C^q[\alpha_s(\mu)]
\,.\end{align}
%%%
The full $q^2$ and $\mu$ dependence of $C_q$ can be reconstructed by solving \eq{Cq_RGE}
(either order by order in $\alpha_s$ or to all orders to obtain its resummed expression).

The cusp and noncusp anomalous dimensions $\Gamma_\cusp^q(\alpha_s)$ and $\gamma_C^q(\alpha_s)$
in \eq{Cq_RGE} are already scalar series. Following our conventions for anomalous
dimensions in \sec{tnp_normalization}, we parameterize
%%%
\begin{equation} \label{eq:tnps_gamma}
\Gamma(\alpha_s) \equiv 2\Gamma_\cusp^q(\alpha_s)
\,,\qquad
\gamma_\mu(\alpha_s) \equiv 2\gamma_C^q(\alpha_s)
\,,\end{equation}
%%%
in terms of corresponding TNPs $\theta_n^\Gamma$ and $\theta_n^{\gamma_\mu}$.

The remaining nontrivial part of $C_q$ we need to parameterize is the $q^2$ and $\mu$-independent
constant term, which is not predicted by \eq{Cq_RGE} and effectively acts as the
boundary condition for solving the differential equation.
We can formally define it as the matching coefficient evaluated at the canonical
scale $\mu^2 = -q^2$,
%%%
\begin{equation} \label{eq:tnps_H}
c_q(\alpha_s) \equiv C_q(q^2, \mu^2 = -q^2)
\,.\end{equation}
%%%
By choosing the canonical scale proportional to $q^2$, the perturbative
series for $c_q(\alpha_s)$ becomes a scalar series with $q^2$ and $\mu$ independent
coefficients. Here, $c_q(\alpha_s)$ is equal to $c_{q\bar qV}(\alpha_s)$
in \sec{tnp_normalization}, so we parameterize it directly in terms of TNPs $\theta_n^H$,
where the label is meant to remind us that they come from the hard function.

Note that the matching coefficient is defined in a certain renormalization scheme, for which
we use the standard \MSbar\ scheme here. Together with the canonical scale choice, which also
determines the form of the logarithm in \eq{Cq_RGE}, this defines the reference scheme for
the anomalous dimensions and constant term and their TNPs.

%~~~~~~~~~~~~~~~~~~~~~~~~~~~~~~~~~~~~~~~~~~~~~~~~~~~~~~~~~~~~~~~~~~~~~~~~~~~~~~~
\subsubsection{Soft function}
%~~~~~~~~~~~~~~~~~~~~~~~~~~~~~~~~~~~~~~~~~~~~~~~~~~~~~~~~~~~~~~~~~~~~~~~~~~~~~~~

The TNP parameterization of the soft function $\tS(b_T, \mu, \nu)$ proceeds
analogously to that of the matching coefficient $C_q(q^2, \mu)$ above.
A new element is the soft function's dependence on the additional rapidity
renormalization scale $\nu$, which has dimension one. By dimensional analysis,
the soft function can only depend on two ratios $b_T/\mu$ and $\mu/\nu$, so its
full $b_T$ dependence is determined by its dependence on $\mu$ and $\nu$,
which is now governed by a system of RGE equations,
%%%
\begin{align}\label{eq:S_RGEs}
\mu\frac{\df}{\df\mu} \ln\tS(b_T,\mu,\nu)
&= 4 \Gamma_\cusp^q[\alpha_s(\mu)] \ln\frac{\mu}{\nu} + \tilde\gamma_S[\alpha_s(\mu)]
\,,\nn\\
\nu\frac{\df}{\df\nu} \ln\tS(b_T,\mu,\nu)
&= \tilde\gamma_\nu(b_T,\mu)
\,,\nn\\
\mu \frac{\df}{\df \mu} \tilde\gamma_\nu(b_T, \mu)
&= - 4\Gamma^q_\cusp[\alpha_s(\mu)]
\,.\end{align}
%%%
Here, the rapidity anomalous dimensions $\tilde\gamma_\nu(b_T, \mu)$ has
a more nontrivial dependence on $b_T$, which is in turn determined by its own
$\mu$ dependence governed by its own $\mu$ RGE in the last line.

Solving \eq{S_RGEs} now requires two independent boundary conditions, one for $\gamma_\nu(b_T, \mu)$
and one for $\tS(b_T, \mu, \nu)$ itself. The canonical scale in $b_T$ space is
$\mu = b_0/b_T$ with $b_0 = 2 e^{-\gamma_E} \approx 1.12291$,
which corresponds to $\mu = q_T$ in momentum space.
The soft function scales like a squared $2\to0$ matrix element.
Following our conventions
in \sec{tnp_normalization}, we therefore define the relevant scalar series as
%%%
\begin{align} \label{eq:tnps_S}
\gamma_\nu(\alpha_s) \equiv \frac{1}{2} \tilde\gamma_\nu(b_T, \mu = b_0/b_T)
\,,\nn\\
\tilde s(\alpha_s) \equiv \sqrt{\tS(b_T, \mu = b_0/b_T, \nu = b_0/b_T)}
\,,\end{align}
%%%
which we parameterize in terms of corresponding TNPs $\theta_n^{\gamma_\nu}$ and
$\theta_n^S$.
Note that the reference scheme for the TNPs here corresponds to our choices of
using $b_T$ space and its canonical scale, \MSbar\ renormalization, and rapidity
renormalization~\cite{Chiu:2012ir} with the exponential regulator~\cite{Li:2016axz}.

The other perturbative ingredients we need for the soft function are the cusp and
noncusp anomalous dimensions in the first line of \eq{S_RGEs}. Following our conventions
we would again parameterize $\Gamma(\alpha_s) \equiv 2\Gamma_\cusp^q(\alpha_s)$,
consistent with \eq{tnps_gamma}, and $\gamma_S(\alpha_s) = \tilde\gamma_S(\alpha_s)/2$.
In practice, we do not need TNPs for $\gamma_S(\alpha_s)$,
for reasons we will explain in a moment.

%~~~~~~~~~~~~~~~~~~~~~~~~~~~~~~~~~~~~~~~~~~~~~~~~~~~~~~~~~~~~~~~~~~~~~~~~~~~~~~~
\subsubsection{Beam functions}
%~~~~~~~~~~~~~~~~~~~~~~~~~~~~~~~~~~~~~~~~~~~~~~~~~~~~~~~~~~~~~~~~~~~~~~~~~~~~~~~

The beam function $\tB_i(x, b_T, \mu, \nu/Q)$ only depends on the combination
$\nu/Q$, as indicated by its argument, and thus by dimensional analysis only on
$b_T/\mu$. Its $b_T$ and explicit $Q$ dependence is thus governed by its RGE system,
which is closely analogous to that of the soft function in \eq{S_RGEs},
%%%
\begin{align}\label{eq:B_RGEs}
\mu\frac{\df}{\df\mu} \ln\tB_q(x,b_T,\mu,\nu/Q)
&= 2 \Gamma_\cusp^q[\alpha_s(\mu)] \ln\frac{\nu}{Q} + \tilde\gamma_B[\alpha_s(\mu)]
\,,\nn\\
\nu\frac{\df}{\df\nu} \ln\tB_q(x,b_T,\mu,\nu/\omega)
&= -\frac{1}{2}\tilde\gamma_\nu(b_T,\mu)
\,,\nn\\
\mu \frac{\df}{\df \mu} \tilde\gamma_\nu(b_T, \mu)
&= - 4\Gamma^q_\cusp[\alpha_s(\mu)]
\,.\end{align}
%%%
We need again the cusp and rapidity anomalous dimensions,
which are the same as before in \eq{S_RGEs}, the noncusp beam anomalous dimension $\gamma_B(\alpha_s) \equiv \tilde\gamma_B(\alpha_s)$, and the beam function boundary condition.

We do not need TNPs for the beam and soft noncusp anomalous dimensions for the
following reason. When using the beam and soft function's RGEs to reconstruct their full
fixed-order expressions, we only need the known anomalous dimension coefficients
(for our considered resummation orders). Their TNPs would only enter
in the evolution itself. However, since the beam and soft functions start their evolution at
the same canonical scale $\mu = b_0/b_T$, only their total $\mu$ anomalous dimension
actually enters in the resummation,
which by consistency is equal to minus that of the hard function.
We therefore only need the single noncusp $\mu$ anomalous dimension in \eq{tnps_gamma}.

Importantly, the RGEs do not depend on $x$, which implies that the
additional $x$ dependence factorizes from the $b_T$ and $Q$ dependencies and
only enters via the beam boundary condition, which is now defined at the
canonical scales $\mu = b_0/b_T$ and $\nu = Q$,
%%%
\begin{equation}
\tilde b_i(x, \alpha_s) \equiv \tB_i(x, b_T, \mu = b_0/b_T, \nu/Q = 1)
\,.\end{equation}
%%%
The additional complication for the beam function arises because its $x$ dependence
is not predicted by its RGE, so the beam boundary condition is a general one-dimensional
function of $x$. To further break down this dependence, we calculate its series
coefficients $\tilde b_{i,n}(x)$ in terms of collinear PDFs $f_j(x)$,
%%%
\begin{align} \label{eq:ope_beam_soft_tmd}
\tilde b_{i,n}(x)
&= \sum_j \int\! \frac{\df z}{z} \, \tilde I_{ij,n}(z) \, f_{j}\Bigl( \frac{x}{z} \Bigr)
\,,\end{align}
%%%
where $\tilde I_{ij,n}(z)$ are perturbatively calculable matching
kernels. The $x$ dependence of the beam function is thus determined via
the Mellin convolution of the $x$ dependence of the PDFs and the $z$ dependence
of the matching kernels. Since the $x$ dependence of the PDFs tends to be quite
strong, the mix of contributing PDFs determines the overall size of $\tilde b_{i,n}(x)$
as well as playing an important role in determining its shape in $x$.
The $I_{ij,n}(z)$ are perturbative coefficients, so we can in principle estimate their natural size
as in \sec{tnp_normalization}. (In fact, their moments in $x$ enter into our sample of
matrix-element constants). In contrast, it would be quite difficult to estimate
the natural size of $\tilde b_{i,n}(x)$ directly.
\Eq{ope_beam_soft_tmd} is thus an example where parameterizing a dependence
(here the channel dependence) is beneficial or even necessary for obtaining a
natural-size estimate.

Following our discussion in \sec{parameterization}, if we do not require precise
correlations in $x$, one option would be to only parameterize the integral of $I_{ij,n}(z)$,
with e.g.\ a trivial $z$ dependence $\sim\delta(1-z)$. If we do require
proper correlations in $x$, i.e.\ in $Y$ and/or $\Ecm$, we need to properly
parameterize the $z$ dependence. At the orders we are working their true expressions
are actually known~\cite{Ebert:2020yqt, Luo:2020epw}. Therefore, as a starting
point we parameterize them using their known functional
form in $z$ multiplied by an overall scalar coefficient
%%%
\begin{equation} \label{eq:tnps_B}
\tilde I_{ij,n}(z, \theta_n^{B_{ij}}) = \frac{3}{2}\,\theta_n^{B_{ij}}\, \hat{\tilde I}_{ij,n}(z)
\,,\end{equation}
%%%
where we include a factor of $3/2$ to be conservative and account for the fact that
their true values are typically somewhat below their natural size. Another option
would be to explicitly normalize the $\hat{\tilde I}_{ij,n}(z)$ in some way.

With the TNP parameterization in \eq{tnps_B}, we effectively treat the shape as
exactly known, while the overall normalization
is unknown. We prefer this option to using $\delta(1-z)$, because it means
we have the exact correlations for the overall normalization uncertainty we do consider.
Of course, at the highest known order we cannot do that, and strictly speaking
at lower orders we should not be allowed to use the known shape but include some shape
uncertainties.
In the future, the $z$ dependence of the matching kernels can be explicitly parameterized,
for example by using strategy 2 and expanding them in $\varepsilon = 1-z$, since their $z\to 1$
limit is actually well understood~\cite{Billis:2019vxg}.

The dominant partonic channels are $ij = \{qqV, qg\}$, which start at $\ord{1}$
and $\ord{\alpha_s}$. At higher orders, further singlet channels
$ij = \{q\bar qV, qqS$, $qq\Delta S\}$ appear, whose precise definition
is given in \refcite{Billis:2019vxg}. Since they only give small corrections even
at the order they appear, for our numerical results in \sec{qT_results}
we only consider two TNPs for the beam boundary condition, namely
a single effective $\theta_n^{B_{qq}}$,
%%%
\begin{equation}
\theta_n^{B_{qq}} \equiv \theta_n^{B_{qqV}}
\equiv \theta_n^{B_{q\bar qV}}
\equiv \theta_n^{B_{qqS}}
\equiv \theta_n^{B_{qq\Delta S}}
\,,\end{equation}
%%%
which collectively varies all $qq$ channels together with $\theta_n^{B_{qg}}$
for the $qg$ channel.

In principle, we also have to include the QCD splitting functions, which govern
the evolution of the PDFs, in our counting. In the resummed cross section, the
PDFs enter through the beam functions where they are evaluated at the scale of
the beam function, which means their evolution contributes to the $q_T$ resummation by
resumming single logarithms of $b_T$. That is, they count as a noncusp anomalous
dimension.
Constructing TNP parameterizations for the splitting functions can be
done similarly to the beam function matching kernels by considering their $z\to 1$
and also $z\to 0$ limits.
In fact, in this way TNPs for the four-loop splitting functions have already been
considered in \refcite{McGowan:2022nag} including constraints from their known
moments.
Since varying the splitting functions is rather involved technically, as it requires re-evolving
the PDFs, we refrain from doing so here, and leave this for future work.
Instead, if needed, this source of uncertainty can be probed for now by
conventional $\mu_F$ variations.

%~~~~~~~~~~~~~~~~~~~~~~~~~~~~~~~~~~~~~~~~~~~~~~~~~~~~~~~~~~~~~~~~~~~~~~~~~~~~~~~
\subsubsection{Summary of TNPs}
\label{sec:qT_TNPs_summary}
%~~~~~~~~~~~~~~~~~~~~~~~~~~~~~~~~~~~~~~~~~~~~~~~~~~~~~~~~~~~~~~~~~~~~~~~~~~~~~~~

To summarize, we have a minimum of seven TNPs,
corresponding to seven independent perturbative ingredients and thus
sources of uncertainty: Three anomalous dimensions and four
boundary conditions, which belong to the category of matrix-element constants,
%%%
\begin{equation}
\theta_n^\gamma: \gamma \in \{\Gamma, \gamma_\mu, \gamma_\nu\}
\,,\qquad
\theta_n^\f: \f \in \{H, S, B_{qq}, B_{qg} \}
\,.\end{equation}
%%%

There is actually one piece of perturbative information that we have silently taken for granted
so far: The solution of the RGEs also requires the QCD $\beta$ function, because it governs
the $\mu$ dependence of $\alpha_s(\mu)$, and its TNP would in principle enter in
the resummation at the same loop order as the TNP of the cusp anomalous dimension.
In practice however, while the overall $\mu$ evolution of $\alpha_s(\mu)$ is important,
the higher-order corrections to it tend to be numerically very small. We therefore
continue to treat the $\beta$ function as known to avoid adding significant but
unnecessary complexity.

In addition to the above 7 TNPs (or 6 if we are willing to count the beam function as a single one),
we have 3 (or 4) more once we account for the full set of partonic channels of the beam function
(still without accounting for its functional dependence).
In addition, we have $2$ more once we account for singlet contributions
to the hard function, $1$ more if we also count the $\beta$ function,
and several more once we account for the splitting functions.

Let us contrast this with a scale-variation based approach to estimate perturbative
uncertainties. Even the most sophisticated currently available scale-variation-based
setup~\cite{Billis:2024dqq}
involves 5 scales ($\mu_H$, $\mu_B$, $\mu_S$, $\nu_B$, $\nu_S$) that are being varied,
which thus cannot
begin to capture even the minimal space of theory uncertainties and correlations.
And in most other approaches to $q_T$ resummation even fewer scales are considered.

%===============================================================================
\subsection{Numerical results}
\label{sec:qT_results}
%===============================================================================

%-------------------------------------------------------------------------------
\begin{figure}
\includegraphics[width=0.48\textwidth]{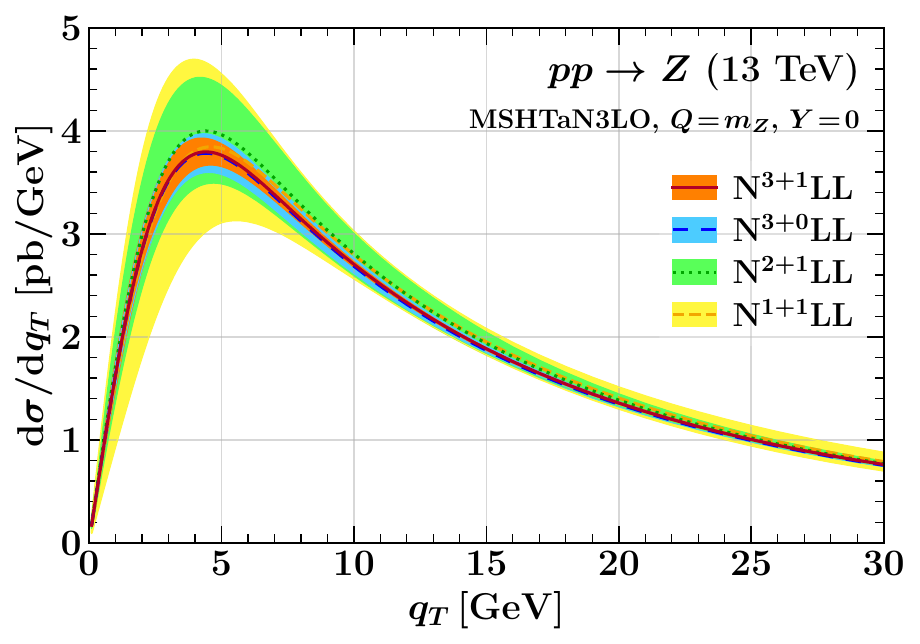}%
\hfill%
\includegraphics[width=0.5\textwidth]{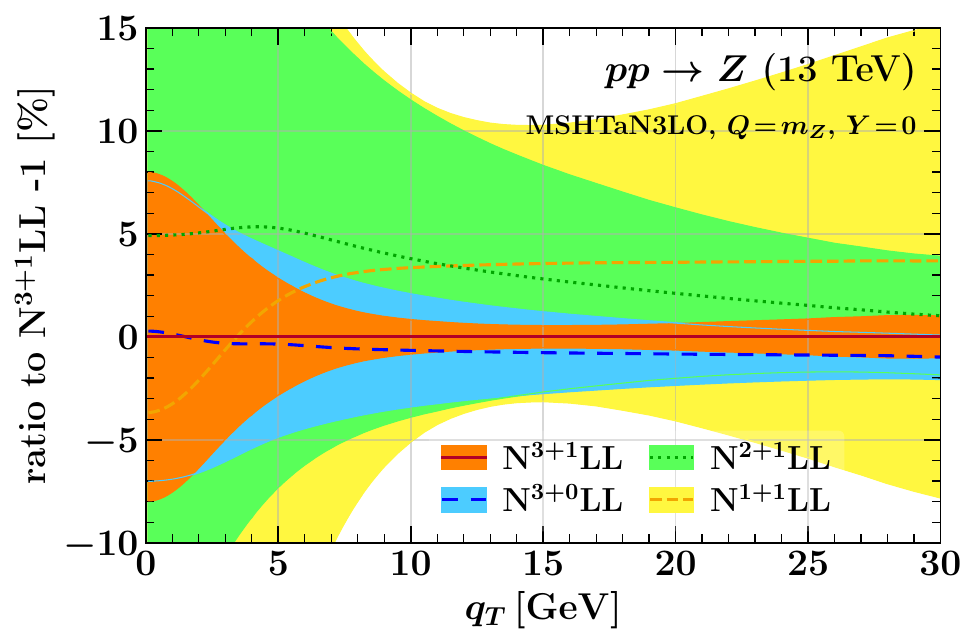}%
\caption{Leading-power $q_T\equiv p_T^Z$ spectrum for inclusive $pp\to Z$ production
at the 13 TeV LHC at N$^{1+1}$LL (yellow), N$^{2+1}$LL (green),
N$^{3+0}$LL (blue), and N$^{3+1}$LL (orange). The results are shown for the absolute
spectrum on the left and as the relative difference to the N$^{3+1}$LL central value
on the right. The bands show the total theory uncertainty at 95\% theory CL.}
\label{fig:spectrum}
\end{figure}
%-------------------------------------------------------------------------------

%-------------------------------------------------------------------------------
\begin{figure}
\hfill%
\includegraphics[width=0.485\textwidth]{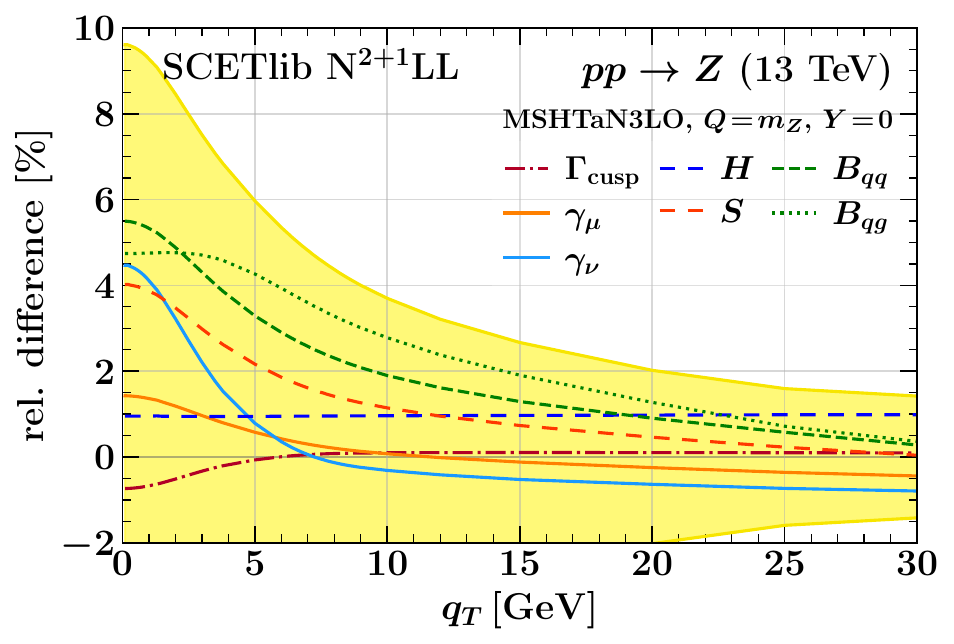}%
\hfill%
\includegraphics[width=0.485\textwidth]{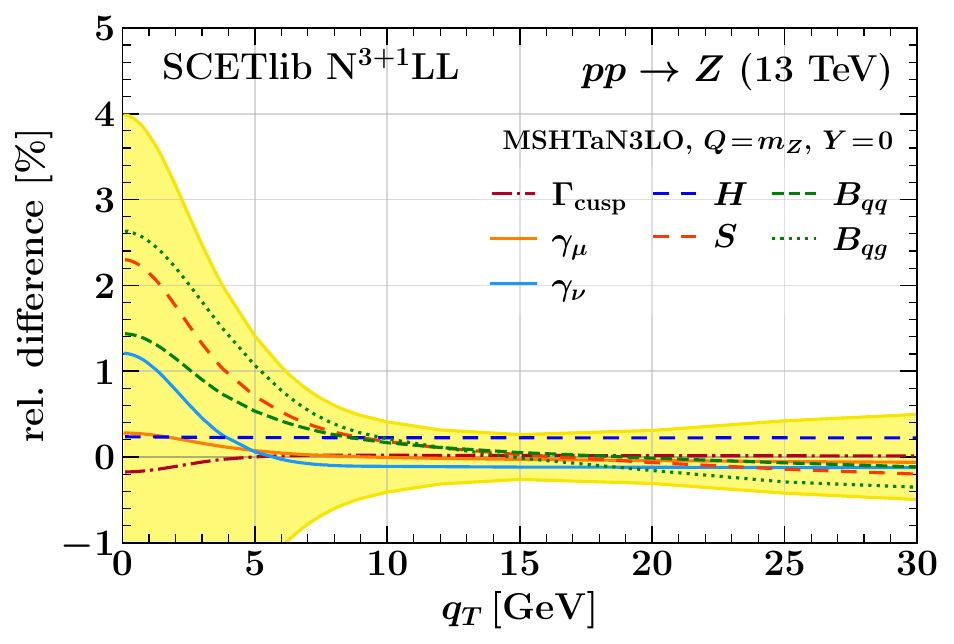}%
\\
\hspace*{\fill}%
\includegraphics[width=0.485\textwidth]{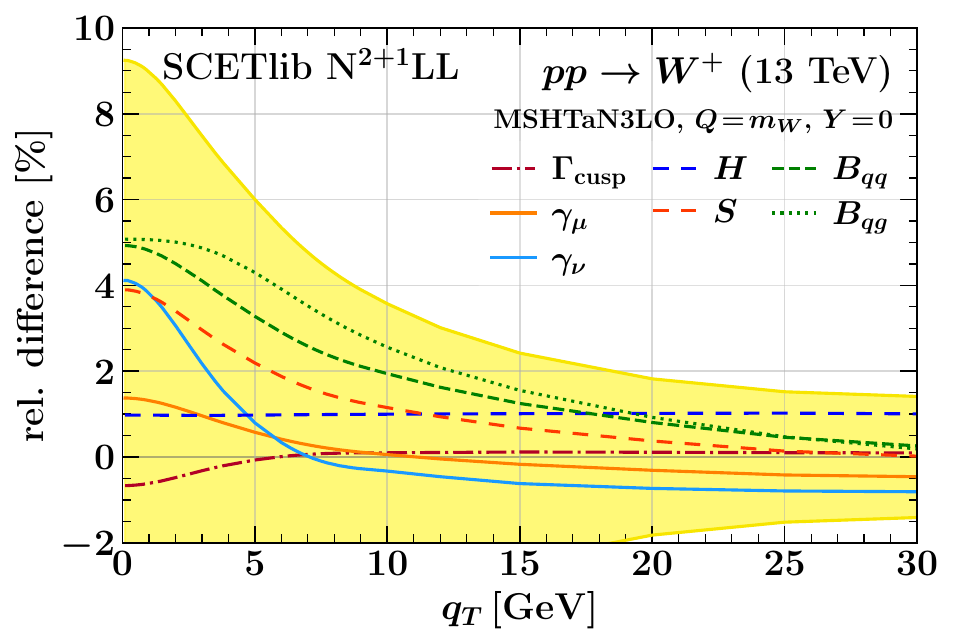}%
\hfill%
\includegraphics[width=0.485\textwidth]{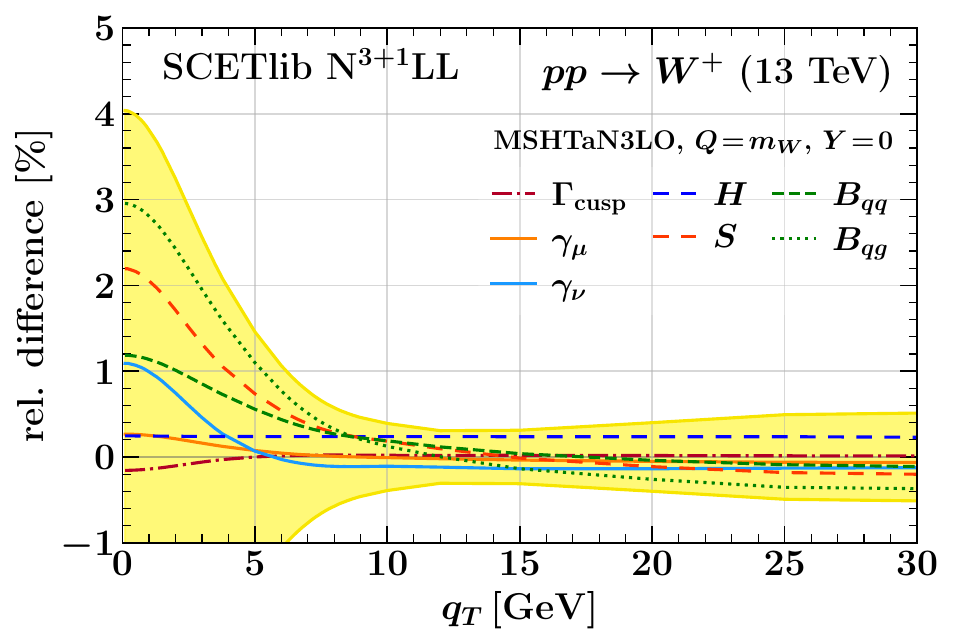}%
\\
\includegraphics[width=0.50\textwidth]{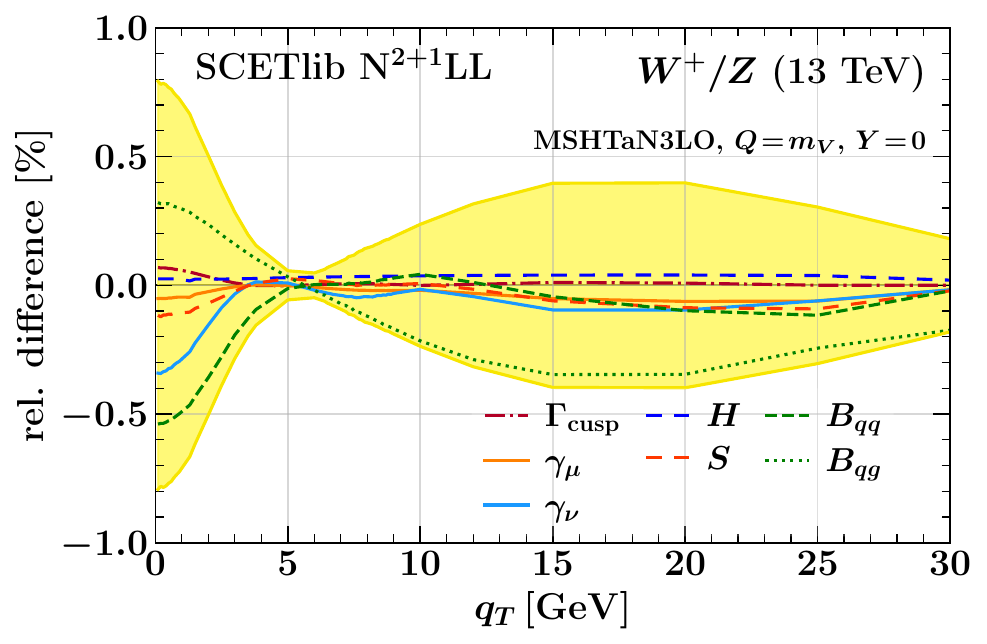}%
\hfill%
\includegraphics[width=0.50\textwidth]{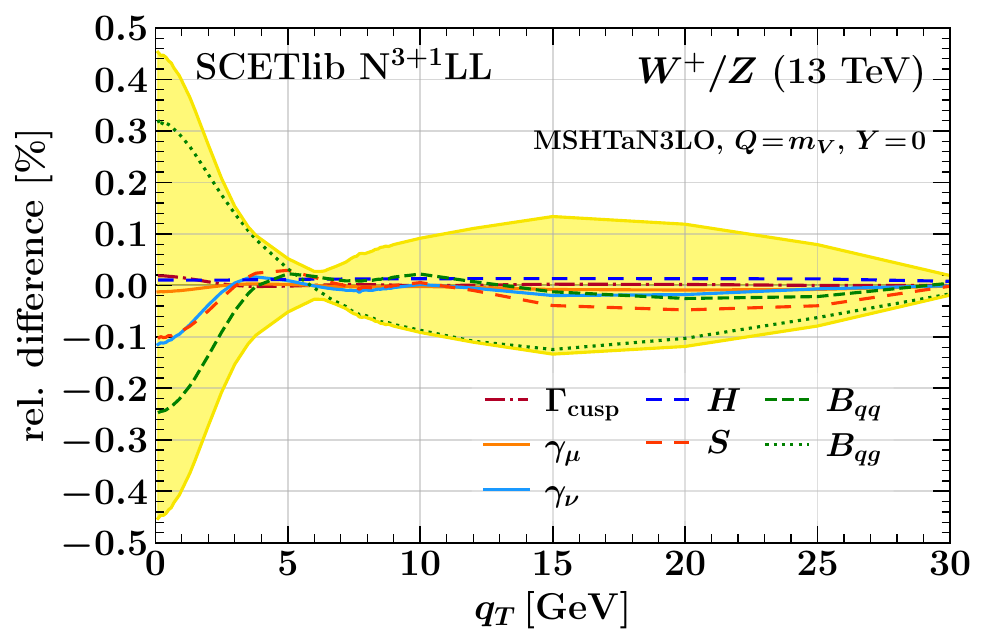}%
\caption{Breakdown of the relative uncertainties in the leading-power $q_T \equiv p_T^{Z,W}$
spectrum at the 13 TeV LHC at N$^{2+1}$LL (left panels) and N$^{3+1}$LL (right panels)
for $pp\to Z$ (top row), $pp\to W^+$ (middle row), and their
ratio (bottom row). The different lines show the impact of varying the corresponding
theory nuisance parameter by $+1$ or $-1$, corresponding to 68\% theory CL.
The yellow band shows their sum in quadrature. See the text for more details.}
\label{fig:impact_ZW_level1}
\end{figure}
%-------------------------------------------------------------------------------

%-------------------------------------------------------------------------------
\begin{figure}
\hfill%
\includegraphics[width=0.485\textwidth]{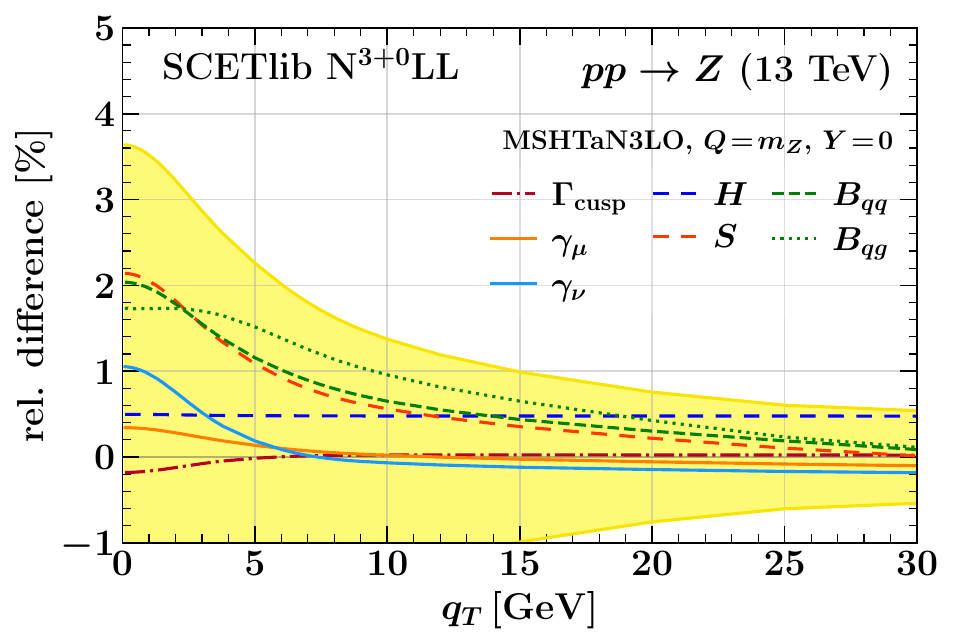}%
\hfill%
\includegraphics[width=0.485\textwidth]{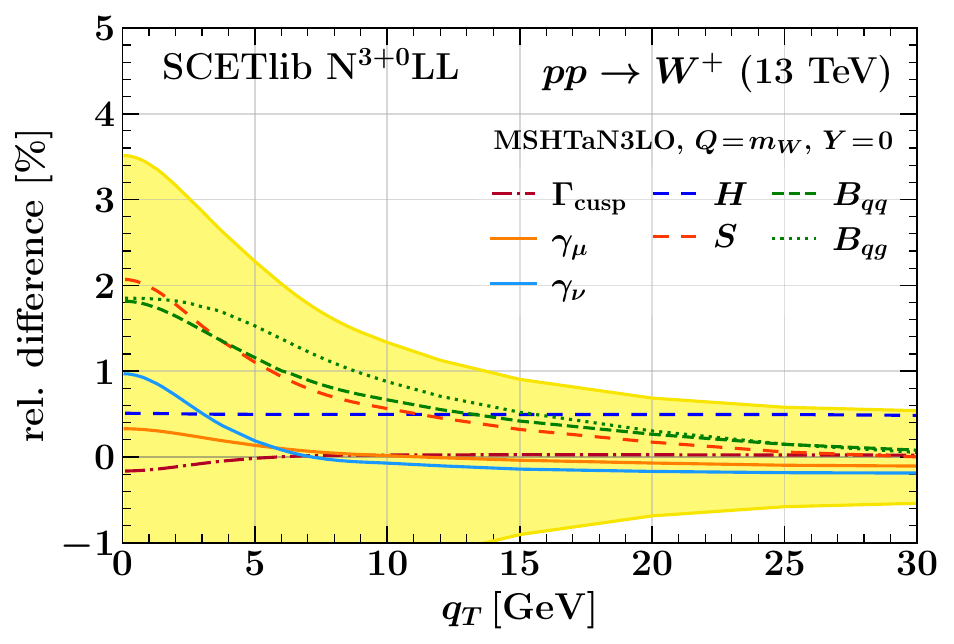}%
\\
\hspace*{\fill}%
\includegraphics[width=0.5\textwidth]{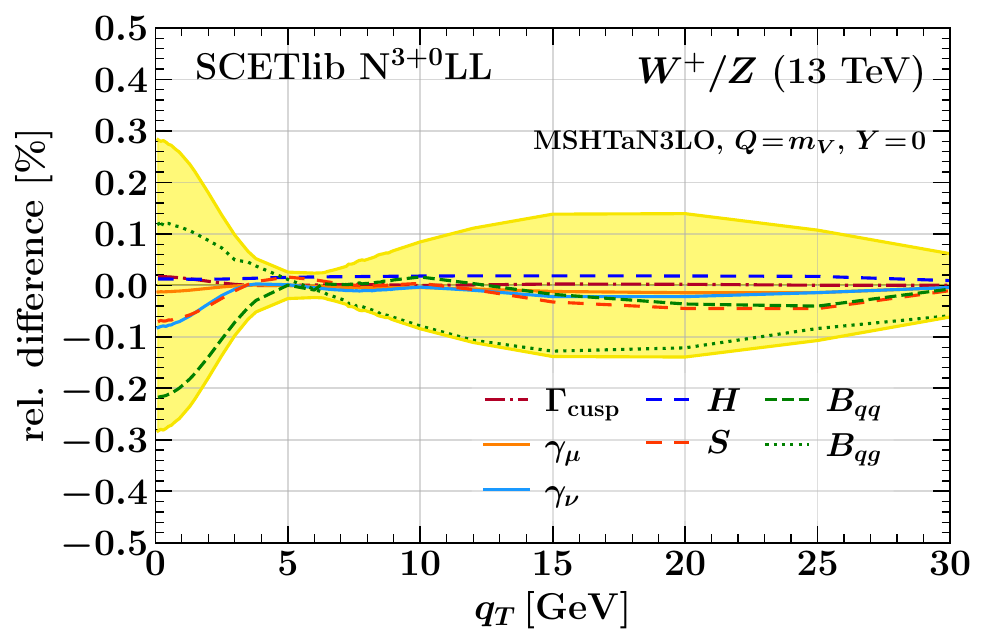}%
\hspace*{\fill}%
\caption{Same as \fig{impact_ZW_level1} but for the approximate implementation
at N$^{3+0}$LL.}
\label{fig:impact_ZW_level0}
\end{figure}
%-------------------------------------------------------------------------------

For our numerical results for the leading-power $q_T$ spectrum, we consider inclusive
$pp\to V$ production with $V = Z,W^\pm$ at the 13 TeV LHC at fixed invariant mass
$Q \equiv \sqrt{q^2} = m_V$ and rapidity $Y = 0$ of the vector boson, unless noted otherwise.
We use the \texttt{MSHT20an3lo}~\cite{McGowan:2022nag} PDF set with $\alpha_s(m_Z) = 0.118$.
All results are obtained with \texttt{SCETlib}~\cite{scetlib} based on its
implementation of $q_T$ resummation up to N$^4$LL~\cite{Billis:2019vxg, Ebert:2020dfc, Billis:2021ecs, Billis:2024dqq}, which we have extended to support the required theory nuisance parameter variations.
Since we only consider the leading-power spectrum without matching to the full fixed-order result
at large $q_T$, we restrict ourselves to $q_T\leq 30\GeV$, where the neglected
$\ord{q_T^2/Q^2}$ and higher power corrections amount to at most a few-percent correction and
the uncertainties associated with the matching procedure are also not yet relevant~\cite{Billis:2024dqq}.

In \fig{spectrum}, we start by presenting the $Z$ $q_T$ spectrum at different subsequent
orders up to N$^{3+1}$LL. The uncertainty bands show the total theory uncertainty at
95\% theory CL from varying all TNPs by $\Delta u_n = \pm 2$.
Note that the N$^{3+0}$LL result is an intermediate order, which
is included for illustration and future reference.

In \fig{impact_ZW_level1} we show the breakdown of the theory uncertainty by
individual TNPs. The impacts on the spectrum from varying the TNPs up or down are
roughly symmetric, so for clarity we always only show the $\Delta u_n = +1$ variation for the anomalous
dimensions, hard function, and soft function, and the $\Delta u_n = -1$ variation for the beam function.
Since each TNP corresponds to an independent source of uncertainty, which furthermore
can be considered Gaussian distributed (see \sec{tnp_scalar}), the correct way
to combine them into a total uncertainty is to add them in quadrature.
This is shown by the light yellow band%
\footnote{%
To be precise, in case of small asymmetries we sum in quadrature the larger of the
up and down variations for each TNP for definiteness.}%
, corresponding to the total uncertainty at 68\% theory CL.
In the top and middle rows we show the results for $Z$ and $W^+$ and in the bottom
row their ratio.

Concerning the point-by-point correlations in the shape of the $q_T$ spectrum, each of the TNP
variations shown by the different lines in \fig{impact_ZW_level1} reflects a
100\% correlated uncertainty component across the $q_T$ spectrum.
We observe that the different components
have quite different shapes and include cases where the correlation is always
positive as well as cases switching sign from correlated to anticorrelated at different
points in $q_T$. Hence, as anticipated, it is not possible to correctly model the correlations
in the $q_T$ spectrum by a 100\% (anti)correlated hypothesis.

Let us briefly compare this to a scale-variation based approach.
By scanning over different scale variations
in order to account for the shape uncertainties, one is precisely scanning over various (more-or-less arbitrary) 100\% (anti)correlated hypotheses. We remind the reader (see \sec{correlations}) that when correlations matter,
incorrect correlation assumptions can have dramatic consequences on the resulting uncertainties.
In \refcite{TNPalphas}, we will show explicitly that accounting for the correct point-by-point
correlations in $q_T$ is
indeed absolutely critical if one wants to exploit precision measurements of the $q_T$ spectrum, particularly its extremely precisely measured shape, for interpretation purposes such as extracting nonperturbative parameters or the strong coupling constant.
All current attempts at doing so, including the recent analysis in \refcite{ATLAS:2023lhg}, are based on scale variations and are thus subject to uncontrolled
correlation assumptions in the underlying perturbative predictions. Hence, the quoted perturbative theory uncertainties in the extracted parameters of interest cannot be taken at face value but must be interpreted with extreme caution.

As expected, the uncertainties are very similar for the closely related $Z$ and $W$ processes,
whose main differences are the different partonic channel combinations and the small difference in their masses. Since the TNPs for both processes are the same, each of the individual
impacts are 100\% correlated between the processes, and as a result cancel in the ratio to
very large extent, roughly by a factor of 10.
We stress that whilst a large degree of cancellation is expected and has been encountered
many times before, we can now correctly quantify it for the first time, and
in particular also its dependence on $q_T$.

In \fig{impact_ZW_level0} we show the same results using the approximate implementation
of the theory nuisance parameters at N$^{3+0}$LL.
This is the setup utilized for the resummed component
of the $q_T$ spectrum in the analysis of \refcite{CMS:2024lrd},
where it is also further matched to the fixed-order NLO$_1$ result for the $V$+1-parton process.
At N$^{3+0}$LL, formally the same $\theta_n$ enter as at N$^{3+1}$LL but their
impacts are only approximately correct. Namely, their shape
is approximated by the corresponding one at N$^{2+1}$LL, while their overall
impact is similar to that at N$^{3+1}$LL.
Whilst the precise shapes of the components differ between N$^{3+1}$LL and N$^{3+0}$LL
their overall qualitative behaviour is similar. The total uncertainties
at N$^{3+0}$LL are similar to but somewhat larger than at full N$^{3+1}$LL.
Notably, the uncertainties on the $W/Z$ ratio, which strongly depend
on the detailed correlations, are very similar to those at N$^{3+1}$LL.
Therefore, we can conclude that the N$^{3+0}$LL result provides a clear improvement
over N$^{2+1}$LL and a reasonable approximation to the more correct N$^{3+1}$LL
result. Although the N$^{3+1}$LL result should be preferred, the approximate N$^{3+0}$LL
result can serve as a viable compromise if the former cannot be utilized for some reason.
One such reason could be the availability of the required fixed-order matching at large $q_T$.
Since N$^{3+0}$LL implements the N$^3$LL structure it can be consistently matched to NLO$_1$,
whereas N$^{3+1}$LL implements the full N$^4$LL structure and therefore requires matching
to NNLO$_1$.

\begin{figure}
\hfill%
\includegraphics[width=0.485\textwidth]{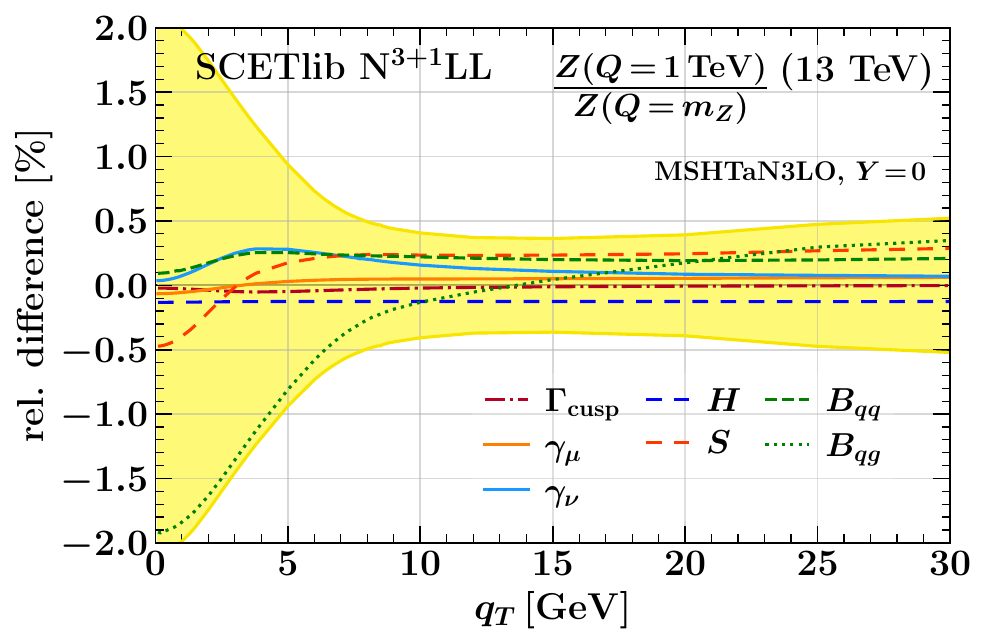}%
\hfill%
\includegraphics[width=0.485\textwidth]{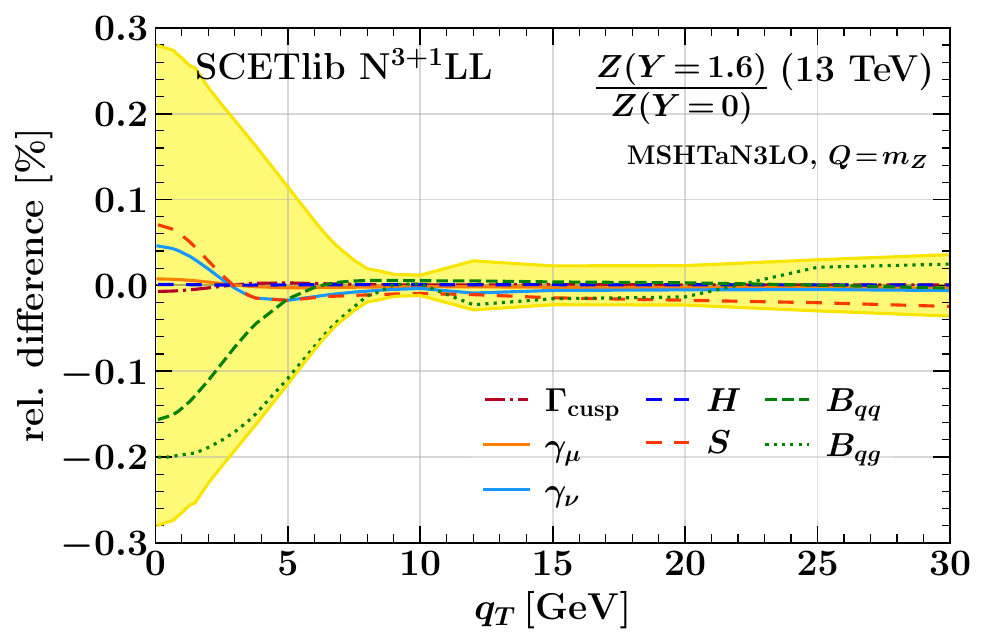}%
\caption{Relative uncertainties in the leading-power $q_T \equiv p_T^{Z}$
spectrum at the 13 TeV LHC at N$^{3+1}$LL for the ratio
of $pp\to Z$ at $Q = 1\TeV$ vs.\ $Q = m_Z$ (left) and $Y = 1.6$ vs.\ $Y = 0$ (right).
The meaning of the curves is the same as in \fig{impact_ZW_level1}.}
\label{fig:other_Z_ratios}
\end{figure}

\begin{figure}
\hfill%
\includegraphics[width=0.485\textwidth]{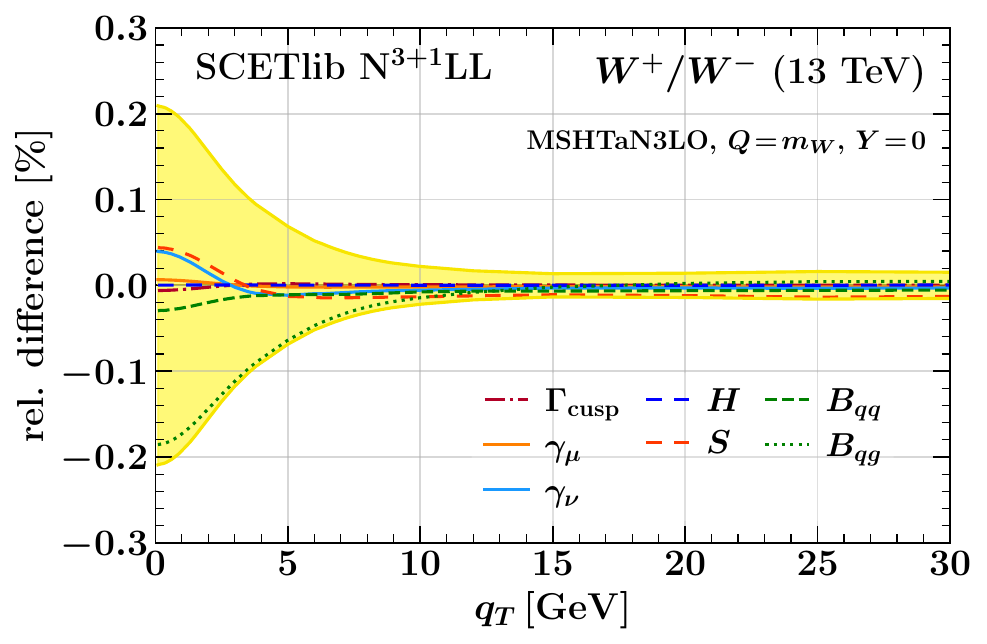}%
\hfill%
\includegraphics[width=0.485\textwidth]{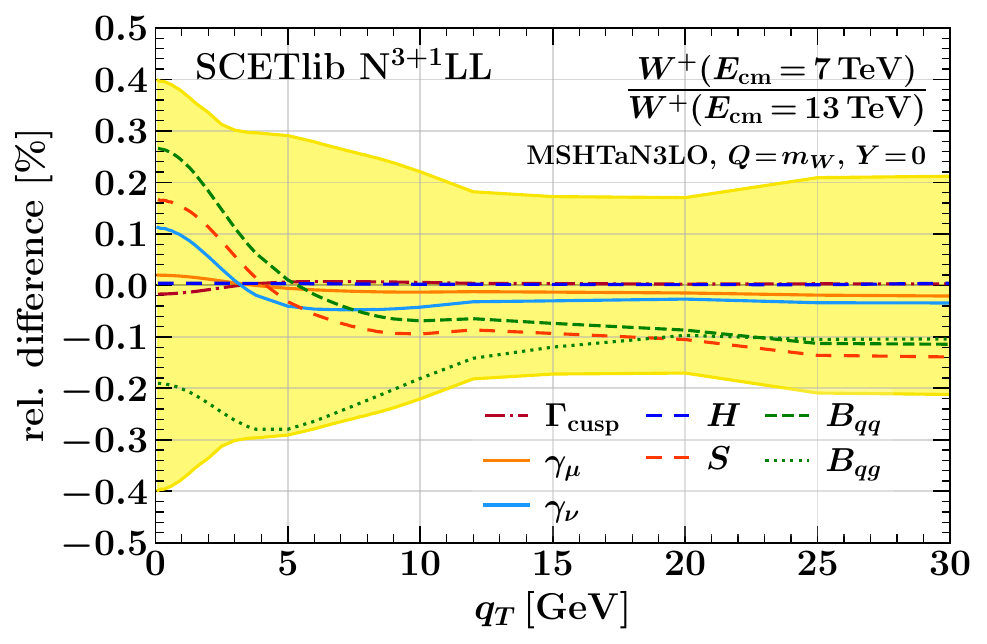}%
\caption{Relative uncertainties in the leading-power $q_T \equiv p_T^{W}$
spectrum at N$^{3+1}$LL for the ratio
of $pp\to W^+$ vs.\ $pp\to W^-$ at the 13 TeV LHC (left) and $pp\to W^+$
at the 7 TeV LHC vs.\ 13 TeV LHC (right).
The meaning of the curves is the same as in \fig{impact_ZW_level1}.}
\label{fig:other_W_ratios}
\end{figure}

In \fig{other_Z_ratios} we show the ratios of the $q_T$ spectra for $Z$ production
at $Q = 1\TeV$ vs.\ $Q = m_Z$ and $Y = 1.6$ vs.\ $Y = 0$.
\Fig{other_W_ratios} shows the ratios of the $q_T$ spectra
for $W^+$ vs.\ $W^-$ and for $W^+$ at 13 TeV vs.\ 7 TeV.
The cancellation of uncertainties is expectedly most pronounced for $W^+/W^-$.
It is weakest but still present for the case of $Q = 1\TeV$ vs.\ $Q = m_Z$.
This is also not unexpected, since the spectrum
mostly depends on $q_T/Q$, so for different $Q$ the $q_T$ spectra are shifted
against each other.

We stress that the primary purpose of the various ratios we show is to easily
visualize the effect of correlations and the resulting degree of cancellations.
When correctly accounting for the theory correlations there is no
difference as far as theory uncertainties are concerned in using the ratio or
the quantities separately.
In a real analysis, one would typically not use ratios but simply perform
a combined analysis of all relevant processes, which constrains the TNPs among
all of them accounting for
all correlations and resulting cancellations. In the limit where one particular
process is much more precisely measured than the others, one can think of it as
effectively acting as a control process to obtain improved predictions for the
others.

%-------------------------------------------------------------------------------
%-------------------------------------------------------------------------------

%-------------------------------------------------------------------------------
%-------------------------------------------------------------------------------

An important observation is that the dominant uncertainties that remain in the ratios
and tend to cancel the least are those due to the beam functions, in particular
for $W/Z$ but also in many cases for the other ratios in \figs{other_Z_ratios}{other_W_ratios}.
This is because the main difference between the processes, which is due to the different
combinations of flavor channels, precisely enters via a different relative mix of
different beam functions.
This motivates a more detailed study of the beam function TNPs.

To conclude this section, we stress again that here we only consider the leading-power contributions.
This is warranted as these are by far the dominant
contributions to the spectrum at small $q_T$, so the precise correlation and resulting
cancellation of their uncertainties is a critical ingredient to any interpretation of
precision measurements of the $q_T$ spectrum, which we are able to properly take into account
for the first time. The uncertainties in the ratios illustrate the level of precision that can
now be reached via the cancellation of the dominant uncertainties in a combined analysis.
At the resulting sub-percent level of precision, many other previously subleading
effects can become equally or more important and must be accounted for to maintain
this level of precision, motivating future work on them. We briefly comment on
these in the next subsection.

%===============================================================================
\subsection{Subleading effects}
\label{sec:qT_subleading}
%===============================================================================

The application of our approach to the leading-power resummed contribution
represents a crucial milestone toward a more complete and comprehensive understanding of
the $q_T$ spectrum. A complete treatment also requires accounting for several
other subdominant effects. These are in particular:
%%%
\begin{itemize}
   \item The neglected subleading power corrections in \eq{power_expansion} starting
   at $\ord{q_T^2/Q^2}$.
   \item Effects due to finite quark masses of $\ord{m_q^2/q_T^2}$.
   \item QED and electroweak effects.
   \item Nonperturbative corrections of $\ord{\lqcd^2/q_T^2}$.
\end{itemize}
%%%
The nonperturbative corrections already have a parametric nature.
In the future, our approach can also be applied systematically to the first three
effects following the methodology developed in the previous sections.
In the absence of their complete TNP-based treatment,
they can still be included from existing results based on conventional methods.
In other words, a TNP treatment of even just the leading-power resummed component is
already extremely valuable, simply because it contributes
by far the dominant uncertainties.

%%%%%%%%%%%%%%%%%%%%%%%%%%%%%%%%%%%%%%%%%%%%%%%%%%%%%%%%%%%%%%%%%%%%%%%%%%%%%%%%
\section{Conclusions}
\label{sec:conclusions}
%%%%%%%%%%%%%%%%%%%%%%%%%%%%%%%%%%%%%%%%%%%%%%%%%%%%%%%%%%%%%%%%%%%%%%%%%%%%%%%%

The theory nuisance parameter approach developed in this paper holds enormous potential
to make perturbative predictions more robust and also more precise:
\begin{itemize}
\item
It allows for the first time to correctly account for theory correlations, which
are important whenever one simultaneously interprets multiple measurements
(including different bins in a spectrum).
\item
The theory uncertainties and correlations are straightforward to propagate,
like any other nuisance parameters,
into fits, Monte-Carlo generators, multivariate analyses, neural networks, etc.
\item
In fits to experimental measurements, it is possible and consistent
to profile the theory nuisance parameters and thereby constrain them,
effectively reducing the theory uncertainties by the measurements, which is not
possible with existing methods.
\item
New structures (e.g.\ partonic channels or additional logarithmic powers) appearing
at higher order are explicitly anticipated and accounted for by the theory uncertainties.
\item
All new, even partial, higher-order information can have immediate phenomenological
impact in reducing theory uncertainties, even if the complete next order is not yet available.
\item
The theory uncertainties have a well-defined and meaningful statistical interpretation.
\end{itemize}

When applied to color-singlet transverse-momentum ($p_T$) resummation,
the theory nuisance parameters allow one to correctly and fully account for the
theory correlations in the shape of the small-$p_T$ spectrum, between different
$Q$ values, partonic channels, hard processes (e.g. $W$ and $Z$ production),
collider energies, and different resummation-sensitive variables
(e.g. $p_T^V$, $p_T^\ell$ near the Jacobian peak, or $\phi^*$).
In this context, our approach opens the door to reaching sub-percent level theoretical precision, which
will be able to match the incredible precision already achieved by experimental
measurements. To fully reach this level of theoretical precision, a variety of subleading
effects must still be accounted for.
We thus hope that our results also provide strong motivation for future work in this direction.
For expedience, they can at first be included using conventional methods, which
does not invalidate the TNP-based treatment of the dominant uncertainties.
More importantly, our approach is also not limited to the dominant resummed contribution.
It can be systematically and incrementally applied also to subleading effects as
they become relevant at any given level of theory precision.

More generally, it will obviously be impossible to equip existing predictions with TNP-based
uncertainties all at once. We should stress that this is also not required by the TNP approach.
To the contrary, a more practical, incremental adoption, focusing on the dominant sources
of uncertainties first is exactly in the spirit of our approach, namely to parameterize
and include the sources of uncertainties in the theory predictions in order of their relevance.

%%%%%%%%%%%%%%%%%%%%%%%%%%%%%%%%%%%%%%%%%%%%%%%%%%%%%%%%%%%%%%%%%%%%%%%%%%%%%%%%
\acknowledgments
I would like to thank Maarten Boonekamp for fruitful discussions that originally
inspired this work. I am very grateful to my experimental colleagues,
Simone Amoroso, Ludovica Aperio Bella, Josh Bendavid, Maarten Boonekamp,
Stefano Camarda, Glen Cowan, Andre David, Daniel Froidevaux, Kenneth Long,
Kerstin Tackmann, and many others, for numerous
discussions and encouragement to pursue this work. I also thank Georgios Billis,
Thomas Cridge, Giulia Marinelli, and Johannes Michel for discussions and
collaboration on related work. I am indebted to Giulia Marinelli (to be
repaid in chocolate muffins) for her generous help in preparing the numerical
results in \sec{qT}.
I thank Lawrence Berkeley National Laboratory for hospitality while
parts of this work were carried out.
This project has received funding from the European Research Council (ERC)
under the European Union's Horizon 2020 research and innovation programme
(Grant agreement 101002090 COLORFREE).
%%%%%%%%%%%%%%%%%%%%%%%%%%%%%%%%%%%%%%%%%%%%%%%%%%%%%%%%%%%%%%%%%%%%%%%%%%%%%%%%

%%%%%%%%%%%%%%%%%%%%%%%%%%%%%%%%%%%%%%%%%%%%%%%%%%%%%%%%%%%%%%%%%%%%%%%%%%%%%%%%
\appendix
%%%%%%%%%%%%%%%%%%%%%%%%%%%%%%%%%%%%%%%%%%%%%%%%%%%%%%%%%%%%%%%%%%%%%%%%%%%%%%%%

%%%%%%%%%%%%%%%%%%%%%%%%%%%%%%%%%%%%%%%%%%%%%%%%%%%%%%%%%%%%%%%%%%%%%%%%%%%%%%%%
\section{Sample of Known Perturbative Series}
\label{app:series_sample}
%%%%%%%%%%%%%%%%%%%%%%%%%%%%%%%%%%%%%%%%%%%%%%%%%%%%%%%%%%%%%%%%%%%%%%%%%%%%%%%%

The QCD matrix-element constants and anomalous dimensions included in the samples
of known perturbative series in \sec{tnp_statistics} are listed in
\tabs{matrix_elements}{anomalous_dimensions}. For definiteness, we give the explicit
1-loop results in the 3rd column. The 4th column shows the included loop orders
and the last column gives the original references (starting at 3 loops for brevity).
As already mentioned in \sec{tnp_distribution}, we have made an effort
to include all matrix-element constants known to four loops
and anomalous dimensions known to four and five loops in QCD. The
included quantities that are known at lower orders are certainly not exhaustive,
and more can be added in the future.

Following the normalization conventions discussed in \secs{tnp_constants}{tnp_anom_dims},
we consider matching coefficients and
jet and beam functions directly, while we consider the square root for decay rates
and soft functions, and also include corresponding factors of $2$ and $1/2$ for the
associated anomalous dimensions.
For the beam function matching kernels, $\tilde I_{ij}(z)$,
we reduce their $z$ dependence by considering their $z^1$ moment (2nd Mellin moment).
For $\tilde I_{qq}(z)$ we also consider its total integral (1st Mellin moment),
which exists since this kernel does not have a $1/z$ singularity. Similarly,
for the QCD splitting functions we consider their lowest moments in $z$, which
in some cases are known to five loops.

For scheme-dependent quantities (decoupling constants, Wilson coefficients, beam,
jet, and soft functions)
we always use the \MSbar\ scheme and canonical logarithms to define their
scalar series for the constant terms or boundary conditions.
That is, the constant terms are defined as the remaining nonlogarithmic terms
when all logarithmic terms are written in terms of the respective canonical
(possibly distributional) logarithms.
These choices define the reference scheme for their corresponding TNPs.
Some explicit examples with more details can be found in \sec{qT_TNPs}.
For the soft functions we only consider the quark functions, since the gluon
ones are closely related to the quark ones by Casimir scaling.
For the threshold and thrust soft functions we consider them both in position
and momentum space, since the translation between spaces causes a significant
reshuffling of the constant terms with the logarithmic terms such that the
constants in either space become largely uncorrelated. On the other hand, for the
$q_T$ or $b_T$ soft function we only consider the $b_T$-space result because
the constant terms in $q_T$ and $b_T$ space appear very strongly correlated.

%-------------------------------------------------------------------------------
\begin{table}
\tabcolsep 4pt
\begin{tabular}{lccll}
\hline\hline
Name & $f$ & $f_1$ & $n$ & refs.
\\\hline
$\alpha_s$ decoupling & $\zeta_\alpha$ & $0$  & $1,2,3,4$ &
\cite{Chetyrkin:1997un, Schroder:2005hy, Chetyrkin:2005ia, Gerlach:2018hen}
\\
quark mass decoupling & $\zeta_m$ & -  & $2,3,4$ &
\cite{Chetyrkin:1997un, Liu:2015fxa, Gerlach:2018hen}
\\
$ggH$ Wilson coefficient & $C_1$ & $-\frac{8}{3}T_F$ & $1,2,3,4$ &
\cite{Chetyrkin:1997un, Schroder:2005hy, Chetyrkin:2005ia, Herzog:2017dtz, Gerlach:2018hen}
\\
$q\bar qH$ Wilson coefficient & $C_2$ & - & $2,3,4$ &
\cite{Chetyrkin:1997un, Liu:2015fxa}
\\
$ggHH$ Wilson coefficient & $C_{HH}$ & $-\frac{8}{3}T_F$ & $3,4$ &
\cite{Grigo:2014jma, Spira:2016zna, Gerlach:2018hen}
\\
$\gamma^*\to q\bar q$ $R$-ratio (nonsinglet) & $\sqrt{R^{ns}_{q\bar qV}}$ & $\frac{3}{2}\,C_F$ & $1,2,3,4$ &
\cite{Gorishnii:1990vf, Surguladze:1990tg, Baikov:2008jh, Baikov:2012zm, Herzog:2017dtz}
\\
$\gamma^*\to q\bar q$ $R$-ratio (singlet) & $\sqrt{R^{s}_{q\bar qV}}$ & - & $3,4$ &
\cite{Gorishnii:1990vf, Surguladze:1990tg, Baikov:2008jh, Baikov:2012zm, Herzog:2017dtz}
\\
$H\to gg$ & $\sqrt{R_{gg}}$ & $\frac{73}{6}\,C_A-\frac{14}{3}\,T_Fn_f$ & $1,2,3,4$ &
\cite{Baikov:2006ch, Moch:2007tx, Herzog:2017dtz}
\\
$H\to q\bar q$ (nonsinglet) & $\sqrt{R_{q\bar qS}}$ & $\frac{17}{2}\,C_F$ & $1,2,3,4$ &
\cite{Chetyrkin:1996sr, Baikov:2005rw, Herzog:2017dtz}
\\
quark vector form factor & $c_{qqV}$ & $(-8+\frac{\pi^2}{6})\,C_F$ & $1,2,3,4$ &
\cite{Baikov:2009bg, Lee:2010cga, Gehrmann:2010ue, Lee:2022nhh}
\\
gluon scalar form factor & $c_{gg}$ & $\frac{\pi^2}{6}\,C_A$ & $1,2,3,4$ &
\cite{Baikov:2009bg, Lee:2010cga, Gehrmann:2010ue, Lee:2022nhh}
\\
quark scalar form factor & $c_{qqS}$ & $(-2+\frac{\pi^2}{6})\,C_F$ & $1,2,3,4$ &
\cite{Gehrmann:2014vha, Chakraborty:2022yan}
\\
quark jet function & $j_q$ & $(7-\pi^2)\,C_F$ & $1,2,3$ &
\cite{Bruser:2018rad}
\\
gluon jet function & $j_g$ & $(\frac{67}{9}-\pi^2)\,C_A-\frac{20}{9}\,T_Fn_f$ & $1,2,3$ &
\cite{Banerjee:2018ozf}
\\
quark EEC jet function & $j_q^{\rm EEC}$ & $(4-\frac{4\pi^2}{3})\,C_F$ & $1,2,3$ &
\cite{Ebert:2020sfi}
\\
gluon EEC jet function & $j_g^{\rm EEC}$ & $(\frac{65}{18}-\frac{4\pi^2}{3})\,C_A-\frac{5}{9}\,T_Fn_f$ & $1,2,3$ &
\cite{Ebert:2020sfi}
\\
$qq$ $b_T$ beam fct (integral) & $\tilde I_{qq,1}$ & $C_F$ & $1,2,3$ &
\cite{Ebert:2020yqt, Luo:2020epw}
\\
$qq$ $b_T$ beam fct ($z^1$ moment) & $\tilde I_{qq,2}$ & $\frac{1}{3}\,C_F$ & $1,2,3$ &
\cite{Ebert:2020yqt, Luo:2020epw}
\\
$qg$ $b_T$ beam fct ($z^1$ moment) & $\tilde I_{qg,2}$ & $\frac{1}{3}\,T_F$ & $1,2,3$ &
\cite{Ebert:2020yqt, Luo:2020epw}
\\
$gg$ $b_T$ beam fct ($z^1$ moment) & $\tilde I_{gg,2}$ & $0$ & $1,2,3$ &
\cite{Ebert:2020yqt, Luo:2020epw}
\\
$gq$ $b_T$ beam fct ($z^1$ moment) & $\tilde I_{gq,2}$ & $\frac{2}{3}\,C_F$ & $1,2,3$ &
\cite{Ebert:2020yqt, Luo:2020epw}
\\
$b_T$ soft function & $\sqrt{\tilde s_q}$ & $-\frac{\pi^2}{6}\,C_F$ & $1,2,3$ &
\cite{Li:2016ctv}
\\
threshold soft fct. (pos. space) & $\sqrt{\tilde s_{\rm thr}}$ & $\frac{\pi^2}{6}\,C_F$ & $1,2,3$ &
\cite{Li:2014afw}
\\
threshold soft fct. (mom. space) & $\sqrt{s_{\rm thr}}$ & $-\frac{\pi^2}{6}\,C_F$ & $1,2,3$ &
\cite{Li:2014afw}
\\
thrust soft fct. (pos. space) & $\sqrt{\tilde s_{\tau}}$ & $-\frac{\pi^2}{2}\,C_F$ & $1,2,3$ &
\cite{Baranowski:2024vxg}
\\
thrust soft fct. (mom. space) & $\sqrt{s_{\tau}}$ & $\frac{\pi^2}{6}\,C_F$ & $1,2,3$ &
\cite{Baranowski:2024vxg}
\\
heavy-light soft fct. & $\sqrt{s_{hl}}$  & $-\frac{\pi^2}{12}\,C_F$ & $1,2,3$ &
\cite{Bruser:2019yjk}
\\\hline\hline
\end{tabular}
\caption{Quantities included in our sample of known matrix-element constants.}
\label{tab:matrix_elements}
\end{table}
%-------------------------------------------------------------------------------

%-------------------------------------------------------------------------------
\begin{table}
\tabcolsep 4pt
\begin{tabular}{lccll}
\hline\hline
Name & $\gamma$ & $\gamma_0$ & $n$ & refs.
\\\hline
QCD $\beta$ function & $-2\beta$ & $-\frac{22}{3}C_A + \frac{8}{3}T_Fn_f$ & $0,1,2,3,4$ &
\cite{Tarasov:1980au, Larin:1993tp, vanRitbergen:1997va, Czakon:2004bu, Baikov:2016tgj, Herzog:2017ohr, Luthe:2017ttg}
\\
$ggH$ Wilson coefficient & $\gamma_t$ & 0 & $0,1,2,3,4$ &
% \cite{xxx}
\\
quark mass & $\gamma_m$ & $-6C_F$  & $0,1,2,3,4$ &
\cite{Tarasov:1982plg, Larin:1993tq, Chetyrkin:1997dh, Vermaseren:1997fq, Baikov:2014qja, Luthe:2016xec}
\\
vector correlator \makebox[0ex][l]{(nonsinglet)}  & $2\gamma^{\rm ns}_{V}$ & $\frac{8}{3} d_F$ & $0,1,2,3,4$ &
\cite{Baikov:2012zm}
\\
vector correlator \makebox[0ex][l]{(singlet)}  & $2\gamma^{\rm s}_{V}$ & - & $3,4$ &
\cite{Baikov:2012zm}
\\
scalar correlator & $2\gamma_{S}$ & $4 d_F$ & $0,1,2,3$ &
\cite{Chetyrkin:1996sr}
\\
$P^+_{\rm ns}(z)$ ($z^1$ moment) & $2\gamma^{\rm ns+}_2$ & $-\frac{16}{3}\,C_F$ & $0,1,2,3,4$ &
\cite{Larin:1993vu, Baikov:2006ai, Velizhanin:2011es, Baikov:2015tea, Moch:2017uml, Herzog:2018kwj, Blumlein:2021enk}
\\
$P^-_{\rm ns}(z)$ ($z^2$ moment) & $2\gamma^{\rm ns-}_3$ & $-\frac{25}{3}\,C_F $ & $0,1,2,3,4$ &
\cite{Larin:1993vu, Baikov:2015tea, Velizhanin:2014fua, Moch:2017uml, Herzog:2018kwj, Blumlein:2021enk}
\\
$P_{gg}(z)$ ($z^1$ moment) & $2\gamma_2^{gg}+2\beta$ & $-\frac{22}{3}\,C_A $ & $0,1,2,3$ &
\cite{Larin:1996wd, Moch:2021qrk}
\\
$P_{qg}(z)$ ($z^1$ moment) & $2\gamma_2^{qg}$ & $\frac{8}{3}\,T_F $ & $0,1,2,3$ &
\cite{Larin:1996wd, Ablinger:2017tan, Moch:2021qrk}
\\
quark cusp & $2\Gamma_\cusp^q$ & $8\,C_F$ & $0,1,2,3,(4)$ &
\cite{Moch:2004pa, Vogt:2004mw, Henn:2019swt, vonManteuffel:2020vjv, Herzog:2018kwj}
\\
gluon cusp & $2\Gamma_\cusp^g$ & $8\,C_A$ & $3$ &
\cite{Moch:2004pa, Vogt:2004mw, Henn:2019swt, vonManteuffel:2020vjv}
\\
tensor current & $\gamma_T$ & $2\,C_F$ & $0,1,2,3$ &
\cite{Gracey:2000am, Baikov:2006ai, Gracey:2022vqr}
\\
HQET heavy-light current & $\gamma_{\rm HQET}$ & $-3\,C_F$ & $0,1,2,3$ &
\cite{Chetyrkin:2003vi, Grozin:2023dlk}
\\
quark threshold PDF & $\gamma_f^q$ & $6\,C_F$ & $0,1,2,3$ &
\cite{Moch:2004pa, Moch:2017uml, Das:2019btv, Blumlein:2021enk}
\\
gluon threshold PDF & $\gamma_f^g - 2\beta$ & $0$ & $0,1,2,3$ &
\cite{Vogt:2004mw, Das:2020adl}
\\
quark collinear & $2\gamma_C^q$ & $-6\,C_F$ & $0,1,2,3$ &
\cite{Moch:2005id, vonManteuffel:2020vjv, Agarwal:2021zft}
\\
gluon collinear & $2\gamma_C^g + 2\beta$ & $0$ & $0,1,2,3$ &
\cite{Moch:2005tm, vonManteuffel:2020vjv, Agarwal:2021zft}
\\
heavy-quark collinear & $2\gamma_C^Q$ & $-4\,C_F$ & $0,1,2$ &
\cite{Grozin:2015kna, Bruser:2019yjk}
\\
quark jet function & $\gamma_J^q$ & $6\,C_F$ & $0,1,2,3$ &
\cite{Bruser:2018rad}
\\
gluon jet function & $\gamma_J^g - 2\beta$ & $0$ & $0,1,2,3$ &
\\
quark soft function & $\gamma_S^q/2$ & $0$ & $0,1,2,3$ &
\cite{Li:2014afw, Moch:2017uml, Das:2019btv}
\\
gluon soft function & $\gamma_S^g/2$ & $0$ & $3$ &
\cite{Li:2014afw, Das:2020adl}
\\
heavy-light soft function & $\gamma^{Q}_S/2$ & $2\,C_F$ & $0,1,2$ &
\cite{Bruser:2019yjk}
\\
quark rapidity & $\tilde\gamma_\nu^q/2$ & $0$ & $0,1,2,3$ &
\cite{Li:2016ctv, Vladimirov:2016dll, Duhr:2022yyp, Moult:2022xzt}
\\
gluon rapidity & $\tilde\gamma_\nu^g/2$ & $0$ & $3$ &
\cite{Li:2016ctv, Vladimirov:2016dll, Duhr:2022yyp, Moult:2022xzt}
\\\hline\hline
\end{tabular}
\caption{Quantities included in our sample of known anomalous dimensions.}
\label{tab:anomalous_dimensions}
\end{table}
%-------------------------------------------------------------------------------

Concerning the anomalous dimensions, the QCD $\beta$ function is defined as
%%%
\begin{align}
\mu\frac{\df\alpha_s(\mu)}{\df\mu}
&= -2\alpha_s\,\beta(\alpha_s)
\qquad\text{with}\qquad
\beta_0 = \frac{11}{3}C_A - \frac{4}{3}T_Fn_f
\,.\end{align}
%%%
Since it is the anomalous dimension of the coupling itself, it clearly plays a
special role. For example, it is the only anomalous dimension whose $n_f$
dependence starts at one loop. Despite its special role, we include it in our collection
for completeness.
A closely related anomalous dimension, which fits more naturally into our collection,
is the anomalous dimension $\gamma_t(\as)$ of the $ggH$ Wilson coefficient that arises
from integrating out the top quark, which is given to all orders by
%%%
\begin{equation}
\gamma_t(\as) = -2\as^2\,\frac{\df}{\df\as} \frac{\beta(\as)}{\as}
\,,\qquad
\gamma_{t\,n} = -2n\beta_n
\,.\end{equation}
%%%
Note that there are several gluonic anomalous dimensions, whose $n_f$ dependence also starts
at one loop. However, this dependence (and similarly the highest
power of $n_f$ at higher orders) is always that of $\beta(\as)$ itself, which we
therefore subtract. This can also be understood from the fact that the corresponding
quantities are always associated with an explicit power of $\alpha_s$ at the lowest
order, or equivalently, the corresponding operators involve a gluon field strength. It
would actually be more natural to always include an appropriate power of the coupling
with the field strength, which would then automatically remove the $\beta(\as)$ piece
from the anomalous dimension.

Consistency of the $e^+e^-$ thrust~\cite{Becher:2008cf, Abbate:2010xh} and
partonic beam-thrust~\cite{Stewart:2009yx} factorization implies
%%%
\begin{align}
4\gamma_C^i(\as) + 2\gamma_J^i(\as) + \gamma_S^i(\as) &= 0
\,,\end{align}
%%%
where $\gamma_S^i(\as)$ is the (noncusp) anomalous dimension of the (beam)thrust
soft function, and we have already used that the anomalous dimensions of the
SCET$_{\rm I}$ inclusive beam and jet function are equal, $\gamma_B = \gamma_J$~\cite{Stewart:2010qs}.
Consistency of color-singlet threshold factorization~\cite{Sterman:1986aj, Catani:1989ne} implies
%%%
\begin{align}
4\gamma_C^i(\as) + 2\gamma_f^i(\as) + \gamma_{\rm thr}^i(\as) &= 0
\,.\end{align}
%%%
Consistency of the generalized threshold factorization~\cite{Lustermans:2019cau} implies
%%%
\begin{align}
4\gamma_C^i(\as) + \gamma_f^i(\as) + \gamma_J^i(\as) &= 0
\,.\end{align}
%%%
We thus have 3 relations for 5 anomalous dimensions, which means only 2 are independent.
In particular, we have
%%%
\begin{align}
\gamma_S^i(\as) &= -\gamma_{\rm thr}^i(\as) = \gamma_f^i(\as) - \gamma_J^i(\as)
\,.\end{align}
%%%
At three loops, $\gamma_C^i$ and $\gamma_f^i$ have been known
first~\cite{Moch:2004pa, Vogt:2004mw, Moch:2005id, Moch:2005tm},
with $\gamma_J^i$, $\gamma_B^i$, $\gamma_S^i$, $\gamma_{\rm thr}^i$ determined from consistency.
Subsequently, $\gamma_{\rm thr}^i$ and $\gamma_J^q$ have been confirmed by independent
explicit calculations~\cite{Li:2014afw, Bruser:2018rad}.
At four loops, $\gamma_C^i$ and $\gamma_{\rm thr}^i$ are fully known~\cite{vonManteuffel:2020vjv, Agarwal:2021zft, Moch:2017uml, Das:2019btv}, where for the latter the coefficients of
some color structures are only known numerically but with sufficient precision
for practical purposes. We use these to obtain $\gamma_f^i$, $\gamma_J^i$, $\gamma_S^i$
at four loops. In particular, doing so determines the remaining
color coefficients in $\gamma_f^i$ that were only available approximately
in \refscite{Das:2019btv, Das:2020adl}, see also \refcite{Duhr:2022cob}.
To our knowledge, the four-loop $\gamma_J^g$ had not been considered in the literature so far.

For the cusp, soft, and rapidity anomalous dimensions, we do not include the
gluon coefficients up to 3-loop order as they are trivially related to the quark ones by a simple overall
Casimir scaling, $\gamma^g_n = C_A/C_F \gamma^q_n$ for $n \leq 2$.
At 4-loop order, $n = 3$, the quark and gluon coefficients are still related by generalized Casimir scaling,
which however no longer relates the coefficients as a whole, so we include both.

%%%%%%%%%%%%%%%%%%%%%%%%%%%%%%%%%%%%%%%%%%%%%%%%%%%%%%%%%%%%%%%%%%%%%%%%%%%%%%%%
\addcontentsline{toc}{section}{References}
\bibliographystyle{jhep}
\bibliography{scetlib,refs,experiments}

\end{document}